\documentclass[12pt]{article}
\usepackage{epsfig}

\def\r{\vec r}

\newcommand{\n}{\nonumber}
\newcommand{\beq}{\begin{equation}}
\newcommand{\eeq}{\end{equation}}
\newcommand{\bea}{\begin{eqnarray}}
\newcommand{\eea}{\end{eqnarray}}
\newcommand{\la}[1]{\label{#1}}
\newcommand{\ba}{\begin{array}}
\newcommand{\ea}{\end{array}}
\newcommand{\ur}[1]{(\ref{#1})}
\newcommand{\urs}[2]{(\ref{#1},\ref{#2})}
\newcommand{\eq}[1]{eq.~(\ref{#1})}
\newcommand{\eqs}[2]{eqs.(\ref{#1}, \ref{#2})}
\newcommand{\Eq}[1]{Eq.~(\ref{#1})}
\newcommand{\Eqs}[2]{Eqs.(\ref{#1}, \ref{#2})}

\newcommand{\half}{{\textstyle{\frac{1}{2}}}}

\newcommand{\Tr}{{\rm Tr}\,}

\newcommand{\thb}[3]{\bar{\eta}^{#1}_{#2 #3}}
\newcommand{\tha}[3]{\eta^{#1}_{#2 #3}}
\newcommand{\YM}{Yang--Mills~}

\newcommand{\Nz}{$N_{CS}=0\;$}
\newcommand{\No}{$N_{CS}=1\;$}
\newcommand{\x}{{\bf x}}
\newcommand{\IIs}{$I$'s and ${\bar I}$'s }
\newcommand{\II}{$I{\bar I}$ }

\newcommand{\idp}{\int\!\frac{d^4p}{(2\pi)^4}}

\def\Dirac#1{#1\hskip-6pt/}
 \def\Dd{\Dirac\partial}
 \def\Dpp{\Dirac p}
 \def\Dk{k\hskip-6pt/}
 
 \def\Dnabla{\nabla\hskip-8pt/}

\def\ni{\noindent}

\begin{document}

\vspace{1.8cm}

\begin{center}
{\Large\bf Instantons at work}\\

\vskip .6true cm

{\large\bf Dmitri\ Diakonov}$^{1,2}$ \\

\vskip .4true cm

$^1$NORDITA, Denmark\\
$^2$St. Petersburg Nuclear Physics Institute, Russia
\end{center}

\begin{abstract} The aim of this review is to demonstrate that there exists a
coherent picture of strong interactions, based on instantons. Starting from the first
principles of Quantum Chromodynamics -- via the microscopic mechanism
of spontaneous chiral symmetry breaking -- one arrives to a quantitative
description of the properties of light hadrons, with no fitting parameters. 
The discussion of the importance of instanton-induced interactions in
soft high-energy scattering is new.  
\end{abstract}
%\eject
\tableofcontents

\section{Introduction}
\setcounter{equation}{0}
\def\theequation{\arabic{section}.\arabic{equation}}  

Strong interactions as described by Quantum Chromodynamics (QCD)
is a remarkable branch of physics where the observable entities
-- hadrons and nuclei -- are very far from quarks and gluons in terms
of which the theory is formulated. To make matters worse, the
scale of strong interactions 1 fm is nowhere to be found in the QCD
Lagrangian. If we restrict ourselves to hadrons `made of'
$u,d,s$ quarks and glue, the masses of those quarks can be to a good
accuracy set to zero. In this so-called chiral limit the nucleon is
just 5\% lighter than in reality. In the chiral limit there is
not a single dimensional parameter in the QCD Lagrangian. The 1 fm
scale surfaces via a mechanism named the `{\em transmutation of dimensions}'.
QCD is a quantum field theory and beeing such it is not defined
without introducing of some kind of ultra-violet cutoff $\mu$. There
is also a dimensionless gauge coupling constant given at that cutoff
$\alpha_s(\mu)$. The dimensionful quantity $\Lambda$ determining the
scale of the strong interactions is the combination of $\mu$ and
$\alpha_s(\mu)$:
\bea
\la{Lambda}
\Lambda &=&
\mu\,\exp\left(-\frac{2\pi}{b_1\alpha_s(\mu)}\right)\,
\left(\frac{4\pi}{b_1\alpha_s(\mu)}\right)^{\frac{b_2}{2b_1^2}}\,
\left(1+O\left(\alpha_s\right)\right), \\
\n\\
\n\\
\la{b12}
b_1&=&\frac{11}{3}N_c-\frac{2}{3}N_f,\qquad 
b_2=\frac{34}{3}N_c^2-\frac{13}{3}N_cN_f+\frac{N_f}{N_c},
\eea
where $N_c=3$ is the number of quark colours and $N_f$ is the number
of acting quark flavours. The ultra-violet cutoff $\mu$ sets in the dimension
of mass but the exponentially small factor makes $\Lambda$ much less than $\mu$.
To ensure that $\Lambda$ is actually independent of the cutoff, one has to add that
$\alpha_s(\mu)$ has to decrease with $\mu$ according to 

\beq
\frac{2\pi}{\alpha_s(\mu)}=b_1\ln\frac{\mu}{\Lambda}
+\frac{b_2}{2b_1}\ln\ln\frac{\mu^2}{\Lambda^2}
+O\left(\frac{1}{\ln\frac{\mu}{\Lambda}}\right).
\la{alpha_s}\eeq
This formula is called `asymptotic freedom': at large scales $\alpha_s$ decreases. 

All physical observables in strong interactions, like the nucleon mass, the pion decay
constant $F_\pi$, total cross sections, etc. are proportioal to $\Lambda$ in the appropriate
power. That is how the strong interactions scale, 1 fm, appears in the theory. 
One of the theory's goals is to get, say, the nucleon mass in the
form of \eq{Lambda} and to find the numerical proportionality
coefficient.  Doing lattice simulations the first thing one
needs to check is whether an observable scales with $\alpha_s$ as
prescribed by \eq{Lambda}. If it does not, the continuum limit is
not achieved. In analytical approaches, getting an observable in the
form of \eq{Lambda} is extremely difficult. It implies doing
non-perturbative physics. The only analytical approach to QCD I know
of where one indeed gets observables through the transmutation of
dimensions is the approach based on instantons, and it will be the
subject of this paper. 

Instantons are certain large non-perturbative fluctuations of the gluon field 
discovered by Belavin, Polyakov, Schwartz and Tyupkin in 1975 \cite{BPST,Pol}, 
and the name has been suggested in 1976 by 't~Hooft \cite{tH}, who made 
a major contribution to the investigation of the instantons properties. The QCD instanton
vacuum has been studied starting from the pioneering works in the end of the seventies
\cite{CDG,Sh1}; a quantitative treatment of the instanton ensemble has been developed
in refs. \cite{IMP,DP1}. The basic ideas of the instanton vacuum are presented in 
section 2.

Instantons are not the only possible large non-perturbative
fluctuations of the gluon field: one can think also of merons,
monopoles, vortices, etc. I briefly review that in section 3 where also certain 
new material on dyons is presented. However, instantons are the best studied
non-perturbative effects. It may happen that they are not the whole truth
but they are definitely present in the QCD vacuum, and they are
working quite effectively in reproducing many remarkable features
of the strong interactions. For example, instantons lead to the formation of the gluon
condensate~\cite{SVZ} and of the so-called topological susceptibility needed to cure the
$U(1)$ paradox~\cite{tH,Akas}.  The most striking success of instantons
is their capacity to provide a beautiful microscopic mechanism of the spontaneous chiral
symmetry breaking~\cite{DP2a,DP2b,DP3,Pob}. Moreover, instantons enable one
to understand it from different angles and using different mathematical formalisms.
These topics are central in the review and are presented in sections 4,5 and~6.

We know that, were the chiral symmetry of QCD unbroken, the
lightest hadrons would appear in parity doublets. The large actual
splitting between, say, $N(\frac{1}{2}^-,1535)$ and
$N(\frac{1}{2}^+,940)$ implies that chiral symmetry is spontaneously
broken as characterized by the nonzero quark condensate
$<\!\bar\psi\psi\!>\simeq -(250\,{\rm MeV})^3$. Equivalently, it means
that nearly massless (`current') quarks obtain a sizable non-slash
term in the propagator, called the dynamical or constituent mass
$M(p)$, with $M(0)\simeq 350\,{\rm MeV}$. The $\rho$-meson has
roughly twice and nucleon thrice this mass, {\it i.e.} are relatively
loosely bound. The pion is a (pseudo) Goldstone boson and is
very light. The hadron size is typically $\sim 1/M(0)$ whereas the size of 
constituent quarks is given by the slope of $M(p)$. In the instanton approach 
the former is much larger than the latter. It explains, at least on the qualitative
level, why constituent quark models are so phenomenologically
successful. 

It should be stressed that literally speaking instantons do not lead
to confinement, although they induce a growing potential for heavy
quarks at intermediate separations \cite{DPP1}; asymptotically it
flattens out \cite{CDG}. However, it has been realized long ago
\cite{Sh1,D1}, that it is chiral symmetry breaking and not
confinement that determines the basic characteristics of 
nucleons and pions as well as their first excitations. After all, 99\% of the mass
around us is due to the spontaneous generation of the quark constituent 
mass. Probably one would need an explicit confinement to get the properties 
of short-living highly excited hadrons. According to a popular wisdom, moving 
a quark away from a diquark system in a baryon generates a string, also 
called a flux tube, whose energy rises linearly with the separation. 
This is expected in the ``pure-glue'' world with no dynamical quarks.
However, in the real world with light quarks and the spontaneous chiral symmetry 
breaking the string energy exceeds the pion mass $m_\pi=140\,{\rm MeV}$ at a modest
separation of about $0.26\,{\rm fm}$, see Fig. 1. At larger separations the would-be linear
potential is screened since it is energetically favourable to tear the string and 
produce a pion. Virtually, the linear potential can stretch to as much as
$0.4\,{\rm fm}$ where its energy exceeds $2m_\pi$ but that can
happen only for a short time of $1/m_\pi$. Meanwhile, the
ground-state baryons are stable, and their sizes are about
$1\,{\rm fm}$. The pion-nucleon coupling is huge, and there seems to be no
suppression of the string breaking by pions. The paradox is that the
linear potential of the pure glue world, important as it might be to
explain why quarks are not observed as a matter of principle, can
hardly have a direct impact on the properties of lightest hadrons.

%%%%%%%%%%%%%%
%% FIGURE 1 %%
%%%%%%%%%%%%%%
\begin{figure}[t]
%\hskip 2true cm
\epsfig{file=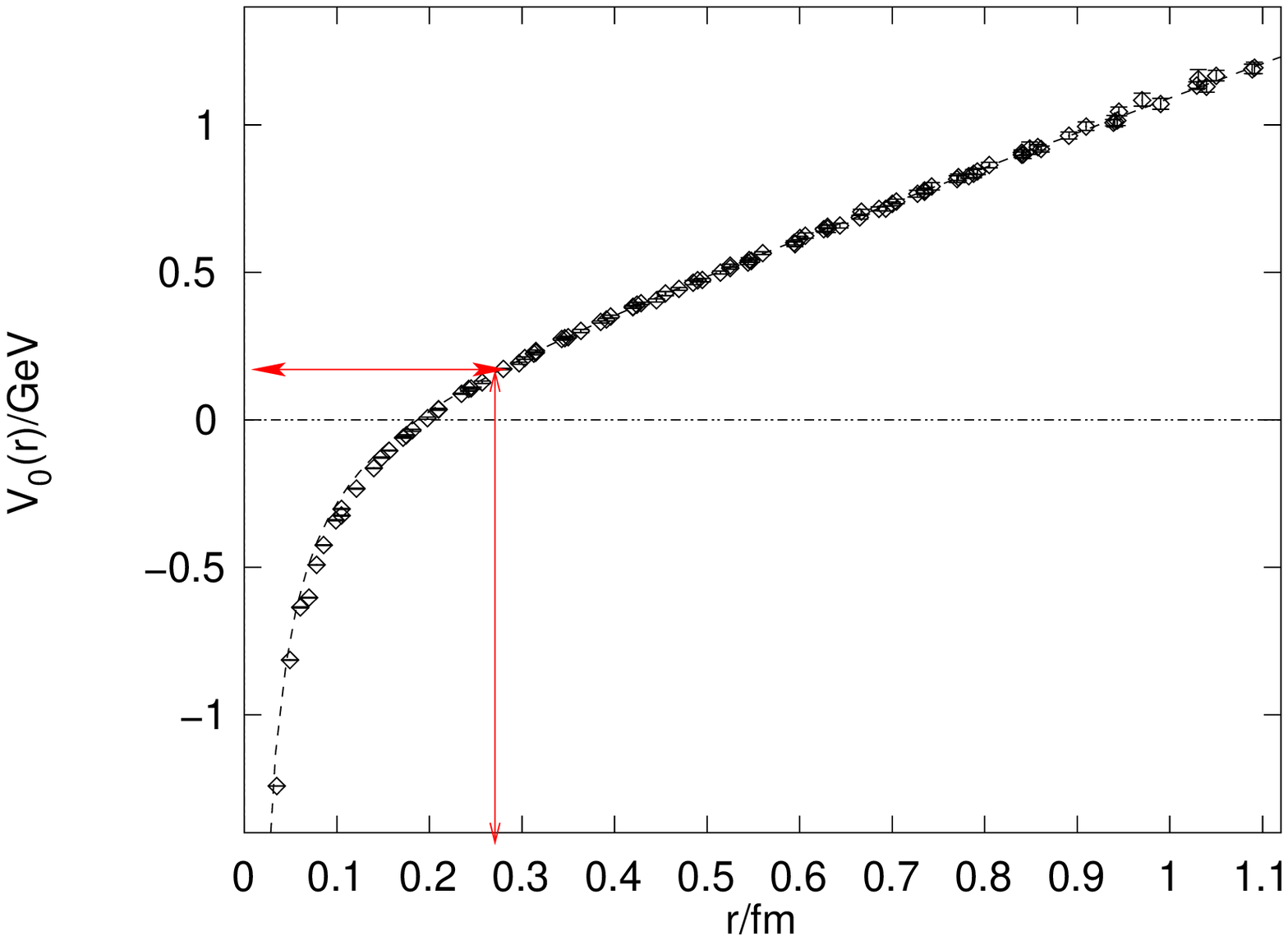,width=6.5cm}
\hspace{.5cm}\epsfig{file=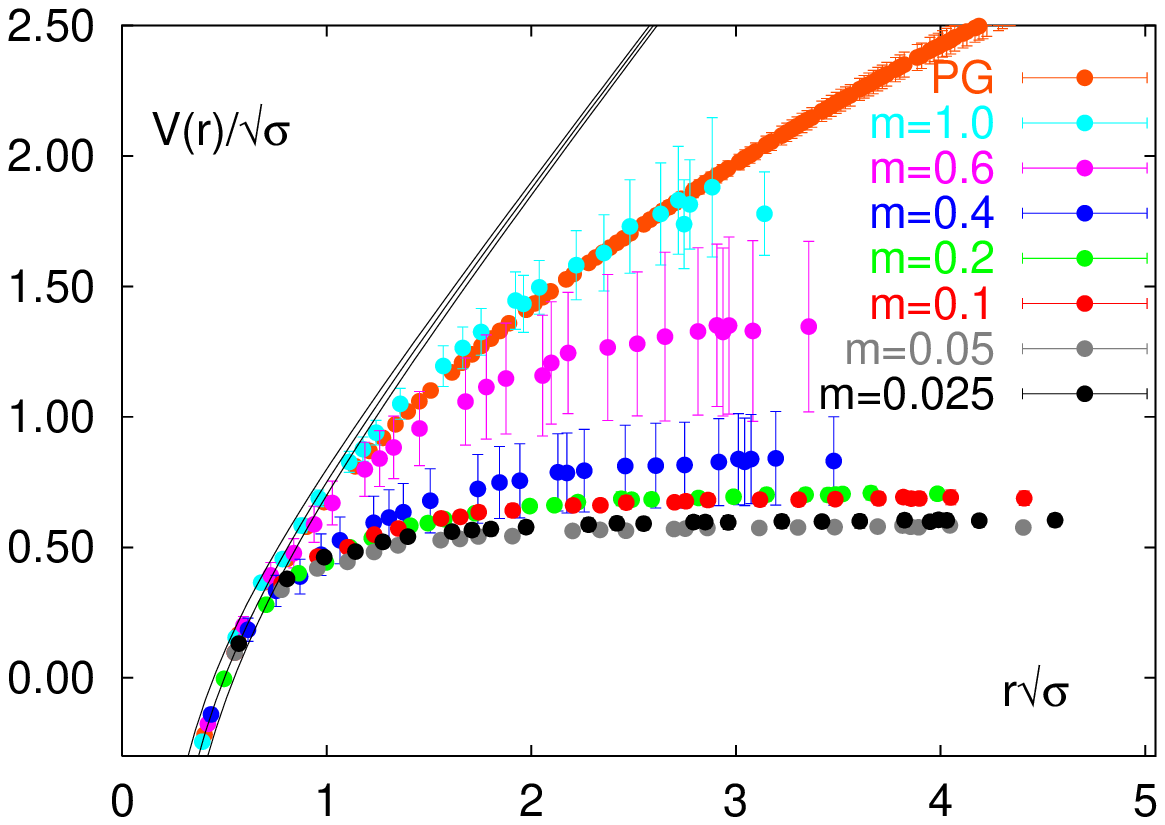,width=7cm}
\vspace{.5cm}
\caption{The lattice-simulated potential between static
quarks in pure glue theory \cite{BSW} exceeds $m_\pi$ at the separation of 
$0.26\,{\rm fm}$ (left). The screening of the linear potential by dynamical quarks
is clearly seen in simulations at high temperatures but below the phase transition 
\cite{KLP} (right). As one lowers the pion mass the string breaking happens at smaller
distances; the scale is $\sqrt{\sigma}\simeq 425\,{\rm MeV}\simeq (0.47\,{\rm fm})^{-1}$.
\label{scrpot}}
\vspace{.01cm}
\end{figure}

Even for highly excited hadrons lying on linear Regge trajectories the situation is not
altogether clear. The usual explanation of resonances lying on linear trajectories 
is that they are rotating confining flux tubes attached to quarks at end points moving with
the speed of light. Why should $350\,{\rm MeV}$ quarks bound by a 
$\sqrt{\sigma}\simeq 425\,{\rm MeV}$ string move with the speed of light is not clear,
but if they do not, the trajectories are not linear. In this picture, the finite (and experimentally
large) width of resonances is due to the same string breaking by meson production. However,
if there is no string in the ground-state nucleon, why should it be excited in a collision? The
lightest degrees of freedom in the real world are pions and one might expect that they are the
first to be excited. An alternative explanation of resonances lying on linear trajectories is that
they are rotating elongated lumps of pion field \cite{DPRegge}, and their decay is due to the
normal pion radiation. It follows then that the dominant decay of a large-J baryon resonance
is a cascade ${\rm Bar}_{J}\to {\rm Bar}_{J-1}+\pi \to {\rm Bar}_{J-2}+\pi\pi\to ...$ 
whereas if it is due to the string breaking it rather has a different pattern 
${\rm Bar}_{J}\to {\rm Bar}_{\sim J/2} +{\rm Mes}_{\sim J/2}$. Studying the decay
patterns of high-J resonances could be illuminating for understanding the relation 
between confining and chiral forces.

Leaving aside the unsettled question of highly excited resonances, 
the situation with the lightest and most important hadrons $\pi, \rho, N, \Delta$...
is, to my mind, clear: it is the spontaneous chiral symmetry breaking (SCSB) rather than the
expected linear confining potential of the pure glue world which is the key to understanding
of their properties. Therefore, since the instanton vacuum describes successfully
the physics of the chiral symmetry breaking, one would expect that instantons do explain the
properties of light hadrons, both mesons and baryons. Indeed, a detailed numerical study of
dozens of correlation functions with different quantum numbers in the instanton medium 
undertaken by Shuryak, Verbaarschot and Sch\"{a}fer~\cite{Sh2} 
(earlier certain correlation functions were computed analytically in refs.~\cite{DP2b,DP3})
demonstrated an impressing agreement with the phenomenology~\cite{Sh3} and with direct
lattice measurements~\cite{CGHN}, see ref.~\cite{Sh4} for a review. In fact, instantons induce
strong interactions between quarks, leading to bound-state baryons with calculable and
reasonable properties. There are specialized reviews on this subject, therefore I touch it only
briefly here (sections 8 and 9).

More recently, there has been much activity in applying instantons to explain
various phenomena in high energy processes including heavy ion collisions. 
For that reason, I have included section 7 which suggests a new point of view on the 
pomeron which might be also related to instantons.  

\vskip 1true cm

\section{What are instantons?}
\setcounter{equation}{0}
\def\theequation{\arabic{section}.\arabic{equation}}  
\subsection{\it Periodicity of the Yang--Mills potential energy}

Being a quantum field theory, QCD deals with the fluctuating gluon and quark
fields. A fundamental fact \cite{Fad,JR} is that the potential energy
of the gluon field is a periodic function in one particular direction
in the infinite-dimensional functional space; in all other directions the
potential energy is oscillator-like. This is illustrated in Fig. 2. 

To observe this periodicity, let us temporarily  work in the
$A_0^a=0$ gauge, called Weyl or Hamiltonian gauge, and forget about
fermions for a while. The remaining pure \YM or ``pure glue" theory
is nonetheless non-trivial, since gluons are self-interacting. For
simplicity I start from the $SU(2)$ gauge group.

The spatial YM potentials $A_i^a(\x, t)$ can be considered
as an infinite set of the coordinates of the system, where $i=1,2,3,\;\;
a=1,2,3$ and $\x$ are ``labels" denoting various coordinates. The
YM action is

\beq
S=\frac{1}{4g^2}\int\! d^4x\; F_{\mu\nu}^aF_{\mu\nu}^a=\int dt
\left(\frac{1}{2g^2}\int\! d^3\x\;{\bf E}^2 - \frac{1}{2g^2}\int\!d^3\x
\;{\bf B}^2 \right)
\la{YMA}\eeq
where ${\bf E}$ is the electric field stregth,

\beq
E_i^a(\x,t)={\dot A}_i^a(\x,t)
\la{E}\eeq
(the dot stands for the time derivative), and ${\bf B}$ is the magnetic
field stregth,

\beq
B_i^a(\x,t)=\frac{1}{2}\epsilon_{ijk}\left(\partial_jA_k^a-
\partial_kA_j^a+\epsilon^{abc}A_j^bA_k^c\right).
\la{B}\eeq

Apparently, the first term in \eq{YMA} is the kinetic energy of the
system of coordinates $\{A_i^a(\x,t)\}$ while the second term is minus
the potential energy being just the magnetic energy of the field.
The simple and transparent form of \eq{E} is the advantage of the Weyl
gauge. Upon quantization the electric field is replaced by the
variational derivative, 
$E_i^a(x) \rightarrow -ig^2\delta/\delta A_i^a(x)$, 
if one uses the `coordinate representation' for the wave 
functional. The functional Schr\"odinger equation for the wave 
functional $\Psi[A_i^a(x)]$ takes the form

\beq
{\cal H}\Psi[A_i]=
\int \!d^3x\left\{-\frac{g^2}{2}\frac{\delta^2}{(\delta A_i^a(x))^2}
+\frac{1}{2g^2}(B_i^a(x))^2\right\}\Psi[A_i]
={\cal E}\Psi[A_i]
\la{schr1}\eeq
where ${\cal E}$ is the eigenenergy of the state in question. The YM
vacuum is the ground state of the Hamiltonian \ur{schr1}, 
corresponding to the lowest energy ${\cal E}$. \\

Let us introduce an important quantity called the Pontryagin index or
the four-dimensional topological charge of the YM fields:

\beq
Q_T=\frac{1}{32\pi^2}\int d^4x\; F_{\mu\nu}^a {\tilde F}_{\mu\nu}^a,
\qquad {\tilde F}_{\mu\nu}^a \equiv \frac{1}{2}
\epsilon_{\mu\nu\alpha\beta}F_{\alpha\beta}^a.
\la{FFd}\eeq
The integrand in \eq{FFd} happens to be a full derivative of the four-vector $K_\mu$:

\beq
\frac{1}{32\pi^2} F_{\mu\nu}^a {\tilde F}_{\mu\nu}^a =\partial_\mu
K_\mu,\qquad
K_\mu=\frac{1}{16\pi^2}\epsilon_{\mu\alpha\beta\gamma}
\left(A_\alpha^a\partial_\beta A_\gamma^a+\frac{1}{3}\epsilon^{abc}
A_\alpha^a A_\beta^b A_\gamma^c\right).
\la{K}\eeq
Therefore, assuming the fields $A_\mu$ are decreasing rapidly enough at
spatial infinity, one can rewrite the 4-dimensional topological charge
\ur{FFd} as

\beq
Q_T=\int d^4x (\partial_0 K_0-\partial_i K_i)=\int dt \frac{d}{dt}
\int d^3\x K_0.
\la{Gauss}\eeq
Introducing the {\it Chern--Simons number}

\beq
N_{CS}=\int d^3\x\;K_0=\frac{1}{16\pi^2}\int d^3\x\:\epsilon^{ijk}
\left(A_i^a\partial_j A_k^a+\frac{1}{3}\epsilon^{abc}
A_i^a A_j^b A_k^c\right)
\la{NCS}\eeq
we see from \eq{Gauss} that $Q_T$ can be rewritten as the difference
of the Chern--Simons numbers characterizing the fields at $t=\pm\infty$:

\beq
Q_T=N_{CS}(+\infty)-N_{CS}(-\infty).
\la{dif}\eeq

The Chern--Simons number of the field has an important property
that it can change by integers under large gauge transformations. Indeed,
under a general time-independent gauge transformation,

\beq
A_i \rightarrow U^\dagger A_iU + iU^\dagger\partial_iU, \;\;\;
A_i\equiv A_i^a\frac{\tau^a}{2},
\la{gt}\eeq
the Chern--Simons number transforms as follows:

\beq
N_{CS}\rightarrow N_{CS}+N_W
+\frac{i}{8\pi^2}\int\!d^3x\,\epsilon^{ijk}\partial_j\,\Tr(\partial_i UU^\dagger A_k).
\la{CSt}\eeq
The last term is a full derivative and can be omitted if, {\it e.g.}, $A_i$ decreases
sufficiently  fast  at spatial infinity. $N_W$ is the winding number of the gauge 
transformation \ur{gt}:

\beq
N_W=\frac{1}{24\pi^2}\int d^3\x\; \epsilon^{ijk}\left[(U^\dagger\partial_iU)
(U^\dagger\partial_jU) (U^\dagger\partial_kU)\right].
\la{wn}\eeq

The $SU(2)$ unitary matrix $U$ of the gauge transformation \ur{gt} 
can be viewed as a mapping from the 3-dimensional space onto the 
3-dimensional sphere of parameters $S^3$. If at spatial infinity we 
wish to have the same matrix $U$ independently of the way we approach 
the infinity (and this is what is usually assumed), then the spatial 
infinity is in fact one point, so the mapping is topologically 
equivalent to that from $S^3$ to $S^3$. This mapping is known to be
non-trivial, meaning that mappings with different winding numbers
are irreducible by smooth transformations to one another. The winding
number of the gauge transformation is, analytically, given by \eq{wn}.
As it is common for topological characteristics, the integrand in \ur{wn}
is in fact a full derivative. For example, if we take the matrix $U(\x)$
in a ``hedgehog" form, $U=\exp[i(r\cdot \tau)/r\, P(r)]$, \eq{wn} 
can be rewritten as

\beq
N_W=\frac{2}{\pi}\int\!dr  \frac{dP}{dr}\sin^2 P = \frac{1}{\pi}
\left[P-\frac{\sin 2P}{2}\right]_0^\infty = \mbox{integer}
\la{wnh}\eeq
since $P(r)$ both at zero and at infinity needs to be multiples of $\pi$
if we wish $U(\r)$ to be unambigiously defined at the origin and
at the infinity.

Let us return now to the potential energy of the YM fields,

\beq
{\cal V}=\frac{1}{2g^2} \int d^3\x \left(B_i^a\right)^2.
\la{potene}\eeq

One can imagine plotting the potential energy surfaces over the
Hilbert space of the coordinates $A_i^a(\x)$. It will be some complicated
mountain country. If the field happens to be a pure gauge, $A_i= 
iU^\dagger\partial_i U$, the potential energy  at such points of the 
Hilbert space is naturally zero. Imagine that we move along the 
``generalized coordinate" being the Chern--Simons number \ur{NCS}, 
fixing all other coordinates whatever they are. Let us take some 
point $A_i^a(\x)$ with the potential energy ${\cal V}$. If we move to 
another point which is a gauge transformation of $A_i^a(\x)$ with a 
winding number $N_W$, its potential energy will be exactly the same as
it is strictly gauge invariant.  However the Chern--Simons 
``coordinate" of the new point will be shifted by an integer $N_W$ 
from the original one. We arrive to the conclusion first pointed out 
by Faddeev \cite{Fad} and Jackiw and Rebbi \cite{JR} in 1976, that 
the potential energy of the YM fields is {\em periodic} in the 
particular coordinate called the Chern--Simons number.

%%%%%%%%%%%%%%
%% FIGURE 2 %%
%%%%%%%%%%%%%%
\begin{figure}[t]
\hskip 3true cm
\epsfig{file=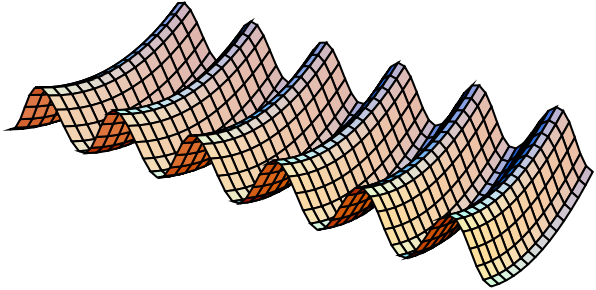,width=8cm,height=5cm}
% \vspace{-.4cm}
\caption{Potential energy of the gluon field is periodic in one direction
and oscillator-like in all other directions in functional space.
\label{periodic}}
\vspace{.4cm}
\end{figure}

\subsection{\it Instantons in simple words}

In perturbation theory one deals with zero-point quantum-mechanical
fluctuations of the YM fields near one of the minima, say, at \Nz.
The non-linearity of the YM theory is taken into account as a
perturbation, and results in series in $g^2$ where $g$ is the gauge
coupling.  In that approach one is apparently missing a possibility for
the system to tunnel to another minimum, say, at \No. The tunneling is
a typical non-perturbative effect in the coupling constant.

Instanton is a large fluctuation of the gluon field in imaginary
(or Euclidean) time corresponding to quantum tunneling from one minimum
of the potential energy to the neighbour one. Mathematically, it was
discovered by Belavin, Polyakov, Schwarz and Tyupkin; \cite{BPST}
the tunneling interpretation was given by V. Gribov, see \cite{Pol}.
The name `instanton' has been introduced by 't~Hooft~\cite{tH} who
studied many of the key properties of those fluctuations. Anti-instantons
are similar fluctuations but tunneling in the opposite direction
in Fig. 2. Physically, one can think of instantons in two ways: on the one hand it is
a tunneling {\em process} occuring in time, on the other hand it is a
localized {\em pseudoparticle} in the Euclidean space.

Following the WKB approximation, the tunneling amplitude can be estimated as $\exp(-S)$, 
where $S$ is the action along the classical trajectory in imaginary time, leading from
the minimum at \Nz at $t=-\infty$ to that at \No at $t=+\infty$.
%\cite{LL}. 
According to \eq{dif} the 4-dimensional topological charge
of such trajectory is $Q_T=1$. To find the best tunneling trajectory
having the largest amplitude one has thus to minimize the YM action
\ur{YMA} provided the topological charge \ur{FFd} is fixed to be unity.
This can be done using the following trick \cite{BPST}. Consider
the inequality

\[
0\leq \int d^4x \left(F_{\mu\nu}^a-{\tilde F}_{\mu\nu}^a\right)^2
\]
\beq
=\int d^4 x \left(2 F^2-2F{\tilde F}\right) =
8g^2S-64\pi^2Q_T,
\la{inequ}\eeq
hence the action is restricted from below:

\beq
S\geq \frac{8\pi^2}{g^2}Q_T =  \frac{8\pi^2}{g^2}.
\la{inequa}\eeq

Therefore, the minimal action for a trajectory with a unity topological
charge is equal to $8\pi^2/g^2$, which is achieved if the trajectory
satisfies the {\em self-duality} equation:

\beq
F_{\mu\nu}^a={\tilde F}_{\mu\nu}^a.
\la{selfdual}\eeq

Notice that any solution of \eq{selfdual} is simultaneously a solution of
the general YM equation of motion $D_\mu^{ab}F_{\mu\nu}^b=0$: that is
because the ``second pair" of the Maxwell equations,
$D_\mu^{ab}{\tilde F}_{\mu\nu}^b=0$, is satisfied identically.

Thus, the tunneling amplitude can be estimated as 
\beq
{\cal A} \sim \exp(-{\rm Action})=\exp\left(-\frac{1}{4g^2}\int\!d^4x\,F_{\mu\nu}^2\right)
=\exp\left(-\frac{8\pi^2}{g^2}\right)=\exp\left(-\frac{2\pi}{\alpha_s}\right).
\la{ta}\eeq
It is non-analytic in the gauge coupling constant and hence instantons are
missed in all orders of the perturbation theory. However, it is not
a reason to ignore tunneling. For example, tunneling of electrons from one
atom to another in a metal is also a non-perturbative effect but we would get
nowhere in understanding metals had we ignored it.

\subsection{\it Instanton configurations}

To solve \eq{selfdual} let us recall a few facts about the Lorentz group
$SO(3,1)$. Since we are talking about the tunneling fields which can only
develop in imaginary time, it means that we have to consider the fields
in Euclidean space-time, so that the Lorentz group is just $SO(4)$ 
isomorphic to $SU(2)\times SU(2)$. The gauge potentials $A_\mu$ belong to the
$(\frac{1}{2},\frac{1}{2})$ representation of the $SU(2)\times SU(2)$
group, while the field strength $F_{\mu\nu}$ belongs to the reducible $(1,0)+(0,1)$
representation. In other words it means that one linear combination
of $F_{\mu\nu}$ transforms as a vector of the left $SU(2)$, and another
combination transforms as a vector of the right $SU(2)$. These
combinations are

\beq
F_L^A=\eta_{\mu\nu}^A(F_{\mu\nu}+\tilde{F}_{\mu\nu}),\qquad
F_R^A=\bar{\eta}_{\mu\nu}^A(F_{\mu\nu}-\tilde{F}_{\mu\nu}),
\la{oneone}\eeq
where $\eta, \bar{\eta}$ are the so-called 't Hooft symbols described
in ref. \cite{tH}, see also below. We see therefore that a self-dual
field strength is a vector of the left $SU(2)$ while its right part is
zero.  Keeping that experience in mind we look for the solution of the
self-dual equation in the form

\beq
A_\mu^a=\thb{a}{\mu}{\nu}\, x_\nu\,\frac{1+\Phi(x^2)}{x^2}.
\la{tHanz}\eeq
Using the formulae for the $\eta$ symbols from ref. \cite{tH} one can
easily check that the YM action can be rewritten as

\beq
S=\frac{8\pi^2}{g^2}\frac{3}{2} \int d\tau
\left[\frac{1}{2}\left(\frac{d\Phi}{d\tau}\right)^2+
\frac{1}{8}(\Phi^2-1)^2\right], \qquad
\tau = \ln\left(\frac{x^2}{\rho^2}\right).
\la{doublewell}\eeq
This can be recognized as the action of the double-well potential whose
minima lie at $\Phi=\pm 1$, and $\tau$ plays the role of ``time"; $\rho$
is an arbitrary scale. The trajectory which tunnels from $1$ at
$\tau=-\infty$ to $-1$ at $\tau = +\infty$ is

\beq
\Phi=-\tanh\left(\frac{\tau}{2}\right),
\la{QMinst}\eeq
and its action \ur{doublewell} is $S=8\pi^2/g^2$, as needed. Substituting
the solution \ur{QMinst} into \ur{tHanz} we get

\beq
A_\mu^a(x)=\frac{2\thb{\mu}{\nu}{a}\rho^2}{x^2(x^2+\rho^2)}.
\la{YMinst}\eeq
The correspondent field strength is

\beq
F_{\mu\nu}^a=-\frac{4\rho^2}{(x^2+\rho^2)^2}\left(\thb{\mu}{\nu}{a}-
2\thb{\mu}{\alpha}{a}\frac{x_\alpha x_\nu}{x^2}-
2\thb{\beta}{\nu}{a}\frac{x_\mu x_\beta}{x^2}\right),\qquad
F_{\mu\nu}^aF_{\mu\nu}^a=\frac{192\rho^4}{(x^2+\rho^2)^4},
\la{fstrength}\eeq
and satisfies the self-duality condition \ur{selfdual}.

The {\em anti-instanton} corresponding to tunneling in the opposite
direct\-ion, from \No to \Nz, satis\-fies the {\em anti}-self-dual
equation, with $\tilde{F}\to -\tilde{F}$; its concrete form is
given by \eqs{YMinst}{fstrength} with the replacement
$\bar{\eta}\to \eta$.

\Eqs{YMinst}{fstrength} describe the field of the instanton in the
singular Lorenz gauge; the singularity of $A_\mu$ at $x^2=0$ is a gauge
artifact: the gauge-invariant field strength squared is smooth at the
origin. Formulae for instantons are more compact in the Lorenz gauge,
and I shall use it further on \footnote{Jackson and Okun \cite{JO} recommend
to call the $\partial_\mu A_\mu=0$ gauge by the name of the Dane Ludvig Lorenz
and not the Dutchman Hendrik Lorentz who certainly used this gauge too 
but several decades later.}.

\subsection{\it Instanton collective coordinates}

The instanton field, \eq{YMinst}, depends on an arbitrary scale
parameter $\rho$ which we shall call the instanton size, while the
action, being scale invariant, is independent of $\rho$. One can
obviously shift the position of the instanton to an arbitrary 4-point
$z_\mu$ -- the action will not change either. Finally, one can rotate
the instanton field in colour space by constant unitary matrices $U$.
For the $SU(2)$ gauge group this rotation is characterized by 3
parameters, {\it e.g.} by Euler angles. For a general $SU(N_c)$ group
the number of parameters is $N_c^2-1$ (the total number of the
$SU(N_c)$ generators) {\em minus} $(N_c-2)^2$ (the number of generators
which do not affect the left upper $2\times 2$ corner where the
standard $SU(2)$ instanton \ur{YMinst} is residing), that is $4N_c-5$.
These degrees of freedom are called instanton orientation in colour
space.  All in all there are

\beq
4\; {\mbox (centre)}\;\;+\;\;1\; {\mbox (size)}\;\;+\;\;
(4N_c-5)\; {\mbox (orientations)}\;\;=\;\;4N_c
\la{collcoo}\eeq
so-called collective coordinates desribing the field of the instanton,
of which the action is independent.

It is convenient to indroduce $2\times 2$ matrices

\beq
\sigma^{\pm}_\mu = (\pm i \overrightarrow{\sigma}, 1),\qquad
x^{\pm}=x_\mu\sigma^{\pm}_\mu,
\la{sigma}\eeq
such that

\beq
2i\tau^a
\tha{\mu}{\nu}{a}=\sigma^+_\mu\sigma^-_\nu
-\sigma^+_\nu\sigma^-_\mu, \qquad
2i\tau^a
\thb{\mu}{\nu}{a}=\sigma^-_\mu\sigma^+_\nu
-\sigma^-_\nu\sigma^+_\mu,
\la{thsig}\eeq
then the instanton field with arbitrary center $z_\mu$, size $\rho$ and
colour orientation $U$ in the $SU(N_c)$ gauge group can be written as

\beq
A_\mu=A_\mu^at^a
=\frac{-i\rho^2U[\sigma^-_\mu(x-z)^+-(x-z)_\mu]U^\dagger}
{(x-z)^2[\rho^2+(x-z)^2]},\qquad \Tr(t^at^b)=\frac{1}{2}\delta^{ab},
\la{instgen}\eeq
or as

\beq
A_\mu^a=\frac{2\rho^2O^{ab}\thb{\mu}{\nu}{b}(x-z)_\mu}
{(x-z)^2[\rho^2+(x-z)^2]},\qquad
O^{ab}=\Tr(U^\dagger t^aU\sigma^b),\qquad O^{ab}O^{ac}=\delta^{bc}.
\la{instgena}\eeq
This is the explicit expression for the $4N_c$-parameter instanton field 
in the $SU(N_c)$ gauge theory, written down in the singular Lorenz gauge. 

\subsection{\it Gluon condensate}

The QCD perturbation theory implies that the fields $A_i^a(\x)$ are performing quantum
zero-point oscillations; in the lowest order these are just plane waves with arbitrary
frequences. The aggregate energy of these zero-point oscillations, $({\bf B}^2+{\bf E}^2)/2$,
is divergent as the fourth power of the cutoff frequency, however for any state one has
$\langle F_{\mu\nu}^2\rangle = 2\langle{\bf B}^2-{\bf E}^2\rangle = 0$, which is just a
manifestation of the virial theorem for harmonic oscillators:  the average potential energy is
equal the kinetic one (I am temporarily in the Minkowski space). One can prove that this
is also true in any order of the perturbation theory in the coupling constant, provided one
does not violate the Lorentz symmetry and the renormalization properties of the theory.
Meanwhile, we know from the QCD sum rules phenomenology that the QCD vacuum
posseses what is called {\em gluon condensate} \cite{SVZ}:  

\beq 
\frac{1}{32\pi^2}\langle F_{\mu\nu}^aF_{\mu\nu}^a\rangle = \frac{1}{16\pi^2}
\langle{\bf B}^2-{\bf E}^2\rangle \simeq (200\; MeV)^4 \;\;>\;0. 
\la{glcond}\eeq  
Instantons suggest an immediate explanation of this basic property of QCD. 
Indeed, instanton is a tunneling process, it occurs in imaginary time; therefore in Minkowski
space one has $E_i^a=\pm iB_i^a$ (this is actually the duality \eq{selfdual}). Therefore, 
during the tunneling ${\bf B}^2-{\bf E}^2$ is positive, and one gets a chance 
to explain the gluon condensate. In Euclidean space the electric field is real as well as the
magnetic one, and the gluon condensate is just the average action density. Let us make a
quick estimate of its value.  Let the total number of instantons and anti-instantons
(henceforth \IIs for short) in the 4-dimensional volume $V$ be $N$. Assuming that the
average separations of instantons are larger than their average sizes (to be justified below),
we can estimate the total action of the ensemble as the sum of invidual actions (see
\eq{inequa}):  

\beq 
\langle F_{\mu\nu}^2\rangle V =\int d^4x F_{\mu\nu}^2 \simeq N\cdot 32\pi^2, 
\la{totact}\eeq 
hence the gluon condensate is directly related to the instanton density in the 4-dimensional
Euclidean space-time:  

\beq 
\frac{1}{32\pi^2}\langle F_{\mu\nu}^aF_{\mu\nu}^a\rangle \simeq \frac{N}{V}
\equiv \frac{1}{{\bar R}^4}. 
\la{glcondest}\eeq In order to get the phenomenological value of the condensate one 
needs thus to have the average separation between pseudoparticles 
\cite{SVZ,Sh1}  

\beq 
{\bar R}\simeq\frac{1}{200\,{\rm MeV}}=1\,{\rm fm}. 
\la{avsep}\eeq  

There is another point of view on the gluon condensate which I describe briefly. In principle,
all information about field theory is contained in the partition function being the functional
integral over the fields. In the Euclidean formulation it is  

\beq 
{\cal Z}=\int DA_\mu \exp\left(-\frac{1}{4g^2}\int d^4x F_{\mu\nu}^2\right)
\stackrel{T\rightarrow\infty}{\longrightarrow} e^{-{\cal E}T}, 
\la{partfu}\eeq 
where I have used that at large (Euclidean) time $T$ the partition function picks up the
ground state or vacuum energy ${\cal E}$. For the sake of brevity I do not write the gauge
fixing and Faddeev--Popov ghost terms. If the state is homogeneous, the energy can be
written as ${\cal E}=\theta_{44}V^{(3)}$ where $\theta_{\mu\nu}$ is the stress-energy tensor
and $V^{(3)}$ is the 3-volume of the system. Hence, at large 4-volumes $V=V^{(3)}T$ the
partition function is ${\cal Z}=\exp(-\theta_{44}V)$. This $\theta_{44}$ includes zero-point
oscillations and diverges badly. A more reasonable quantity is the partition function,
normalized to the partition function understood as a perturbative expansion about the
zero-field vacuum\footnote{The latter can be distinguished from the former by imposing a
condition that it does not contain integration over singular Yang--Mills potentials; recall that
the instanton potentials are singular at the origins.},  

\beq 
\frac{{\cal Z}}{{\cal Z}_{\rm P.T.}}= \exp\left[-(\theta_{44} -\theta_{44}^{\rm P.T.})V\right]. 
\la{pfn}\eeq  
We expect that the non-perturbative vacuum energy density 
$\theta_{44}-\theta_{44}^{\rm P.T.}$ is a negative quantity since we have allowed for
tunneling: as usual in quantum mechanics, it lowers the ground state energy. If the vacuum is
isotropic, one has $\theta_{44}=\theta_{\mu\mu}/4$. Using the trace anomaly,  

\beq 
\theta_{\mu\mu}=\frac{\beta(\alpha_s)}{16\pi\alpha_s^2} \left(F_{\mu\nu}^a\right)^2
\simeq -b\frac{F_{\mu\nu}^2}{32\pi^2}, 
\la{TA}\eeq 
where $\beta(\alpha_s)$ is the Gell-Mann--Low function,  

\beq 
\beta(\alpha_s) \equiv \frac{d\alpha_s(\mu)}{d\ln \mu}
=-b_1\frac{\alpha_s^2(\mu)}{2\pi}-\frac{b_2}{2}\frac{\alpha_s^3(\mu)}{(2\pi)^2}-...,
\la{GML}\eeq 
with $b_{1,2}$ given by \eq{b12}, one gets \cite{DP1}:  

\beq 
\frac{{\cal Z}}{{\cal Z}_{\rm P.T.}}
=\exp\left( \frac{b}{4}V \langle F_{\mu\nu}^2/32\pi^2\rangle_{\rm NP}\right) 
\la{GCdef}\eeq 
where $\langle F_{\mu\nu}^2\rangle_{\rm NP}$ is the gluon field vacuum expectation value
which is due to non-perturbative fluctuations, i.e. the gluon condensate. The aim of any
QCD-vacuum builder is to minimize the vacuum energy or, equivalently, to maximize the
gluon condensate.  It is important that it is a renormalization-invariant quantity,  
meaning that its dependence on the ultraviolet cutoff $\mu$ and the bare charge
$\alpha_s(\mu)$ given at this cutoff is such that it is actually cutoff-independent. Such a
combination is called $\Lambda$, see \eq{Lambda}. The gluon condensate has
to be proportional to $\Lambda^4$ by dimensions.   

The fact that the vacuum energy or, equivalently, the gluon condensate is a
re\-normaliz\-ation-invari\-ant quant\-ity leads to an infinite number of low-energy
theorems \cite{NSVZ}. Translated into the instanton-vacuum language, the
renormalizability of the QCD implies that the probability that there are $N$
\IIs in the vacuum is \cite{DP1,DPW}

\beq 
P(N) \sim \exp\left[-\frac{b}{4}\left(\ln\frac{N}{\langle N\rangle} -1\right)\right],
\la{DN}\eeq 
where $\langle N\rangle\simeq V\langle F_{\mu\nu}^aF_{\mu\nu}^a\rangle/(32\pi^2)$
is the {\em average} number of \IIs.  

\subsection{\it One-instanton weight}

The notion ``instanton vacuum" implies that one assumes that the QCD partition function 
\ur{partfu} is mainly saturated by an ensemble of interacting \IIs, together with quantum
fluctuations about them. Instantons are necessarily present in the QCD vacuum if only
because they lower the vacuum energy with respect to the purely perturbative (divergent)
one. The question is whether they give the dominant contribution to the gluon condensate,
and to other basic quantities. To answer this question one has to compute the partition
function \ur{partfu} assuming that it is mainly saturated by instantons, and to compare the
obtained gluon condensate with the phenomenological one.

The starting point of this calculation \cite{DP1,DPW} is the contribution of one isolated
instanton to the partition function \ur{partfu}, or the one-instanton weight. We have already
estimated the tunneling amplitude in \eq{ta} but it is not sufficient: the prefactor is very
important. To the 1-loop accuracy, it has been first computed by 't Hooft \cite{tH} for the
$SU(2)$ colour group, and generalized to arbitrary $SU(N)$ by Bernard \cite{B}.  

The general field can be decomposed as a sum of a classical field of an instanton
$A_\mu^I(x,\xi)$ where $\xi$ is a set of $4N_c$ collective coordinates characterizing 
a given instanton (see \eq{instgen}), and of a presumably small quantum field $a_\mu(x)$:  

\beq 
A_\mu(x)=A_\mu^I(x,\xi) + a_\mu(x). 
\la{genfi}\eeq 
There is a subtlety in this decomposition due to the gauge freedom: an interested reader is
addressed to ref. \cite{DP1} where this subtlety is treated in detail.  The action is  

\beq
{\rm Action}=\frac{1}{4g^2}\int d^4x\; F_{\mu\nu}^2=\frac{8\pi^2}{g^2} 
+\frac{1}{g^2}\int d^4x\; D_\mu F_{\mu\nu}a_\nu 
+\frac{1}{2g^2}\int d^4x\; a_\mu W_{\mu\nu}a_\nu + O(a^3). 
\la{quaform}\eeq  
Here the term linear in $a_\mu$ drops out because the instanton field satisfies the equation 
of motion. The quadratic form $W_{\mu\nu}$ has $4N_c$ zero modes related to the fact that
the action does not depend on $4N_c$ collective coordinates. This brings in a divergence 
in the functional integral over the quantum field $a_\mu$ which, however, can and should 
be qualified as integrals over the collective coordinates: centre, size and orientations.
Formally the functional integral over $a_\mu$ gives  

\beq 
\frac{1}{\sqrt{\det\;W_{\mu\nu}(A^I)}}, 
\la{funcdet}\eeq 
which must be {\it i}) normalized (to the determinant of the free quadratic form, i.e. with no
background field), {\it ii}) regularized (for example by using the Pauli--Villars method), 
and {\it iii}) accounted for the zero modes. Actually one has to compute a ``quadrupole"
combination,  

\beq 
\left[\frac{\det^\prime W \; \det (W_0+\mu^2)}{\det W_0 \; \det (W+\mu^2)}\right]^{-\frac{1}{2}},
\la{quadrup}\eeq 
where $W_0$ is the quadratic form with no background field and $\mu^2$ is the 
Pauli--Villars mass playing the role of the ultraviolet cutoff; the prime reminds that the zero
modes should be removed and treated separately. The resulting one-instanton contribution to
the partition function (normalized to the free one) is \cite{tH,B}:  
\bea
\la{1instw}
\frac{{\cal Z}_{\rm 1-inst}}{{\cal Z}_{\rm P.T.}} &=&\int\! d^4z_\mu\int\! d\rho\int\!
d^{4N_c-5}U\, d_0(\rho), \\
d_0(\rho)&=&\frac{C(N_c)}{\rho^5}\left[\frac{2\pi}{\alpha_s(\mu)}\right]^{2N_c}
(\mu\rho)^{\frac{11}{3}N_c}\exp\left(-\frac{2\pi}{\alpha_s(\mu)}\right). 
\la{1instw1}\eea  
The fact that there are all in all $4N_c$ integrations over the collective coordinates
$z_\mu,\rho,U$ reflects $4N_c$ zero modes in the instanton background.  
The numerical coefficient $C(N_c)$ depends implicitly on the regularization scheme used. 
In the Pauli--Villars scheme exploited above \cite{B}  

\beq 
C(N_c)=\frac{4.60\exp(-1.68N_c)}{\pi^2(N_c-1)!(N_c-2)!}. 
\la{CNc}\eeq 
If the scheme is changed, one has to change the coefficient $C(N_c) \rightarrow
C^\prime(N_c) = C(N_c)\cdot(\Lambda/\Lambda^\prime)^b$. One has \cite{HH}: 
$\Lambda_{\rm P.V.} = e^{\frac{1}{22}}\Lambda_{\overline{\rm MS}} 
= 40.66\,e^{-\frac{3\pi^2}{11N_c^2}}\,\Lambda_{\rm lat} =...$  

\Eq{1instw1} cannot yet be expressed through the 2-loop renormalization-invariant
combination $\Lambda$ \ur{Lambda} as it is written to the 1-loop accuracy only.  
In the 2-loop approximation the instanton weight is given by \cite{VZNS,DP1}  
\bea
\la{d02}
d_0(\rho)\!\!&=&\!\!\! \frac{C(N_c)}{\rho^5}\beta(\rho)^{2N_c} \exp\left[-\beta^{\rm II}(\rho)
\!+\!\left(2N_c\!-\!\frac{b_2}{2b_1}\right)\!\frac{b_2}{2b_1}
\frac{\ln\beta(\rho)}{\beta(\rho)}+O\left(\frac{1}{\beta(\rho)}\right)\right] \\
\n
&\sim & \frac{1}{\rho^5}(\Lambda\rho)^{\frac{11}{3}N_c}, 
\eea 
where $\beta(\rho)\equiv 2\pi/\alpha_s(\rho)$ and $\beta^{\rm II}(\rho)$ are the inverse
charges to the 1-loop and 2-loop accuracy, respectively (not to be confused with the
Gell-Mann--Low function!):  

\bea
\la{beta2} 
\beta^{\rm II}(\rho)&=&\beta(\rho)+\frac{b_2}{2b_1}\ln\frac{2\beta(\rho)}{b_1},\\
\la{beta1}
\beta(\rho)&=&b_1\ln\frac{1}{\Lambda\rho},\qquad b_1=\frac{11}{3}N_c,\qquad
b_2=\frac{34}{3}N_c^2.
\eea  
These equations express the one-instanton weight $d_0(\rho)$ through the 
cutoff-independent combination $\Lambda$ \ur{Lambda}, and the instanton size $\rho$.   
This is how the `transmutation of dimensions' occurs in the instanton calculus and
how $\Lambda$ enters into the game. Henceforth all dimensional quantities will be expressed
through $\Lambda$, which is very much welcome. 

Notice that  the integral over the instanton sizes in \eq{1instw} diverges as a high power of
$\rho$ at large $\rho$: this is of course the consequence of asymptotic freedom. It means that
individual instantons tend to swell. This circumstance plagued the instanton calculus for
many years. If one attemts to cut the $\rho$ integrals ``by hand", one violates the
renormalization properties of the YM theory, as mentioned in the previous section. Actually
the size integrals appear to be cut from above due to instanton interactions. 

\subsection{\it Instanton ensemble} 

To get a volume effect from instantons one needs to consider an \II ensemble, with their total
number $N$ proportional to the 4-dimensional volume $V$. Immediately a mathematical
difficulty arises: any superposition of \IIs is not, strictly speaking, a solution of the equation 
of motion, therefore, one cannot directly use the semiclassical approach of the previous
section. One way to overcome this difficulty is to use a variational principle \cite{DP1}.
Its idea is to use a modified YM action for which a chosen \II ansatz {\em is} a saddle point.
Exploiting the convexity of the exponent one can prove that the true vacuum energy is 
{\em less} than that obtained from the modified action. One can therefore use variational
parameters (or even functions) to get a best upper bound for the vacuum energy. It is not the
Rayleigh-Ritz but rather the Feynman variational principle since the method has been
suggested by Feynman in his famous study of the polaron problem. The gauge theory is more
difficult, though: one has not to loose either gauge invariance or the renormalization
properties of the YM theory. These difficulties were overcome in ref. \cite{DP1}, see also
\cite{DPW}. It should be kept in mind that  we are dealing with ``strong interactions", meaning
that all dimenionless quantities are generally speaking of the order of unity -- there are no
small parameters in the theory. Therefore, one has to use certain approximate methods, and
the variational principle is among the best. Todays direct lattice investigation of the \II
ensemble seem to indicate that we have obtained rather accurate numbers in this difficult
problem.  

In the variational approach, the normalized (to perturbative) and regularized YM partition
function takes the form of a partition function for a grand canonical ensemble of interacting
pseudoparticles of two kind, \IIs:  

\beq 
\frac{{\cal Z}}{{\cal Z}_{\rm P.T.}} \geq \sum_{N_+,N_-}\frac{1}{N_+!} \frac{1}{N_-!}
\prod_n^{N_+ + N_-} \int\! d^4z_n d\rho_n dU_I\; d_0(\rho_n)\; \exp(-U_{\rm int}), 
\la{GPF}\eeq 
where $d_0(\rho)$ is the 1-instanton weight \ur{d02}. The integrals are over the collective
coordinates of (anti)instantons: their coordinates $z$, sizes $\rho$ and orientations given by
$SU(N_c)$ unitary matrices $U$; $dU$ means the Haar measure normalized
to unity. The instanton interaction potential $U_{\rm int}$ (to be discussed below) depends on
the separation between pseudoparticles, $z_m-z_n$, their sizes $\rho_{m,n}$ and their
relative orientations $U_mU_n^\dagger$.  In the variational approach the interaction between
instantons arise from {\em i}) the defect of the classical action, {\em ii}) the non-factorization
of quantum determinants and {\em iii}) the non-factorization of Jacobians when one passes to
integration over the collective coordinates. All three factors are ansatz-dependent, but there
is a tendency towards a cancellation of the ansatz-dependent pieces. Qualitatively, in any
ansatz the interactions between \IIs resemble those of molecules: at large separations there is
an attraction, at smaller separations there is a repulsion. It is very important that the
interactions depend on the relative orientations of instantons: if one averages over
orientations (which is the natural thing to do if the \II medium is in a disordered phase; if not,
one would expect a spontaneous breaking of both Lorentz and colour symmetries \cite{DP1}),
the interactions seem to be repulsive at any separations.  

In general, the mere notion of the instanton interactions is notorious for being ill-defined since
instanton + antiinstanton is not a solution of the equation of motion. Such a configuration
belongs to a sector with topological charge zero, thus it seems to be impossible to distinguish
it from what is encountered in perturbation theory. The variational approach uses brute force
in dealing with the problem, and the results appear to be somewhat dependent on the ansatz
used. Thanks to the inequality for the vacuum energy mentioned above, we still get quite a
useful information. However, recently a mathematically unequivocal definition of the instanton
interaction has been suggested, based on the one hand on analyticity and unitarity
\cite{DPol} and on the other hand on certain singular solutions of the YM equations of
motion \cite{DP4}.  Both definitions cut off automatically contributions of the perturbation
theory. The first three leading terms for the interaction potential at large separations has been
computed by the two very different methods \cite{DPol,DP4} with coinciding results. At
smaller separations one observes a strong repulsion \cite{DP4}.  

At this point I should mention certain experience one gains from a simpler 2-dimensional
so-called $CP^N$ model, also possessing instantons as classical Euclidean solutions.
Contrary to the $4d$ YM theory, the instanton measure in that model is known 
exactly \cite{FFS,BL}. In the dilute limit the instanton measure reduces to the product of
integrals over instanton sizes, positions and orientations, as in \eq{GPF}. The exact measure,
however, is written in terms of the so-called `instanton quarks'  which  does not suppose that
instantons are dilute. The statistical mechanics of \IIs in this model has been studied in ref.
\cite{DM} both by analytical methods and by numerical simulations. Although the `instanton
quark' parametrization allows for complete `melting' of instantons and is quite opposite in spirit
to the dilute-gas ansatz, it has been observed that, owing to a combination of purely
geometric and dynamic reasons, the vast majority of `instanton quarks' form neutral clusters
which can be identified with well-separated instantons. Of course, there is always a fraction
of overlapping instantons in the  vacuum, however, it is small even in the $2d$ case; in the
$4d$ YM case both reasons mentioned above are expected to be even stronger.      

Summing up the discussion, I would say that today there exists no evidence that a
variational calculation with the simplest sum ansatz used in ref. \cite{DP1} is qualitatively or
even quantitatively incorrect, therefore I will cite the numerics from those calculations in what
follows. The main finding \cite{DP1,DPW} is that the \II ensemble \ur{GPF} stabilizes at a
certain density related to the $\Lambda$ parameter (there is no other dimensional quantity in
the theory!)

%%%%%%%%%%%%%%
%% FIGURE 3 %%
%%%%%%%%%%%%%%
\begin{figure}[t]
\hskip 1.5true cm
% \epsfxsize=25pc % will enlarge or reduce the postscript figures
% %based on the xsize
\epsfbox{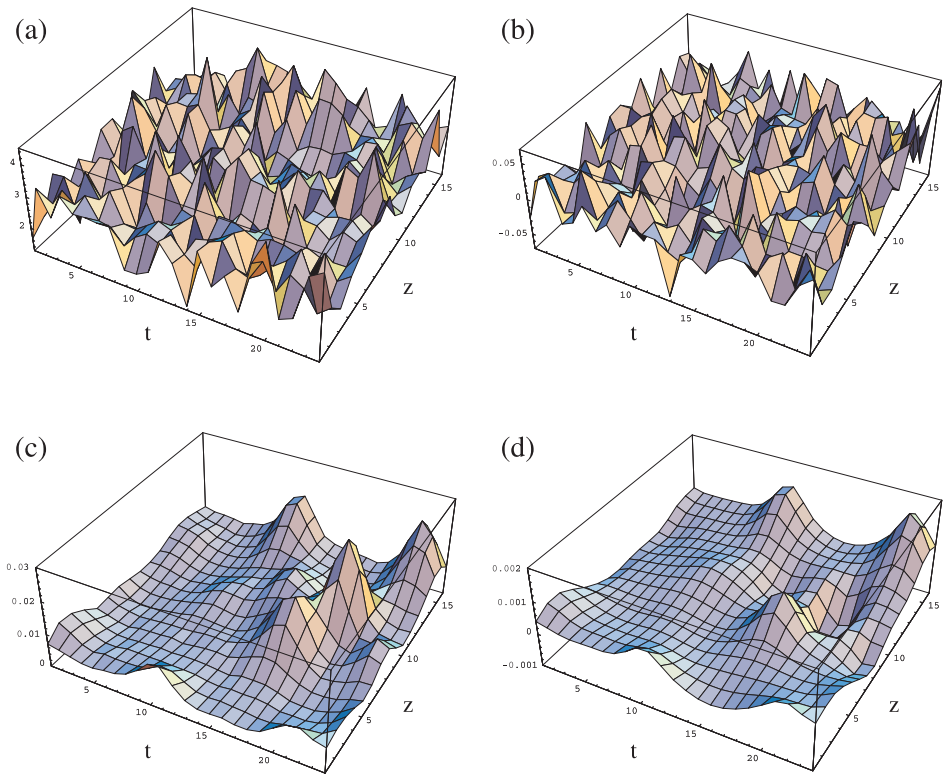} % postscript image file name
% \vspace{-.5cm}
\caption{``Cooling'' the normal zero-point oscillations reveals large fluctuations of the gluon
field, which were identified with instantons and anti-instantons with random positions and
sizes~\cite{CGHN}. The left column shows the action density and the right column shows the
topological charge density for the same snapshot.
\label{fig:negele1}}
%\vspace{-.5cm}
\end{figure}
   
\beq
\frac{N}{V} \simeq \langle F_{\mu\nu}^2/32\pi^2\rangle
\simeq \frac{1}{V}\langle Q_T^2 \rangle\geq (0.75 \Lambda_{\overline{\rm MS}})^4. 
\la{numval}\eeq 
The average instanton size and the average separation between instantons are,
respectively,
\bea
\la{rhoav}
\bar\rho &\simeq & 0.48/\Lambda_{\overline{\rm MS}} \simeq 0.35\,{\rm fm},\\
\la{Rav}
\bar R &=& \left(\frac{N}{V}\right)^{-\frac{1}{4}}
\simeq 1.35/\Lambda_{\overline{\rm MS}} \simeq 0.95\,{\rm fm},
\eea
if one uses $\Lambda_{\overline{\rm MS}}=280\,{\rm MeV}$ as it follows
from the DIS data. Earlier, very similar charactersitics, $\bar\rho=\frac{1}{3}\,{\rm fm},\;
\bar R=1\,{\rm fm},$ have been suggested by Shuryak \cite{Sh1} from studying
the phenomenological applications of instantons. 

Instanton interactions lead to the modification of the (divergent) size distribution
function $d_0(\rho)$ \ur{d02} by a distribution decreasing at large $\rho$. The use of the
variational principle yields a Gaussian cutoff for large sizes \cite{DP1,DPW}:

\beq
d_0(\rho) \to  d(\rho)=d_0(\rho) \,\exp\left(-\,{\rm const.}\sqrt{\frac{N}{V}}\,\rho^2\right).
\la{gauss}\eeq
In fact, it is a rather narrow distribution peaked around $\bar\rho$ \ur{rhoav}; therefore
for practical estimates in what follows I shall just replace all instantons by the average-size
one. 

It should be said that, strictly speaking, nothing can prevent some instantons to be
anomalously large and overlapping with other. For overlapping instantons the
notion of size distribution becomes senseless. The question is quantitative: how often
and how strong do instanstons overlap. Given the estimate \urs{rhoav}{Rav}, it
seems that the majority of instantons in the vacuum ensemble are well-isolated. \\

In the recent years instantons have been intensively studied by direct numerical simulations
of gluon fields on the lattice, using various configuration-smoothing methods 
\cite{CGHN,deF,DGHK}. A typical snapshot of gluon fluctuations in the vacuum
is shown in Fig. 3 borrowed from ref. \cite{CGHN}. Naturally, it is heavily dominated by
normal perturbative UV-divergent zero-point oscillations of the field. However, after
``cooling'' down these oscillations one reveals a smooth background field which
was shown in ref. \cite{CGHN} to be nothing but an ensemble of instantons and
anti-instantons with random positions and sizes~\footnote{Quite recently a more
involved fluctuation-smearing procedure carried on the lattice has indicated 
that instantons might have an additional structure, see the next section.}.  
The lower part of Fig. 3 is what is left of the upper part after ``cooling'' that particular
configuration. The average sizes and separations of instantons found vary somewhat
depending on the concrete smearing method used. Ref. \cite{CGHN} gives the following
values

\beq
\bar\rho\simeq 0.36\,{\rm fm},\qquad 
\bar R=(N/V)^{-\frac{1}{4}}\simeq 0.89\,{\rm fm},
\la{N}\eeq
which are not far from the estimate from the variational principle. The ratio,

\beq
\frac{\bar \rho}{\bar R} \simeq \frac{1}{3}, 
\la{pf}\eeq 
seems to be more stable: it follows from phenomenological \cite{Sh1}, variational
\cite{DP1,DPW} and lattice \cite{CGHN,deF,DGHK} studies. It means that the {\em packing
fraction}, i.e. the fraction of the 4-dimensional volume occupied by instantons appears to be
rather small, $\pi^2 \bar\rho^4 /\bar R^4 \simeq 1/8$. This small packing fraction of the
instantons gives an {\it a posteriori} justification for the use of the semi-classical methods. 
As I shall show in the next sections, it also enables one to identify adequate degrees of
freedom to describe the low-energy QCD.

\section{Non-instanton semiclassical configurations}
\setcounter{equation}{0}
\def\theequation{\arabic{section}.\arabic{equation}}  

Instantons induce certain potential between static quarks in the pure glue theory,
which depends on the instanton size distribution \cite{CDG,DPP1,DPpot}. For fast
convergent size distributions like that given by \eq{gauss} the potential first rises 
as function of the interquark separation but asymptotically it flattens out. If the size
distribution happens to fall off as $d(\rho)\sim 1/\rho^3$ at large $\rho$ one gets
a linear infinitely rising potential \cite{DPpot}. However, such a size distribution means
that large instantons inevitably overlap, which means that the ``center, size, orientations''
collective cooredinates are not the most adequate. Standard instantons do not induce an
infinitely rising linear potential between static quarks in the pure glue theory. This observation
stimulates the search for other semiclassical gluonic objects that could be responsible for
confinement. Among them are {\em merons}~\cite{AFF,CDG2}, {\em calorons with non-trivial
holonomy}~\cite{KvB,LL}, {\em BPS monopoles or dyons}~\cite{Bog,PS}, 
{\em vortices}~\cite{tHvor,Ma}, etc. I describe briefly these objects below.\\

\ni
\underline{Merons}\\

\ni
Meron is a  self-dual solution of the YM equation of motion \cite{AFF},

\beq
A^a_\mu=\bar\eta^a_{\mu\nu}\,\frac{x_\nu}{x^2},
\la{Ameron1}\eeq
which is interesting because at fixed time its spatial components reminds the
field of a $3d$ monopole of size $t$:

\beq
A^a_i=\epsilon_{aij}\,\frac{r_j}{r^2+t^2}.
\la{Ameron2}\eeq
However, its action diverges logarithmically both in the UV and IR regions;
therefore such configurations are not encountered in the vacuum. Callan, Dashen
and Gross \cite{CDG2} suggested to consider a regularized meron pair,

\beq
A^a_\mu=\bar\eta^a_{\mu\nu}\left[\frac{(x-z_1)_\nu}{(x-z_1)^2+\rho_1^2}
+\frac{(x-z_2)_\nu}{(x-z_2)^2+\rho_2^2}\right],
\la{A2merons}\eeq
whose action is finite, and at $|x|\to\infty$ it becomes the field of an instanton. 
\Eq{A2merons} corresponds to the field of a monopole pair which has been created, 
moved to a separation $R=|z_1-z_2|$ and then annihilated after the time $R$. Merons 
can be viewed as ``half-instantons''. The idea of Callan, Dashen and Gross was that
instantons could dissociate into a gas of (regularized) merons, which could lead
to confinement. Recently, a relation of merons to monopoles and vortices has
been discussed -- see ref. \cite{MN} where such relation has been studied  
numerically, and references to earlier work therein. So far merons have not been
identified from lattice configurations. Although the idea that meron configurations
may well prove to be relevant to confinement in $4d$, I have certain reservations
as confinement is also a property of the $3d$ pure YM theory where there are
no merons. Therefore, if confinement is due to some semiclassical ``objects'',
$3d$ and $2d$ field configurations may be of more relevance.\\

\ni
\underline{Dyons and calorons with non-trivial holonomy} \\

\ni
Dyons or Bogomolnyi--Prasad--Sommerfeld (BPS) monopoles are self-dual 
solutions of the YM equations of motion with {\em static} ({\it i.e.} time-independent)
action density, which have both the magnetic and electric field at infinity decaying as
$1/r^2$. Therefore these objects carry both electric and magnetic charges. In the 
$3\!+\!1$-dimensional $SU(2)$ gauge theory there are in fact  two types of self-dual
dyons~\cite{LY}: $M$ and $L$ with (electric, magnetic) charges $(+,+)$ and $(-,-)$, and two
types of anti-self-dual dyons $\bar M$ and $\bar L$ with charges $(+,-)$ and $(-,+)$,
respectively. Their explicit fields can be found {\it e.g.} in ref.~\cite{DPSUSY}. In $SU(N_c)$
theory there are $2N_c$ different dyons~\cite{LY,DHK}: $M_1,M_2,...M_{N_c-1}$ ones with
charges counted with respect to $N_c-1$ Cartan generators and one $L$ dyon with charges
compensating those of $M_1...M_{N_c-1}$ to zero, and their anti-self-dual counterparts.    

Speaking of dyons one implies that the Euclidean space-time is compactified in the
`time' direction whose inverse circumference is temperature $T$, with the usual periodic
boundary conditions for boson fields. However, the temperature may go to zero, in which
case the $4d$ Euclidean invariance is restored.

Dyons' essence is that the $A_4$ component of the dyon field tends to a constant value at
spatial infinity. This constant $A_4$ can be eliminated by a time-dependent gauge
transformation. However then the fields violate the periodic boundary conditions,
unless $A_4$ has quantized values corresponding to trivial holonomy, see below.
Therefore, in a general case one implies that dyons have a non-zero value of
$A_4$ at spatial infinity.

A single dyon can be considered in whatever gauge. The simplest is the ``hedgehog'' gauge
in which the dyon field is smooth everywhere. However, if one wishes to have a vacuum
filled by dyons, one has to take more than one dyon. Two and more dyons can be put
together only if all colour components of $A_4^a$ at infinity are the same for all dyons. 
Before assembling ``hedgehogs'' together, their ``hair'' has to be ``gauge-combed''
in such a way that $A_4^at^a$ at infinity is a constant matrix, which can well be chosen to
be diagonal, {\it i.e.} belonging to the Cartan subalgebra. The gauge meeting this requirement  
is called ``stringy'' gauge as it necessarily has a pure-gauge string-like singularity starting
from the dyon center. For explicit formulae see {\it e.g.} the Appendix of ref. \cite{DPSUSY}. In
other words two or more dyons need to have the same orientation in colour space. This
orientation is preserved throughout the $3d$ volume. The $A_4$ component of the YM field
plays here  the role of the Higgs field in the adjoint representation. The spatial size of $M$
dyons is $1/|A_4^3|$ (for the $SU(2)$ group) and that of $L$ dyons is $1/|2\pi T-A_4^3|$
where $T$ is the Euclidean temperature. The mere notion of the ensemble of dyons (or
monopoles) implies that colour symmetry is in a sense spontaneously broken. Of course,
once colour is aligned, one can always randomize the colour orientation by an arbitrary
point-dependent gauge transformation, just as the direction of the Higgs field can be
randomized but that does not undermine the essence of the Higgs effect. 

This is very distinct from instantons which can have arbitrary colour orientations in the
ensemble. Mathematically, one can see it from the number of zero modes. For example, for
the $SU(2)$ group instantons have 4 translation, 1 dilatation and 3 orientation zero modes,
see subsection 2.4. A dyon (in any gauge group) has only 4 zero modes out of which 3 are
spatial translations and 1 has a double description: one can call it the spurious translation in
the `time' direction or, better, a global $U(1)$ phase. The dyon solution has no collective
coordinates that can be identified with colour orientation. This fact is not accidental
but related precisely to that dyons have a non-vanishing field $A_4$ at infinity. One
can formally introduce an `orientation' degree of freedom but the corresponding zero
mode will be not normalizable. In the instanton case all fields decay fast enough at
infinity to make the orientation mode normalizable and hence physical.  

The fact that dyons have non-zero $A_4$ at spatial infinity means also that the
Polyakov line (also called the {\em holonomy}) is in general non-trivial at infinity, {\it i.e.}
does not belong to the group center, unless $A_4$ assumes certain quantized values:

\beq
P=\left.{\rm P}\,\exp\left(i\int_0^{1/T}\!dt\,A_4\right)\right|_{x\to\infty} \notin {\bf Z}(N_c).
\la{Pol1}\eeq 

Evaluating the effects of quantum fluctuations about a dyon proved to be a difficult task,
much more difficult than that about instantons. The most advanced calculation is in
ref. \cite{Zar} where quantum determinants have been computed in some limiting case.
Probably the technique developed therein allows one to read off the result for determinants
in a general case but this has not been done. One thing is clear, however, without
calculations: the quantum determinants about dyons strongly diverge in the IR region!
The point is, at non-zero temperatures the quantum action contains, {\it inter alia}, 
a potential-energy term \cite{GPY,NW} 

\beq
V_{\rm pert}=\left.\frac{1}{3T(2\pi)^2}\phi^2(2\pi T-\phi^2)\right|_{{\rm mod}\; 2\pi T},
\qquad \phi=\sqrt{A_4^aA_4^a}\qquad[{\rm for}\,SU(2)\,{\rm group}].
\la{Tpot}\eeq 
As follows from \eq{Pol1} the trace of the Polyakov line is related to $\phi$ as

\beq
\half\,\Tr\,P=\cos\frac{\phi}{2T}.
\la{TrP}\eeq
The zero energy of the potential corresponds to $P=\pm 1$, {\it i.e.} to the trivial
holonomy. If a dyon has $\phi\neq 2\pi T n$ at spatial infinity the corresponding 
quantum action is positive-definite and proportional to the $3d$ volume. Therefore, dyons
with non-trivial holonomy seem to be strictly forbidden: quantum fluctuations about them have
an unacceptably large action! \\

There are two known generalizations of instantons at non-zero temperatures. One is
the periodic instanton of Harrington and Shepard \cite{HS} studied in detail in ref.~\cite{GPY}.
These periodic instantons, also called {\em calorons}, have trivial holonomy at spatial infinity,
therefore they are not suppressed by the above quantum potential energy. The vacuum made
of those instantons has been investigated, using the variational principle, in ref.~\cite{DMir}. 

The other generalization has been constructed recently by Kraan and van Baal~\cite{KvB}
and Lee and Lu~\cite{LL}; it has been named {\em caloron with non-trivial holonomy}.
I shall call it for short the KvBLL caloron. It is also a self-dual solution of the YM equations of
motion with a unit topological charge. The fascinating feature of this construction is that it can
be viewed as ``made of'' one $L$ and one $M$ dyon, with total zero electric and magnetic
charges. Although the action density of isolated $L$ and $M$ dyons does not depend on
time, their combination in the KvBLL solution is generally non-static: the $L,M$ ``consituents''
show up not as $3d$ but rather as $4d$ lumps. When the temperature goes to zero, 
these lumps merge, and the KvBLL caloron reduces to the usual instanton (as
does the standard Harrington--Shepard caloron), {\em plus} corrections of the order of $T$.
However, the holonomy remains fixed and non-trivial at spatial infinity. It means that quantum
fluctuations strongly suppress individual KvBLL calorons at any temperature, as they
suppress single $L,M$ dyons. \\

It is very interesting that precisely these objects determine the physics of the
{\em supersymmetric} YM theory where in addition to gluons there are gluinos, {\it i.e.}
Majorana (or Weyl) fermions in the adjoint representation. Because of supersymmetry,
the boson and fermion determinants about $L,M$ dyons cancel exactly, so that
the perturbative potential energy \ur{Tpot} is identically zero for all temperatures (actually to
all loops). Therefore, in the supersymmetric theory dyons are openly allowed. 
[To be more precise, the cancellation occurs when periodic conditions for gluinos
are imposed, so it is the compactification in one (time) direction that is implied, rather 
than physical temperature which requires antiperiodic fermions. However, I shall still call 
it ``temperature''.] Further on, it turns out \cite{DHKM,DHK} that dyons generate a
non-perturbative potential having a minimum at $\phi=\pi T$, {\em i.e.} where the perturbative
potential would have the maximum. This value of $A_4$ corresponds to the holonomy
$\Tr\,P=0$ at spatial infinity, which is the ``most non-trivial''; as a matter of fact it is one of the
confinement's requirements. However, it implies that $A_4$ has to lie in some direction in
colour space thus breaking the colour group to the maximal Abelian subgroup, at least at
small compactification circumference \cite{DPSUSY}.  

In the supersymmetric YM theory there is a non-zero gluino condensate. It is analogous to
the quark condensate in QCD, which we consider in the next section. However, contrary to
QCD, in the supersymmetric case the gluino condensate can be computed {\em exactly} 
and expressed - via the transmutation of dimensions - through the scale $\Lambda$. Its
correct value is reproduced by saturating the gluino condensate by $L,M$ dyons' zero
fermion modes \cite{DHKM}. On the contrary, the saturation of the (square of) gluino
condensate by instanton zero modes gives the wrong result, namely $\sqrt{4/5}$ that of the
correct value \cite{NSVZsusy}. The paradox is that both derivations are seemingly clean. 

This striking 20-year-old paradox has been recently resolved \cite{DPSUSY} by the
observation that the (square of) gluino condensate must be computed not in the instanton
background but in the background of exact solutions ``made of'' $LL,MM$ and $LM$
dyons. The first two are the double-monopole solutions and the last one is the KvBLL
caloron. As the temperature goes to zero, the $LL$ and $MM$ solutions have locally
vanishing fields, whereas the KvBLL $LM$ solution reduces to the instanton field up to a
locally vanishing difference. Therefore, naively one would conclude that the dyon
calculation of the gluino condensate, which is a local quantity, should be equivalent to the
instanton one, but it is not! The fields vanishing as the inverse size of the system have
a finite effect on such a local quantity as the gluino condensate! This is quite unusual. 
The crucial difference between the (wrong) instanton and the (correct) dyon calculations 
is in the value of the Polyakov loop, which remains finite. In the supersymmetric YM theory  
configurations having $\Tr\,P=0$ at infinity are not only allowed but dynamically preferred
as compared to those with $P=\pm 1$. In non-supersymmetric theory it looks as if it was
the opposite. \\

An intriguing question is whether the perturbative potential energy \ur{Tpot} which forbids
individual dyons in the pure YM theory can be overruled by non-perturbative contributions
of an {\em ensemble} of dyons. I think that there is such a chance. Let us consider the case
of non-zero temperatures, and let us imagine an ensemble of $L,M,\bar L,\bar M$ dyons. 
Assuming that the temperature is high enough one can suppose that dyons are dilute
and hence {\it i}) it is actually an ensemble of static objects, {\it ii}) their interaction
can be neglected as compared to their own action. The $M,\bar M$ action is
$\frac{8\pi^2}{g^2}\frac{|\phi|}{2\pi T}$ whereas the $L,\bar L$ action is   
$\frac{8\pi^2}{g^2}\frac{|2\pi T-\phi|}{2\pi T}$ where $\phi=\sqrt{A_4^aA_4^a}$ and
$A_4^a$ is the common value of the dyons' field at spatial infinity. The weight of one dyon in
the semiclassical approximation is $\exp(-{\rm Action})$, times the integral over
collective coordinates $\int\!d^3z$, times something which makes it dimensionless -- 
usually it is the solution's size. Therefore, I would write the partition function of the dyon 
gas as

\bea
\nonumber
Z_{\rm dyon}&=&\sum_{N_L,N_{\bar L},N_M,N_{\bar M}}
\frac{1}{N_L!N_{\bar L}!N_M!N_{\bar M}!}\,
\left[\int\!d^3z |\phi|^3
\exp\left(-\frac{8\pi^2}{g^2}\frac{|\phi|}{2\pi T}\right)\right]^{N_M+N_{\bar M}} \\
\nonumber
&\times & 
\left[\int\!d^3z |2\pi T-\phi|^3
\exp\left(-\frac{8\pi^2}{g^2}\frac{|2\pi T-\phi|}{2\pi T}\right)\right]^{N_L+N_{\bar L}} \\
&=&\!\!\exp\!\left[2V|\phi|^3
\exp\!\left(\!-\frac{8\pi^2}{g^2}\frac{|\phi|}{2\pi T}\right)+2V|2\pi T\!-\!\phi|^3
\exp\!\left(\!-\frac{8\pi^2}{g^2}\frac{|2\pi T\!-\!\phi|}{2\pi T}\right)\!\right]\!.
\la{Zdyon}\eea
This is naturally a very crude estimate. Interactions have been neglected. Exact quantum
determinants are unknown. However, it is known that they make the gauge
coupling run, {\it i.e.} their role is basically to replace the coupling constants in \eq{Zdyon}
by the running coupling at the temperature scale:
$\frac{8\pi^2}{g^2}\to \left(\frac{\Lambda}{\pi T}\right)^{22/3}$. The overall numerical
factor cannot be determined from such simple considerations -- it must be computed. 
When evaluating quantum determinants one implies that they are normalized
not to the free ones but those in constant background field of $A_4$, to make
the normalized determinants IR finite. The determinants in the constant $A_4$ field
are known to produce the perturbative potential \ur{Tpot}.

An educated guess for the non-perturbative potential induced by dyons is therefore

\beq
V_{\rm dyon}=-c\,\left[|\phi|^3
\left(\frac{\Lambda}{\pi T}\right)^{\frac{22}{3}\frac{|\phi|}{2\pi T}}
+|2\pi T-\phi|^3\left(\frac{\Lambda}{\pi T}\right)^{\frac{22}{3}\frac{|2\pi T-\phi|}{2\pi T}}\right],
\la{Vdyon}\eeq
to which the perturbative potential  \ur{Tpot} must to be added. Let me note that
in the supersymmetric theory there is a similar non-perturbative potential \cite{DHKM}
(but in that case it is an exact result) and the perturbative potential is absent. 

%%%%%%%%%%%%%%
%% FIGURE 4 %%
%%%%%%%%%%%%%%
\begin{figure}[t]
\hskip 1.5true cm
% \epsfxsize=25pc % will enlarge or reduce the postscript figures
% %based on the xsize
\epsfbox{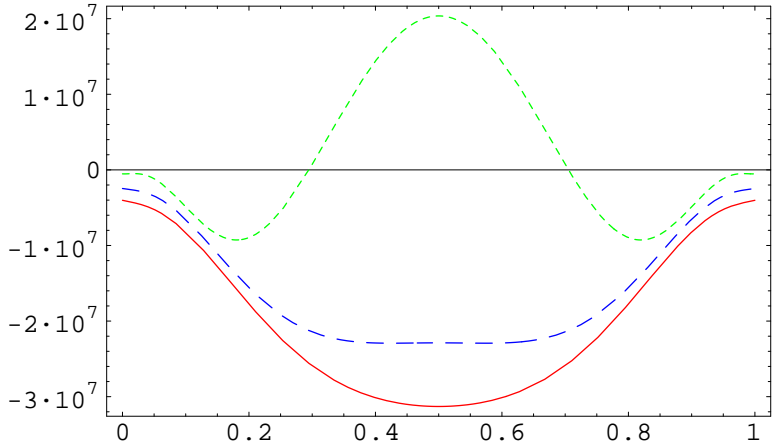} % postscript image file name
% \vspace{-.5cm}
\caption{An illustration of how dyons can induce the confinement-deconfinement
phase transition. We plot the sum of the perturbative \ur{Tpot} and dyon-induced \ur{Vdyon}
potentials as function of $\phi/2\pi T$. We choose the constants $\Lambda=280\,{\rm MeV}$,
$c=2$. The curves corresponds to temperatures $T=250\,{\rm MeV}$ (solid), 
$T=280\,{\rm MeV}$ (long-dashed) and $T=400\,{\rm MeV}$ (short-dashed). 
In this example, the phase transition is at $T_c=280\,{\rm MeV}$. 
\label{fig:dyonpot}}
%\vspace{-.5cm}
\end{figure}

The perturbative potential \ur{Tpot} has minima at $\phi=0,2\pi T$ corresponding
to trivial holonomy $P=\pm 1$, while the non-perturbative potential \ur{Vdyon} has the
minimum just in the middle, at $\phi = \pi T$ corresponding to the non-trivial holonomy
$\Tr\,P=0$, which is one of the requirements of confinement.  Clearly, at very large
temperatures the perturbative potential wins, and the system settles in the perturbative
vacuum with trivial holonomy. At temperatures below certain critical $T_c={\rm
const.}\,\Lambda$ the non-perturbative potential prevails, dyons with the asymptotic value of
$A^3_4=\pi T$ are preferred, and the system presumably goes into the confinement phase. 
For the particular model of the dyon-induced potential \ur{Vdyon} it is a second-order
phase transition, see Fig.  4. For $SU(3)$ and higher groups both the perturbative and
non-perturbative potentials depend on more variables than just $\phi$, and one can well
imagine their interplay leading to a first order transition. I have presented here what might be
a crude microscopic model of the confinement-deconfinement transition, based on dyons. 

It should be noted that recent lattice simulations support  the presence of KvBLL calorons 
(or, more generally, dyons) near the phase transition point ~\cite{IMMPSV,Gatt}.  

As the temperature lowers, the ensemble of $L,\bar L, M,\bar M$ dyons becomes more
dense, and these objects (which have time-independent action densities when they are
well-isolated) do not represent a static field anymore: the action density will have $4d$
lumps. Their statistical mechanics must be most intriguing! As I mentioned before, at
$T\to 0$ the KvBLL calorons tend to instantons, with their $L,M$ `constituents' merged 
together. This is the most delicate moment. I do not know what prevents them from
glueing up completely into normal instantons, and what guarantees that non-trivial holonomy 
is dynamically preferred. 

There are problems with the dyon scenario at high temperatures as well. Dyons survive for a
while above $T_c$ although the minima of the potential shown in Fig. 4 
correspond to the holonomy approaching $P=\pm 1$. Eventually, the $4d$ theory at high
temperatures becomes the $3d$ theory.  We believe that confinement is also a property of the
pure-glue $3d$ YM theory but in the true $3d$ case there are no dyons as there is no
$A_4$ component of the field, which is critical for the dyon construction, and we do not
know of any other classical solutions there. \\

\ni
\underline{Monopoles and vortices}\\

\ni
A monopole is, by definition, a $3d$ field configuration whose magnetic field is
decaying as $1/r^2$ at large distances. A center vortex is a $2d$ field configuration
such that a Wilson loop winding about it takes the value of one of the non-trivial elements of
the group center ${\bf Z}(N_c)$. A simple dimensional analysis based on a rescaling of the
assumed solutions' size shows that there are no such solutions of the classical YM equations
apart from BPS dyons described above but in the true $3d$ case they are absent. It should
be noted however that many non-trivial classical solutions appear when one imposes twisted
instead of periodic boundary conditions~\cite{GA}. 

In the lattice community a completely different definition of monopoles and vortices
is used, for practical reasons. They are defined through gauge fixing, see a recent
review by Greensite \cite{Green}. Monopoles and vortices identified on the lattice do 
not necessarily correspond to any semiclassical field configurations. Lattice monopoles
are points in $3d$ or lines in $4d$ space while lattice center vortices are lines
in $3d$ and surfaces in $4d$. If there is any materialistic meaning behind them 
both monopoles and vortices need to be `thick', that is to have on the average a finite radius
of the order of $1/\Lambda$ since it is the only physical scale in the YM theory. This hints
that they need not be classical solutions (the more so that there are none) but rather local
minima of the effective action with quantum corrections included.  

To give an example, in high-temperature $2\!+\!1$ dimensional pure YM theory a `thick'
center vortex solution has beed found as a local minimum of the classical plus 1-loop
quantum action~\cite{Dvort}. In $3d$ and $4d$ the 1-loop effective action about a
vortex has been computed in ref.~\cite{DMvort} and the corresponding local minima
found, too. So far no similar analysis has been performed for $3d$ monopoles,
which would be of some interest.  Although the radius of the vortex has been found 
in units of $1/\Lambda$, the size appears to be too large and the minimum is very shallow. 
Later on it was discovered that vortices have a negative mode~\cite{Bord} {\it i.e.} are
unstable. These studies are rather discouraging for attempts to attribute any semiclassical
meaning to vortices. It should be added that nothing is known about the statistical mechanics
of vortices from the theory side, nor even is the object itself defined.  \\   

The idea that some semiclassical objects may be relevant for confinement faces 
certain difficulty of a general nature. It is widely believed that the string tension between
static quarks is stable in the limit of large number of colours $N_c$. This is supported by
lattice observations at medium $N_c$~\cite{LT,DDPRV}. Assuming the simplest scenario 
that confinement is driven by weakly interacting center vortices populating the YM vacuum, 
one finds then that the density of vortices piercing any given $2d$ plane should
rise linearly with $N_c$~\cite{DDD}. It means that at large enough $N_c$ vortices have to
overlap. What happens then is not clear. If they remain independent despite overlap, it
should be clarified how it happens. If they are not independent but interact strongly, the
simple argument why vortices lead to the area behaviour of large Wilson loops is lost. 

In fact {\em all} semiclassical objects face a problem at large $N_c$, and the reason
is quite general: they possess a finite action density $\Tr\,F_{\mu\nu}^2$. If semiclassical
objects saturate the non-perturbative gluon condensate (see subsection 2.5) their density
has to rise linearly with $N_c$, since the gluon condensate does so, as it follows from 
$N_c$ counting rules. This is a universal problem for instantons, dyons and vortices: they
have to overlap at large $N_c$. A way to prevent these objects from complete ``melting''
is as follows. Start with instantons. The use of the variational principle~\cite{DP1,DPW}
confirms the expectation that instanton density is $O(N_c)$ whereas their average
size is $O(1)$, {\it i.e} they have to overlap at large $N_c$. However, instanton is basically
an $SU(2)$ object, and it has been noticed by T. Sch\"afer \cite{SNc} that instantons
residing in different ``corners'' of the large $SU(N)$ group space do not ``talk'' to each
other, meaning that they might well be statistically independent. A more precise meaning
of it is suggested by dyons. Same as instantons, dyons have finite action and finite size,
however there are precisely $N_c$ different types of dyons (see above), therefore each 
type of dyons has a finite density at large $N_c$ although different types have to overlap.
A dyon is an $SU(2)$ object built about one of the Cartan generators of the $SU(N)$ group,
${\rm diag}(0...0,1,-1,0...0)$. Clearly, only dyons that are nearest neighbours in colour
space do interact, however $O(N_c)$ types of dyons are transparent for any given type. 
Therefore, there is no harm if dyons of the not-nearest-neighbour types sit on top of each
other. Dyons from commuting $SU(2)$ subgroups do not interact either in the classical or in
the quantum sense. Probably a similar argument can be given for vortices, as there are
$N_c$ types of them.     

Thus, the transition to infinite $N_c$ in terms of semiclassical objects may be in fact quite
smooth. It is very important to understand this transition clearly since lattice simulations
indicate \cite{LT,DDPRV} that already $N_c\!=\!2$ is not so far away from infinity. 
Apparently, confinement should be explained by the same mechanism both at small and
large $N_c$. Also, the asymptotic area law for large spatial Wilson loops is believed to be
the property of all theories starting from $2\!+\!1$ dimensions and to 4 dimensions, and it
would be natural if it is driven basically by the same mechanism. 

At the same time it should be stressed again that at large $N_c$ any semiclassical picture
implies that objects are very dense, and hence their colour fields are fast varying. 
It indicates that the semiclassical methodology is probably not the best way to understand
confinement. Personally, I think that passing to dual gauge-invariant variables \cite{Ddual}
which do not need to oscillate violently, is a more promising strategy in solving this awful
problem. 

\vskip 1.5true cm 

\section{Chiral symmetry breaking by instantons}
\setcounter{equation}{0}
\def\theequation{\arabic{section}.\arabic{equation}}  

\subsection{\it Chiral symmetry breaking by definition}

The QCD Lagrangian with $N_f$ massless flavours is known to
posses a large global symmetry, namely a symmetry under $U(N_f)\times
U(N_f)$ independent rotations of left- and right-handed quark fields.
This symmetry is called {\em chiral} \footnote{The word was coined by
Lord Kelvin in 1894 to describe moleculas not superimposable on its
mirror image.}. Instead of rotating separately the 2-component Weyl
spinors corresponding to left- and right-handed components of
quark fields, one can make independent vector and axial $U(N_f)$
rotations of the full 4-component Dirac spinors -- the QCD lagrangian
is invariant under these transformations too.

Meanwhile, axial transformations mix states with different
P-parities.  Therefore, were that symmetry exact, one would observe
parity degeneracy of all states with otherwise the same quantum
numbers.  In reality the splittings between states with the same
quantum numbers but opposite parities are huge. For example, the
splitting between the vector $\rho$ and the axial $a_1$ meson is $(1260
- 770)\simeq 500\,{\rm MeV}$; the splitting between the nucleon and its
parity partner is even larger:  $(1535 - 940)\simeq 600\,{\rm MeV}$.

The splittings are too large to be explained by the small bare or
current quark masses which break the chiral symmetry from the
beginning. Indeed, the current masses of light quarks are: $m_u \simeq
4\,{\rm MeV},\;\;m_d\simeq 7\,{\rm MeV},\;\;m_s\simeq 150\,{\rm MeV}$. The 
conclusion one can draw from these numbers is that the chiral symmetry
of the QCD Lagrangian is broken down {\em spontaneously}, and very
strongly. Consequently, one should have light (pseudo) Goldstone
pseudoscalar hadrons -- their role is played by pions which indeed are
by far the lightest hadrons.

The order parameter associated with chiral symmetry breaking is
the so-called {\em chiral} or {\em quark condensate}:

\beq
\langle\bar\psi\psi\rangle\simeq -(250\;MeV)^3.
\la{chcond}\eeq
It should be noted that this quantity is well defined only for
massless quarks, otherwise it is somewhat ambiguous. By definition,
this is the quark Green function taken at one point; in momentum space
it is a closed quark loop:

\beq
\langle\bar\psi\psi\rangle=-N_c\int\!\frac{d^4k}{(2\pi)^4i}\,\Tr\frac{Z(k)}{M(k)-\Dk}.
\la{chcond1}\eeq

If the quark propagator is massless and has only the `slash' term, the trace over the spinor
indices in the loop gives an identical zero. Therefore, chiral symmetry breaking implies that
a massless (or nearly massless) quark develops a non-zero dynamical mass $M(k)$, {\it i.e.}
a `non-slash' term in the propagator. There are no reasons for this quantity to be a constant
independent of the momentum; moreover, we understand that it should anyhow vanish at
large momentum. Sometimes it is called the constituent quark mass, however a
momentum-dependent {\em dynamical quark mass} $M(k)$ is a more adequate term which I
shall use below. 

The spontaneous generation of the dynamical quark mass (equivalent to the spontaneous
chiral symmetry breaking, SCSB) is the most important feature of QCD being key to the 
whole hadron phenomenology. The theory's task is to get $M(k)$ in the form 

\beq
M(k)=\Lambda f(k/\Lambda)
\la{M1}\eeq
where $\Lambda$ is the renormalization-invariant combination \ur{Lambda}
and $f$ is some function. Instantons enable one to get $M(k)$ in the needed
form and to find the function. But first let us derive some general relations.

We start by writing down the QCD partition function. Functional integrals are well defined in
Euclidean space which is obtained by the following formal substitutions of Minkowski space
quantitites:  
\bea
\n
 ix_{M0}&=&x_{E4},\quad x_{Mi}=x_{Ei},\quad A_{M0}=iA_{E4},\quad A_{Mi}=A_{Ei}, \\
i\bar\psi_M&=&\psi_E^\dagger,\quad \gamma_{M0}=\gamma_{E4},\quad
\gamma_{Mi}=i\gamma_{Ei},\quad\gamma_{M5}=\gamma_{E5}. 
\la{ME}\eea  
Neglecting for brevity the gauge fixing and Faddeev--Popov ghost terms, the QCD partition
function with quarks can be written as  
\bea
\n
{\cal Z}&=&\int\! DA_\mu D\psi D\psi^\dagger\, \exp\left[-\frac{1}{4g^2}\int
F_{\mu\nu}^2+\sum_f^{N_f} \int\psi_f^\dagger(i\Dnabla+im_f)\psi_f\right] \\
&=& \int\! DA_\mu \exp\left[-\frac{1}{4g^2}\int F_{\mu\nu}^2\right]
 \prod_f^{N_f}\det(i\Dnabla + im_f). 
\la{partfuq}\eea  
The chiral condensate of a given flavour $f$ is, by definition,  

\beq 
\langle\bar\psi_f\psi_f\rangle_M=-i\langle\psi_f^\dagger\psi_f\rangle_E
=-\frac{1}{V}\frac{\partial}{\partial m_f} \left(\ln {\cal Z}\right)_{m_f\rightarrow 0}. 
\la{ccdef}\eeq  
The Dirac operator has the form  

\beq 
i\Dnabla=\gamma_\mu(i\partial_\mu+A_\mu^{\rm I\bar I}+a_\mu) 
\la{Dirop}\eeq 
where $A_\mu^{\rm I\bar I}$ denotes the classical field of the \II ensemble and 
$a_\mu$ is a presumably small field of quantum fluctuations about that ensemble, which I
shall neglect as it has little impact on chiral symmetry breaking. Integrating over $DA_\mu$ 
in \eq{partfuq} means averaging over the \II ensemble with the partition function \ur{GPF},
therefore one can write  

\beq 
{\cal Z} = \overline{\det(i\Dnabla+im)} 
\la{zav}\eeq 
where I temporarily restrict the discussion to the case of only one flavour for simplicity.
Because of the $im$ term the Dirac operator in \ur{zav} is formally not Hermitian; however the
determinant is real due to the following observation. Suppose we have found the eigenvalues
and eigenfunctions of the Dirac operator,  

\beq 
i\Dnabla\Phi_n=\lambda_n\Phi_n, 
\la{eig}\eeq 
then for any $\lambda_n \neq 0$ there is an eigenfunction
$\Phi_{n^\prime}=\gamma_5\Phi_n$ whose eigenvalue is $\lambda_{n^\prime}=-\lambda_n$. 
This is because $\gamma_5$ anticommutes with $i\nabla$. Owing to this the fermion
determinant can be written as  
\bea
\n
\det(i\Dnabla+im) &=& \prod_n(\lambda_n+im)=\sqrt{\prod(\lambda_n^2+m^2)}
=\exp\left[\frac{1}{2}\sum_n\ln(\lambda_n^2+m^2)\right] \\
&=&\!\exp\left[\frac{1}{2}\int_{-\infty}^\infty\! d\lambda\;\nu(\lambda) \ln(\lambda^2+m^2)\right],
\quad \nu(\lambda)\equiv \sum_n\delta(\lambda-\lambda_n), 
\la{fermdet}\eea 
where I have introduced the {\em spectral density} $\nu(\lambda)$ of the Dirac operator
$i\nabla$. Note that the last expression is real and even in $m$, which is a manifestation of
the QCD chiral invariance.  Differentiating \eq{fermdet} in $m$ and putting it to zero one gets
according to the general \eq{ccdef} a formula for the chiral condensate:  

\bea 
\n
\langle\bar\psi\psi\rangle &=& -\frac{1}{V}\frac{\partial}{\partial m} \left[\frac{1}{2}\int
d\lambda\; \overline{\nu(\lambda)}\ln(\lambda^2+m^2)\right]_{m\rightarrow 0} \\
&=&\left.-\frac{1}{V}\int_{-\infty}^\infty d\lambda\;
\overline{\nu(\lambda)}\frac{m}{\lambda^2+m^2}\right|_{m\rightarrow 0} 
\la{BC1}\eea 
where $\overline{\nu(\lambda)}$ means averaging over the instanton ensemble together with
the weight given by the fermion determinant itself. The latter, however, may be cancelled in
the so-called quenched approximation where the back influence of quarks on the dynamics
is neglected. Theoretically, this is justified at large $N_c$.  Naively, one would think that the
r.h.s. of \eq{BC1} is zero at $m\rightarrow 0$. That would be correct for a finite-volume system
with a discrete spectral density. However, if the volume goes to infinity faster than $m$ goes
to zero (which is what one should assume in the thermodynamic limit) the second factor  
in the integrand becomes a representation of a $\delta$-function,

\beq 
\frac{m}{\lambda^2+m^2}\stackrel{m\rightarrow 0}{\longrightarrow} 
{\rm sign}(m)\,\pi\,\delta(\lambda), 
\la{thl}\eeq 
so that one obtains \cite{DP2a}:  

\beq 
\langle\bar\psi\psi\rangle=-\frac{1}{V}{\rm sign}(m) \pi\overline{\nu(0)}. 
\la{BC2}\eeq  
It is known as the Banks--Casher relation \cite{BC}. The chiral condensate is thus
proportional to the averaged spectral density of the Dirac operator at zero eigenvalues. 

The appearance of the sign function is not accidental: it means that at small $m$ QCD
partition function depends on $m$ non-analytically:  

\beq 
\ln{\cal Z}= V(c_0+\pi\overline{\nu(0)}|m|+c_2m^2\ln(|m|)+...). 
\la{nonan}\eeq  
The fact that the partition function is even in $m$ is the reflection of the original invariance of
the QCD under $\gamma_5$ rotations; the fact that it is non-analytic in the
symmetry-breaking parameter $m$ is typical for spontaneous symmetry breaking.
A generalization of the above formulae to the case of several quark flavours
can be found in ref. \cite{D2}.
 
%\newpage
\subsection{\it Physics: quarks hopping from one instanton to another} 

Below I follow refs. \cite{DP2a,DP2b}.

The key observation is that the Dirac operator in the background field of one (anti) instanton
has an exact zero mode with $\lambda=0$ \cite{tH}. It is a consequence of a general
Atiah--Singer index theorem; in our case it is guaranteed by the unit Pontryagin index or the
topological charge of the instanton field. These zero modes are 2-component Weyl spinors:
{\em right}-handed for instantons and {\em left}-handed for antiinstantons. Explicitly, the zero
modes are ($\alpha=1...N_c$ is the colour and $i,j,k=1,2$ are the spinor indices):  
\bea
\n
\left[\Phi_R(x-z_1)\right]^\alpha_i&=&\phi(x-z_1,\rho_1)(x-z_1)^+_{ij}
U_{1k}^\alpha\epsilon^{jk}, \\
\n
\left[\Phi_L(x-z_2)\right]^\alpha_i&=&\phi(x-z_2,\rho_2)(x-z_2)^-_{ij}
U_{2k}^\alpha\epsilon^{jk}, \\
 \phi(x,\rho)&=&\frac{\rho}{\pi(2x^2)^{1/2}(x^2+\rho^2)^{3/2}}.
\la{zm}\eea 
Here $z_{1\mu},\rho_1,U_1$ are the center, size and orientation of an instanton
and $z_{2\mu},\rho_2,U_2$ are those of an antiinstanton, respectively, $\epsilon^{jk}$ is the
$2\times 2$ antisymmetric matrix.  

For infinitely separated $I$ and $\bar I$ one has thus two degenerate states with exactly zero
eigenvalues. As usual in quantum mechanics, this degeneracy is lifted through the
diagonalization of the Hamiltonian, in this case the `Hamiltonian' is the full Dirac operator. 
The two ``wave functions" which diagonalize the ``Hamiltonian" are the sum and the difference
of the would-be zero modes, one of which is a 2-component left-handed spinor, and the
other is a 2-component right-handed spinor. The resulting wave functions are 4-component
Dirac spinors; one can be obtained from another by multiplying by the $\gamma_5$ matrix. 
As the result the two would-be zero eigenstates are split symmetrically into two
$4$-component Dirac states with {\em non-zero} eigenvalues equal to the overlap integral
between the original states:  

\bea
\nonumber
\lambda&=&\pm |T_{\rm I\bar I}|,\\
T_{\rm I\bar I}\!\!&=&\!\!\!\!\int\!\! d^4x\,
\Phi_1(x\!-\!z_1,U_1)^\dagger(\!-i\Dd)\Phi_2(x\!-\!z_2,U_2)\!
\stackrel{z_{12}\rightarrow\infty}{\longrightarrow}\! 
-\frac{2\rho_1\rho_2}{z_{12}^4}\Tr(U_1^\dagger U_2z_{12\,\mu}\sigma_\mu^+). 
\la{overl}\eea  
We see that the splitting between the would-be zero modes falls off as the third power of the
dist\-ance be\-tween $I$ and $\bar I$; it also depends on their relative orientation. The fact
that two levels have eigenvalues $\pm\lambda$ is in perfect agreement with the $\gamma_5$
invariance mentioned in the previous section.  

When one adds more \IIs each of them brings in a would-be zero mode.
After the diagonalization they get split symmetrically with respect to the $\lambda=0$ axis. 
Eventually, for an \II ensemble one gets a continuous band spectrum with a spectral density
$\nu(\lambda)$ which is even in $\lambda$ and finite at $\lambda=0$. 

Let the total number of \IIs in the 4-dimensional volume $V$ be $N$. The spread $\kappa$ of
the band spectrum of the would-be zero modes is given by their {\em average overlap}
\ur{overl}:  

\beq
\kappa^2=\frac{N}{V}\int\!d^4z_{12}dU_{12}\, |T_{12}|^2=
\frac{N\rho^4}{VN_c}\int\!\frac{d^4k}{(2\pi)^4}\,\frac{F^4(k\rho)}{k^2}
=6.62107\,\frac{N\rho^2}{VN_c}
\la{kappa}\eeq
where 

\beq
F(k\rho)=2t\left[I_0(t)K_1(t)-I_1(t)K_0(t)-\frac{1}{t}I_1(t)K_1(t)\right]_{t=\frac{k\rho}{2}}
\stackrel{k\to\infty}{\longrightarrow}\frac{6}{(k\rho)^3},\qquad
F(0)=1,
\la{F}\eeq
is the Fourier transform of the instanton zero mode \ur{zm}; the modified Bessel functions
are involved here. Numerically, if one takes the instanton density $N/V=(1\,{\rm fm})^{-4}$
and the average instanton size $\rho=\frac{1}{3}\,{\rm fm}$ (the old Shuryak's values
consistent with the variational estimate through $\Lambda$, see the previous section) one
obtains $\kappa=100\,{\rm MeV}$. 

In the random instanton ensemble, one gets the following spectral density of the
Dirac operator \cite{DP2b}:

\beq
\nu(\lambda)=\frac{N}{\pi\kappa}\sqrt{1-\frac{\lambda^2}{4\kappa^2}}
\la{specdens}\eeq
From \eq{BC2} one immediately finds the value of the chiral condensate:

\beq
\langle\bar\psi\psi\rangle = -\frac{1}{R^2\rho}\sqrt{\frac{N_c}{6.62}}
=-(253\,{\rm MeV})^3
\la{cond}\eeq
which is quite close to the phenomenological value \footnote{Contrary to the
gluon condensate, the chiral or quark condensate is somewhat dependent on the scale
where one estimates it. The above number refers to the scale given by the average instanton
size, that is 600 MeV.}. 

%%%%%%%%%%%%%%%%%%%%%%%%%%%%%%%%%%%
%%    FIGURE 5
%%%%%%%%%%%%%%%%%%%%%%%%%%%%%%%%%%%
\begin{figure}[t]
\setlength\epsfxsize{6.5cm}
\centerline{\epsfbox{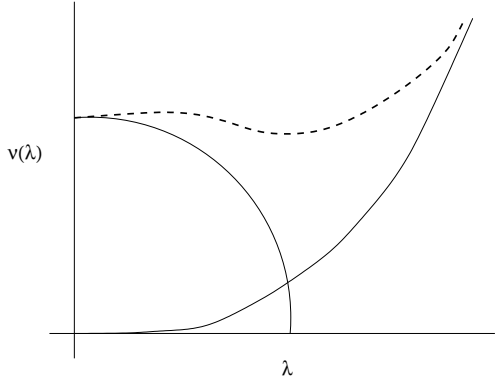}}
\caption{Schematic eigenvalue distribution of the Dirac operator.  The
solid lines are the zero mode and free contributions, the dashed line an
estimate of the full spectrum.}
\label{sdfig}
\vspace{.5 cm}
\end{figure} 

I would like to stress that the chiral consensate is not linear in the instanton density $N/V$
what one would naively expect but rather proportional to its square root (the gluon
condensate is, naturally, linear).  If the instanton density goes to zero the spectral density of
the Dirac operator tends to a $\delta$-function at zero eigenvalues. This is what one expects  
from the zero modes in the infinitely-dilute limit. 

\Eq{specdens} is known as the Wigner semicircle spectrum. For high eigenvalues
$\lambda \gg \kappa$ the spectral density is asymptotically given by that of  free massless
quarks:  

\beq
\nu(\lambda) \approx \frac{N_c}{4\pi^2}\lambda^3\,.
\la{freespdens}\eeq
Schematically, the combination of the low- and high-energy spectra are shown in Fig. 5
where the intereference with the intermediate modes with $\lambda\geq 2\kappa$ has
been ignored.

We see thus that the spontaneous chiral symmetry breaking by instantons
can be interpreted as a delocalization of the ``would-be'' zero modes, induced by the
background instantons, resulting from quarks hopping between them \cite{DP2a,DP2b}. 
Imagine random impurities (atoms) spread over a sample with final density, such that each
atom has a localized bound state for an electron. Due to the overlap of those localized
electron states belonging to individual atoms, the levels are split into a band, and the
electrons become delocalized. That means conductivity of the sample, the so-called
Mott--Anderson conductivity. In our case the localized zero quark modes of individual
instantons randomly spread over the volume get delocalized due to their overlap, which
means chiral symmetry breaking. 

This analogy between chiral symmetry breaking in QCD and the problem of electrons in
condensed matter systems with random impurities goes even further \cite{DP2b}. 
The acquisition of a dynamical mass by a quark is fully analogous to the appearance in the
Green function of an electron in a metal with impurities of a finite relaxation time 
(but in our case this time depends on the momentum). The appearance of the massless
pole in the pseudoscalar channel corresponding to the Goldstone pion is analogous to the
formation of a diffusion mode in the density-density correlation function.  For the recent
development of these and related ideas, see refs. \cite{JNPZ,OTV,Now} and references
therein. 

Recently, the instanton mechanism of the SCSB has been scrutinized by
direct lattice methods \cite{N2,Da,Net}. At present there is one group
\cite{Net} challenging the instanton mechanism. However, the density
of alternative `local structures' found there explodes as the lattice
spacing decreases, and this must be sorted out first. Studies by other
groups \cite{N2,Da} support  the mechanism described above.

\subsection{\it Quark propagator and dynamical quark mass}

The spectral density of the Dirac operator, averaged over the instanton vacuum,
carries very limited information, although one can already see that chiral symmetry is
spontaneously broken. In fact, one can compute analytically much more complicated 
correlation functions in the instanton vacuum, such as the quark propagator and
correlators of mesonic currents \cite{DP2b}. What is difficult to calculate analytically
can be done by numeric simulations of the instanton ensemble \cite{Sh2}.

%%%%%%%%%%%%%%%%%%%%%%%%%%%%%%%%%%%
%%    FIGURE 6
%%%%%%%%%%%%%%%%%%%%%%%%%%%%%%%%%%%
\begin{figure}[t]
\setlength\epsfxsize{6.5cm}
\centerline{\epsfbox{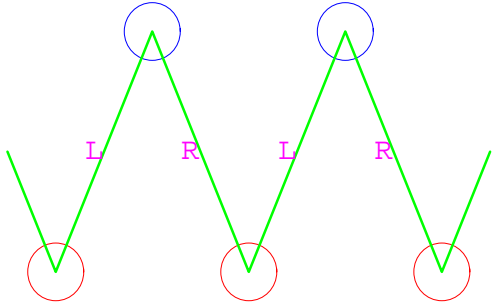}}
\caption{Quarks hopping from instantons to anti-instantons and {\it vice versa} flip helicity. 
An infinite number of such jumps generates the dynamical mass $M(p)$. Actually, one
has to take into account that quarks can `return' to the same pseudoparticle infinitely
many times \cite{DP2b}.}
\label{prop1}
\vspace{.5 cm}
\end{figure} 

Each time a quark `hops' from one random instanton to an anti-instanton 
(and {\it vice versa}) it has to change its helicity, because instanton's zero mode
is right-handed while the anti-instanton's one is left-handed, see the schematic drawing in
Fig. 6. Delocalization implies quarks make an infinite number of such jumps. An infinite
number of helicity-flip transitions generates a non-slash term in the quark propagator, i.e. the
dynamically-generated mass $M(p)$, see Fig. 7. It implies the spontaneous chiral symmetry
breaking.

Mathematically, one has to consider the quark propagator in the gluon background
being the superposition of an infinite number of \IIs, and then average the propagator
over the positions, sizes and orientations of instantons according to their partition function
\ur{GPF}. This is a hopeless task, unless one exploits the fact that the packing fraction
of instantons is small. The actual expansion parameter is 
$\alpha=\pi^2\rho^4N/(VN_c)\sim 1/20$ which is not so bad. In the leading order in that
parameter one can derive a closed equation for the quark propagator averaged over the
ensemble. Its solution has the form \cite{DP2b,Pob}

\beq
G(p)=Z(p)\,\frac{\Dpp +iM(p^2)}{p^2+M^2(p^2)}.
\la{qprop}\eeq
The `wave function renormalization' factor $Z(p)$ differs from unity by a function proportional
to the above small parameter $\alpha$, and this difference will be neglected. The dynamical
quark mass $M(p)$ is, on the contrary, proportional to the {\em square root} of the
packing fraction:

\beq
M(p^2)={\rm const.}\,\sqrt{\frac{\pi^2N\rho^2}{VN_c}}\,F^2(p\rho),
\la{Mp}\eeq 
with the function $F(t)$ given by \eq{F}; it is related to the Fourier transform of the
zero mode \footnote{It has been known from the perturbative analysis of the 1970's that
asymptotically $M(p)\sim (\alpha_s/4\pi)<\!\bar\psi\psi\!>/p^2$ whereas \eq{Mp}
gives at large virtuality $M(p)\sim 1/p^6$. This is because perturbative gluons are
ignored in the instanton derivation. At very large $p$ the perturbative regime $\sim 1/p^2$
has to take over.}. 

The overall numerical constant is found from the self-consistency or gap 
equation \cite{DP2b}:

\beq
4N_c\int \frac{d^4 p}{(2\pi)^4}\frac{M^2(p)}{M^2(p)+p^2}
=\frac{N}{V}.
\la{selfcons}\eeq
For the `standard' values of the instanton ensemble, $N/V=(1\,{\rm fm})^{-1},\;
\rho=\frac{1}{3}\,{\rm fm}$, one gets at zero momentum $M(0)=345\,{\rm MeV}$. 
The dynamical mass \ur{Mp} is plotted in Fig. 7 on top of the recent lattice data
for this quantity obtained by an extrapolation to the chiral limit \cite{Mlat}. 

%%%%%%%%%%%%%%%%%%%%%%%%%%%%%%%%%%%
%%    FIGURE 7
%%%%%%%%%%%%%%%%%%%%%%%%%%%%%%%%%%%
\begin{figure}[t]
\setlength\epsfxsize{7.5cm}
\centerline{\epsfbox{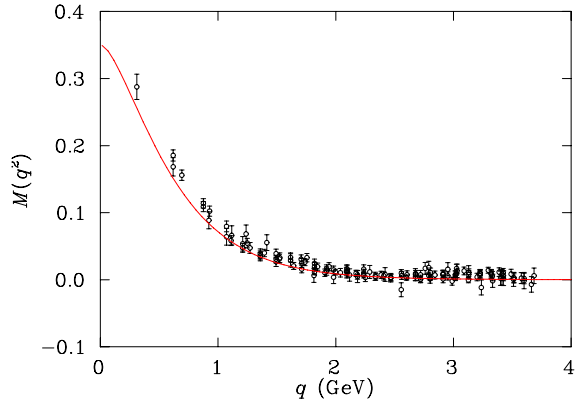}}
\caption{Dynamical quark mass $M(p)$ \cite{DP2b} as function of quark virtuality. The
scattered points are lattice data obtained by exprapolation to the continuum and
chiral limits \cite{Mlat}. Courtesy P.~Bowman.}
\label{Mofp_lat}
\vspace{.5 cm}
\end{figure}

Knowing the quark propagator one is able to compute the chiral condensate directly
without referring to the Banks--Casher relation. By definition, the chiral condensate
is the quark propagator taken at one point; in momentum space it is a closed quark loop:

\beq
-\langle\bar\psi\psi\rangle_{\rm Mink}=i\langle\psi^\dagger\psi\rangle_{\rm Eucl}=
4N_c\idp\,\frac{M(p)}{p^2+M^2(p)}= {\rm const.}\,\sqrt{\frac{N\,N_c}{\pi^2\,V\,\rho^2}}.
\la{chcond2}\eeq
Putting in the `standard' instanton ensemble parameters one gets the same value of the
condensate as before: $\langle\bar\psi\psi\rangle = -(253\,{\rm MeV})^3$. 

Furthermore, using the small packing fraction as an expansion parameter one can also
compute \cite{DP2b} more complicated quantities like 2- or 3-point mesonic correlation
functions of the type  

\beq 
\langle J_A(x)J_B(y)\rangle, \qquad \langle J_A(x)J_B(y)J_c(z)\rangle, \qquad
J_A=\bar\psi\Gamma_A\psi 
\la{mescorr}\eeq 
where $\Gamma_A$ is a unit matrix in colour but an arbitrary matrix in flavour and spin.
Instantons influence the correlation functions in two ways: {\it i}) the quark and antiquark
propagators get dressed and obtain the dynamical mass, as in \eq{qprop}, {\it ii}) quark and
antiquark may scatter simultaneously on the same pseudoparticle; that leads to certain
correlations between quarks or, in other words, to effective quark interactions. These
interactions are strongly dependent on the quark-antiquark quantum numbers: they are strong
and attractive in the scalar and especially in the pseudoscalar and the axial channels, and
rather weak in the vector and tensor channels.  I shall derive these interactions in the next
section, but already now we can discuss the pseudoscalar and the axial isovector channels.
These are the channels where the pion shows up as an intermediate state.  Since we have
already obtained chiral symmetry breaking by studying a single quark propagator in the
instanton vacuum, we are doomed to have a massless Goldstone pion in the appropriate
correlation functions. However, it is instructive to follow how does the Goldstone theorem
manifest itself in the instanton vacuum. It appears that technologically it follows from a kind of
detailed balance in the pseudoscalar channel (such kind of equations are encountered in
perturbative QCD where there is a delicate cancellation between real and virtual gluon
emission). Since we have a concrete dynamical realization of chiral symmetry breaking we
can not only check the general Ward identities of the partially conserved
axial currents (which work of course) but we are in a position to find quantities whose values
do not follow from general relations. One of the most important quantities is the $F_\pi$
constant: it can be calculated as the residue of the pion pole. One obtains \cite{DP2b,D3}:  

\bea
\n
F^2_\pi &\approx &
4N_c\int \frac{d^4 p}{(2\pi)^4}\,\frac{M^2(p)}{[M^2(p)+p^2]^2} \\
&=&{\rm const.}\cdot\frac{N}{V}\,
\bar\rho^2\,\ln\frac{\bar R}{\bar\rho} \approx
(100\,{\rm MeV})^2\qquad {\it vs.}\;\; (93\;{\rm MeV})^2\quad ({\rm exper.}).
\la{Fpi}\eea

This is an instructive formula. The point is, $F_\pi$ is anomalously small in the strong
interactions scale which, in the instanton vacuum, is given by the average size of
pseudoparticles, $1/\bar\rho \simeq 600\;MeV$. The above formula says that $F_\pi$ is
down by the packing fraction factor $(\bar\rho/\bar R)^2\simeq 1/9$. It can be said that
$F_\pi$ measures the diluteness of the instanton vacuum. However it would be wrong to say
that instantons are in a dilute gas phase -- the interactions are crucial to stabilize the medium
and to support the known renormalization properties of the theory, therefore they are rather in
a liquid phase, however dilute it may turn to be.  By calculating three-point correlation
functions in the instanton vacuum it is possible to determine {\it e.g.} the charge radius of the
pion as the Goldstone excitation \cite{DP2b}:  

\beq 
\sqrt{r_\pi^2}\simeq \frac{\sqrt{N_c}}{2\pi F_\pi}\simeq (340\;{\rm MeV})^{-1} \qquad
{\it vs.}\;\;(310\;{\rm MeV})^{-1}\;\;({\rm exper.}). 
\la{chra}\eeq  

In flavour-singlet pseudoscalar channel the instanton-induced interactions
are not strong attraction as in the non-singlet channel. Therefore, the $\eta^\prime$
meson is not a Goldstone boson: the famous $U_A(1)$ problem is solved by instantons,
as anticipated at the very beginning of the instanton era by 't Hooft \cite{tH}. 
Moreover, in the limit $N_f/N_c\rightarrow 0$ instantons reproduce \cite{DP3,DPW}
the theoretical Witten--Veneziano formula \cite{W,V} for the singlet $\eta^\prime$ mass, 
as given by

\beq
m_{\eta^\prime}^2=\frac{2N_f<\!Q_T^2\!>/V}{F_\pi^2}
\la{etap}\eeq
where $<\!Q_T^2\!>/V=<\!(N_+-N_-)^2\!>/V$ is the topological susceptibility. In the 
instanton vacuum it is related to the difference between the number of \IIs. It should be
stressed that \eq{etap} is correct in the chiral limit, and that the topological susceptibility 
is that of the pure-glue world, without quarks. As emphasized in subsection 2.7, 
the instanton vacuum is described by the {\em grand} canonical ensemble of \IIs, 
with the fluctuating number of pseudoparticles $N_\pm$. In the $CP$-conserving
vacuum, {\it i.e.} for the `instanton angle' $\theta=0$, one finds the equal averages 
$<\!N_+\!>=<\!N_-\!>\sim V$ from a saddle-point equation. At $\theta\neq 0$ 
the saddle-point values for $<\!N_\pm\!>$ are complex conjugate to each 
other~\cite{DP1,DPW}. The square of the difference between the numbers of
\IIs is behaving in the normal thermodynamic way,  $<\!(N_+-N_-)^2\!>\sim V$,
and gives rise to the topological susceptibility $<\!Q_T^2\!>$. 

When the back influence of quarks on the instanton ensemble is taken into
account, which is a $O(N_f/N_c)$ effect, the average $<\!(N_+-N_-)^2\!>$
gets dynamically suppressed since at $N_+\neq N_-$ the number of 
left- and right-handed zero modes are not equal, and the fermion determinant
goes to zero in the chiral limit~\cite{NVZ,LS,DPW}. This is in accordance with
the general anomalous Ward identites in the $U_1(A)$ channel \cite{V}. 

The Witten--Veneziano formula \ur{etap} is an idealization at $N_c\to\infty$,
$m\to 0$. For non-zero quark masses there is a singlet-octet mixing \cite{V,DE}
resulting in physical $\eta,\eta^\prime$ mesons. Actually the mixing angle
appears to be rather small -- about $10^\circ$. 
        
Let me note that all quantities exhibit the natural behaviour in the number of colours 
$N_c$~\cite{DP3,SNc}:
\bea
\n
\langle F_{\mu\nu}^2\rangle \sim \frac{N}{V} &=& O(N_c),\qquad 
\langle \bar\psi \psi\rangle = O(N_c),\qquad F_\pi^2= O(N_c), \\
\bar\rho &=& O(1),\qquad M(0)=O(1),\qquad \sqrt{r_\pi^2} =O(1),\;\; {\rm etc.} 
\la{laNc}\eea  

A systematic numerical study of various correlation functions in the instanton vacuum has
been performed by Shuryak, Verbaarschot and Sch\"afer \cite{Sh2}, see also the review
\cite{Sh4}. In all cases considered the results agree well or very well with experiments and
phenomenology. 

%\vskip 2true cm 

\section{Instanton-induced interactions}
\setcounter{equation}{0}
\def\theequation{\arabic{section}.\arabic{equation}}  

There are two philosophically different but mathematically equivalent ways of
computing observables in the instanton vacuum. The first is to compute an observable
for a given configuration of \IIs and then average over the ensemble \cite{DP2b}. The 
second is opposite: first average over the ensemble \cite{DP3}. 
Since two or more quarks can  scatter off the same instanton or anti-instanton, averaging
over their positions and orientations in colour space induce certain correlations between
quarks, which one can also call the interactions. The result is an effective action for quarks
which contains instanton effects in induced multi-quark interactions.  The
would-be zero modes serve here as a bridge, passing information from the
instanton vacuum to the effective quarks through the induced vertex.  The
consequent interactions are vertices involving $2N_f$ quarks, commonly cited
as 't Hooft interactions since he was the first to specify the proper quantum numbers.

It is convenient to decompose the 4-component Dirac bi-spinors
describing quark fields into left- and right-handed Weyl spinors
which we denote as

\beq
\psi^{f\alpha i}_{L(R)},\;\;\;\;\;\psi^\dagger_{L(R)f\alpha i},
\la{psi}\eeq
where $f=1...N_f$ are flavour, $\alpha=1...N_c$ are colour
and $i=1,2$ are spinor indices. Let us introduce the
't Hooft-like $2N_f$-fermion vertices generated by \IIs,
which we denote by $Y_{N_f}^{(\pm)}$, respectively. These vertices
are obtained by explicit averaging over (anti)instanton orientation
matrices $U^\alpha_i$ and over the instanton size distribution
$d(\rho)$ \ur{gauss}. Averaging over instanton positions in $4d$ Euclidean
space--time produces the overall conservation of momenta of quarks
entering the vertex $Y$, hence it is convenient to write down
the quark interaction vertex in the momentum space. There are
formfactor functions $F(k\rho)$ \ur{F} associated with the Fourier
transform of the fermion zero modes of one instanton, attached to each
quark line entering the vertex. The $2N_f$-fermion vertex induced
by an instanton  (see Fig. 8) is, in momentum space,

%%%%%%%%%%%%%%%%%%%%%%%%%%%%%%%%%%%
%%    FIGURE 8
%%%%%%%%%%%%%%%%%%%%%%%%%%%%%%%%%%%
\begin{figure}[t]
\setlength\epsfxsize{5cm}
\centerline{\epsfbox{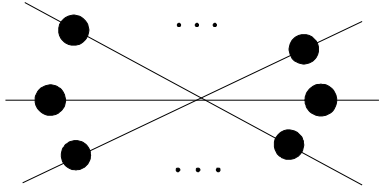}}
\caption{Instanton-induced $2N_f$-quark vertex. The black
blobs denote the formfactor functions $F(k\rho)$ attached to each
quark leg.}
\label{nucfig1}
\end{figure} 
\vskip .5true cm

\[
Y_{N_f}^+=\int\!d\rho\:d(\rho)\int\!
dU\prod_{f=1}^{N_f}\left\{\int\!\frac{d^4k_f}{(2\pi)^4}\: \left[2\pi\rho
F(k_f\rho)\right] \int\!\frac{d^4l_f}{(2\pi)^4}\: \left[2\pi\rho
F(l_f\rho)\right]\right.
\]
\beq
\left.\cdot(2\pi)^4\delta(k_1+...+k_{N_f}-l_1-...-l_{N_f})
\cdot U_{i^\prime_f}^{\alpha_f}U_{\beta_f}^{\dagger j^\prime_f}
\epsilon^{i_f i^\prime_f}\epsilon_{j_f j^\prime_f}
\left[i\psi_{Lf\alpha_fi_f}^\dagger(k_f)\psi_L^{f\beta_fj_f}(l_f)
\right]\right\}.
\la{Y0}\eeq
For the $Y^-$ vertices induced by $\bar I$'s one has to replace
left-handed Weyl spinors $\psi_L, \psi^\dagger_L$ by
right-handed ones, $\psi_R, \psi^\dagger_R$.
Using these vertices one can write down the partition function
to which QCD is reduced at low momenta, as a functional integral
over quark fields \cite{DP3,D2,DPW}:

\beq
{\cal Z} =\int\!\!D\psi D\psi^\dagger
\exp\left(\int\!d^4x\sum_{f=1}^{N_f}\bar\psi_f i\Dd \psi^f\right)
\left(\frac{Y_{N_f}^+}{VM_1^{N_f}}\right)^{N_+}
\left(\frac{Y_{N_f}^-}{VM_1^{N_f}}\right)^{N_-}
\la{Z0}\eeq
where $N_\pm$ are the number of \IIs in the whole $4d$ volume
$V$. The volume factors in the denominators arise because of
averaging over individual instanton positions, and certain mass
factors $M_1^{N_f}$ are put in to make \eq{Z0} dimensionless.
Actually, the mass parameter $M_1$ plays the role of separating
high-frequency part of the fermion determinant in the instanton
background from the low-frequency part considered here.
Its concrete value is irrelevant for the derivation
of the low-energy effective action performed below; in fact it is
established from smooth matching of high- and low-frequency
contributions to the full fermion determinant in the instanton
vacuum \cite{DP2a,DP2b}.

Having fermion interactions in the pre-exponent of the partition
function is not convenient: one should rather have the
interactions in the exponent, together with the kinetic energy
term. This can be achieved by rewriting \eq{Z0} with the help
of additional integration over Lagrange multipliers
$\lambda_\pm$:

\[
{\cal Z} =
\int\!\frac{d\lambda_\pm}{2\pi}\int\!\!D\psi D\psi^\dagger
\exp\left\{N_+\left(\ln\frac{N_+}{\lambda_+VM_1^{N_f}}-1\right)
+N_-\left(\ln\frac{N_-}{\lambda_-VM_1^{N_f}}-1\right)\right.+
\]
\beq
\left.+\int\!d^4x
\sum_{f=1}^{N_f}\bar\psi_f i\Dd \psi^f
+\lambda_+Y_{N_f}^++\lambda_-Y_{N_f}^-
\right\}\!.
\la{Z1}\eeq
Since $N_\pm\sim V\rightarrow\infty$ integration over
$\lambda_\pm$ can be performed by the saddle-point method; the
result is \eq{Z0} we started from.

As seen from \eq{Z1}, $\lambda_\pm$ plays the role of the coupling
constant in the many-quark interactions. It is very important
that their strength is not pre-given but is, rather, determined
self-consistently from the fermion dynamics itself; in
particlular, the saddle-point values of $\lambda_\pm$ depend on
the phase quarks assume in the instanton vacuum. In the chiral symmetry 
broken phase the values of $\lambda_\pm$,
as determined by a saddle-point equation, appear to be real.

To get the $2N_f$-fermion vertices \ur{Y0} in a closed form one
has to explicitly integrate over instanton orientations in colour
space. For the $2N_f$-fermion vertex one has to
average over $N_f$ pairs of $(U, U^\dagger)$. In particular,
one has:

\[
\int dU=1,\;\;\;\;\int dU\:U_i^\alpha U_\beta^{\dagger j}
=\frac{1}{N_c}\delta_\beta^\alpha\delta_i^j,\;\;\;\;\;
\int dU\:U_{i_1}^{\alpha_1}U_{i_2}^{\alpha_2}
U_{\beta_1}^{\dagger j_1}U_{\beta_2}^{\dagger j_2}
\]
\beq
=\frac{1}{N_c^2-1}\left[\delta^{\alpha_1}_{\beta_1}
\delta^{\alpha_2}_{\beta_2}
\left(\delta^{j_1}_{i_1}\delta^{j_2}_{i_2}-\frac{1}{N_c}
\delta^{j_1}_{i_2}\delta^{j_2}_{i_1}\right)
+\delta^{\alpha_1}_{\beta_2}\delta^{\alpha_2}_{\beta_1}
\left(\delta^{j_1}_{i_2}\delta^{j_2}_{i_1}-\frac{1}{N_c}
\delta^{j_1}_{i_1}\delta^{j_2}_{i_2}\right)\right],
\;\;\;{\rm etc.}
\la{aver}\eeq

\vskip .5true cm

In the simplest case of one quark flavour, $N_f=1$, the ``vertex'' \ur{Y0} is just a mass term
for quarks,

\beq
Y^\pm_1=\frac{i}{N_c}\int\frac{d^4k}{(2\pi)^4}
\int\!d\rho\:\nu(\rho)[2\pi\rho F(k\rho)]^2
\left[\psi_\alpha^\dagger(k)\frac{1\pm\gamma_5}{2}\psi^\alpha(k)
\right]\!,
\la{Y1}\eeq
with a momentum dependent dynamically-generated mass $M(k)$ given by

\beq
M(k)=\frac{\lambda}{N_c}\int\!d\rho\:\nu(\rho)\left[2\pi\rho
F(k\rho)\right]^2 \approx\frac{\lambda}{N_c}\left[2\pi\bar\rho
F(k\bar\rho)\right]^2.
\la{momdep}\eeq
In fact, it is exactly the same dynamical mass \ur{Mp} as obtained in refs. \cite{DP2b,Pob}
by first considering the quark propagator in a given configuration of \IIs and then
averaging over the ensemble. This result comes as follows.   

In order to find the overall scale $\lambda$ of the dynamical mass
one has to put \ur{Y1} into \eq{Z1}, integrate over
fermions, and find the minimum of the free energy with respect to
$\lambda_\pm$.  At $\theta=0$ the QCD vacuum is $CP$ invariant so that
$N_+=N_-=N/2$ and consequently $\lambda_+=\lambda_-=\lambda$
\footnote{Fluctuations of the topological charge, $N_+-N_-$, leading to
the topological susceptibility (related to the solution of
the $U(1)$ problem) has been considered in refs.\cite{DP3,DPW}.}.
In this case the $\gamma_5$ term in $Y^\pm$ gets cancelled, and the
exponent of the partition function \ur{Z1} reads:

\[
-N\ln \lambda+\int\!d^4x\int\!\frac{d^4k}{(2\pi)^4}
\Tr\ln\left\{\Dk+i\frac{\lambda}{N_c}[2\pi\bar\rho
F(k\bar\rho)]^2 \right\}
\]
\beq =-N\ln \lambda+2N_cV\int\!\frac{d^4k}{(2\pi)^4}
\ln\left\{k^2+\left(\frac{\lambda}{N_c}[2\pi\bar\rho F(k\bar\rho)]^2
\right)^2\right\}.
\la{Z11}\eeq

Differentiating it with respect to $\lambda$ and using \eq{momdep} one gets the gap equation
\eq{selfcons}  which is in fact a requirement on the overall scale of the constituent quark
mass $M(k)$; its momentum dependence is anyhow given by \eq{momdep}.  Since the
momentum integration in \eq{selfcons} is well convergent and is actually
cut at momenta $k\sim 1/\bar\rho$, the saddle-point value of the
Lagrange multiplier $\lambda$ is of the order of $\sqrt{N_cN/V}/\bar\rho$. The steepness of
the saddle-point integration in $\lambda$ is guaranteed by the volume, hence the use of the
saddle-point method is absolutely justified. Note that \eq{Y1} reproduces the massive quark
propagator \ur{qprop}, hence the chiral condensate is, as before, given by \eq{chcond2}. \\

For two flavours, $N_f=2$  averaging \eq{Y1} over the instanton orientations with
the help of \eq{aver} gives a nontrivial 4-fermion interaction.
It is, of course, non-local: a formfactor function $F(k\rho)$ is
attributed to each fermion entering the vertex; in addition
it should be averaged over the sizes of instantons. The non-locality
is thus of the order of the average instanton size in the vacuum.
One has \cite{DP3,D3}:

\[
Y_2^+=\frac{i^2}{N_c^2-1}\int\frac{d^4k_1d^4k_2d^4l_1d^4l_2}
{(2\pi)^{12}}\delta(k_1+k_2-l_1-l_2)
\]
\[
\cdot\int\!d\rho\:\nu(\rho)\:(2\pi\rho)^4
F(k_1\rho)F(k_2\rho)F(l_1\rho)F(l_2\rho)
\]
\[
\cdot\frac{1}{2!}\epsilon^{f_1f_2}\epsilon_{g_1g_2}\left\{
\left(1-\frac{1}{2N_c}\right)[\psi_{Lf_1}^\dagger(k_1)\psi_L^{g_1}(l_1)]
[\psi_{Lf_2}^\dagger(k_2)\psi_L^{g_2}(l_2)]\right.
\]
\beq
\left. +\frac{1}{8N_c}
[\psi_{Lf_1}^\dagger(k_1)\sigma_{\mu\nu}\psi_L^{g_1}(l_1)]
[\psi_{Lf_2}^\dagger(k_2)\sigma_{\mu\nu}\psi_L^{g_2}(l_2)]\right\}\!.
\la{Y2}\eeq
For the $\bar I$-induced vertex $Y^-$ one has to replace
left-handed components by right-handed ones. In all square
brackets summation over colour is understood. Note that the
last-line (tensor) term is suppressed at large $N_c$; it,
however, is crucial at $N_c=2$ to support the actual $SU(4)$
chiral symmetry in that case \cite{DP3}. The antisymmetric
$\epsilon^{f_1f_2}\epsilon_{g_1g_2}$ structure demonstrates that
the interactions have a determinant form in the two flavours.
Using the identity

\beq
2\epsilon^{f_1f_2}\epsilon_{g_1g_2}=\delta_{g_1}^{f_1}\delta_{g_2}^{f_2}
-(\tau^A)_{g_1}^{f_1}(\tau^A)_{g_2}^{f_2}
\la{id}\eeq
and adding the $\bar I$-induced vertex $Y_2^-$
one can rewrite the leading-$N_c$ (first) term of \eq{Y2} as

\beq
(\psi^\dagger\psi)^2+(\psi^\dagger\gamma_5\psi)^2
-(\psi^\dagger\tau^A\psi)^2-(\psi^\dagger\tau^A\gamma_5\psi)^2
\la{NJLf}\eeq
which resembles closely the interaction of Nambu--Jona-Lasinio \cite{NJL} model. 
It should be stressed though that in contrast to that {\em at hoc} model the
full 't~Hooft interaction {\it i}) violates the $U_A(1)$ symmetry in a very definite
manner, {\it ii}) has a fixed interaction strength related to the density of instantons and {\it iii})
contains an intrinsic ultraviolet cutoff due to the formfactor functions $F(k\rho)$. In addition, 
at $N_c=2$ it correctly preserves the actual Pauli--G\"ursey $SU(4)$ chiral 
symmetry~\cite{DP3}. It should be added that the naive addition of a nonzero current quark
mass to the NJL Lagrangian fails to reproduce several known low-energy Ward identities, as
well as the phenomenologically-known coefficients in the Gasser--Leutwyler chiral
Lagrangian (the terms containing $m^2$ and $m\cdot p^2$). The microscopic instanton
approach preserving all symmetries of QCD is capable to correctly incorporate nonzero
quark masses, and it does so in a rather nontrivial way \cite{Mus}. \\

Higher number of flavours, $N_f\geq 3$, have been considered in ref. \cite{DP3,D3} 
where also the bosonization of the instanton-induced interactions has been performed.
For three quark flavours four-fermion interactions cannot be sufficient, as the $U_A(1)$
symmetry is not broken without a 6-fermion interaction involving all three flavours.  
In this case the four-fermion interaction is not `fundamental' but arises from the 
$2N_f$-fermion 't~Hooft vertex after one integrates out, say, the strange quark.\\

I have been often asked: Why mention instantons when using four-fermion interactions? 
Isn't it just the Nambu--Jona-Lasinio model? The answer is that the NJL model is not QCD: 
if it is the correct low-energy effective theory it should be derived from QCD. 
The model has been proposed a decade before the advent of QCD and formulated in
terms of nucleons, not quarks, and time has come to learn what is QCD suggesting instead 
of the NJL model. A concrete form of NJL-like interactions of quarks, with specific formfactors
and strength related to $\Lambda$, is derived from QCD via instantons. As indicated above,
it preserves correctly the symmetries of QCD. Phenomenologically, it is successful.
Instantons enable one to understand clearly the domain of applicability of the low-energy
effective theory, its limitations and ways to generalize it, for example if one wishes 
to take non-zero current quark masses or to include perturbative gluons, see below. I can
admit that the effective four-fermion interactions with some couplings and formfactors may
follow from the microscopic QCD via another, non-instanton, formalism. Had there been an
alternative derivation (from dyons, vortices or whatever) then we could compare the effective
models following from different microscopic considerations and choose the one which fits
experiments better. We do not even know if potential alternative mechanisms of the SCSB
could lead to anything resembling the NJL model at low momenta. Today there exists only
one derivation of the NJL-like model and it is from instantons. Therefore, unless the situation
changes, I would refer to instantons when four-fermion interactions are used as an effective
low-energy theory. \\

There has been an interesting application of the instanton-induced interactions
to the problem of a phase transition from the usual chiral-broken phase to the
so-called colour superconducting phase. The point is, the instanton-induced vertex
\ur{Y2}, being Fierz-transformed to the diquark channel (as contrasted to the quark-antiquark 
one), gives an attraction for two quarks with the $L=0,\,S=0,\,{\rm isospin}=0$ 
quantum numbers \cite{BLa,DFL}. At $N_c=2$ the strength of this attraction is 
exactly the same as in the $\bar qq$ channel owing to the $SU(4)$ symmetry mentioned
above \cite{DP3}. At $N_c\geq 3$ the attraction in the $\bar q q$ channel prevails, leading
under normal circumstances to the $\bar q q$ condensation which breaks chiral symmetry
\cite{DFL}. However, at sufficiently high baryon number density the relative strengths
of the effective attraction in the $\bar q q$ and $qq$ channels reverse: it becomes
favourable for the diquarks to condense \cite{RSSV,ARW}. Since diquarks are in a
triplet state for the $SU(3)$ colour, diquark condensation means the spontaneous
breaking of colour symmetry, hence the term `colour superconductivity'. On the contrary, 
the chiral symmetry gets restored at high matter density. It is a first order phase transition. 

The instanton-induced phase transition has been studied in some detail in ref. \cite{CD1} 
along the lines of the present section. To that end one needs to generalize the formfactor
functions of the instanton-induced vertex to nonzero chemical potentials. The diquark
condensation at high matter density mimics the Higgs phenomenon and gives rise to
`Meissner' masses of gluons. To find the gluon masses one needs to construct
a conserved N\"other current from the non-local 't~Hooft vertex or, in other words, 
to find out how to generalize 't~Hooft vertex to include perturbative gluons into it~\cite{CD2}. 
  
%\vskip 2true cm 
   
\section{Effective Chiral Lagrangian}
\setcounter{equation}{0}
\def\theequation{\arabic{section}.\arabic{equation}}  

It is very important that the dynamical quark mass is parametrically
much less than $1/\bar\rho$: the dimensionless quantity

\beq
\left(M\bar\rho\right)^2\sim\frac{\pi^2\bar\rho^4 N}{VN_c}\;\ll\;1
\la{p}\eeq
is suppressed by the packing fraction of instantons in the vacuum. The whole approach to
the instanton vacuum implies that instantons are on the average relatively dilute and that this
packing fraction is numerically small. At low momenta $k\le 1/\bar\rho=600\,{\rm MeV}$ 
there are exactly two degrees of freedom left: quarks with a dynamical mass 
$M\ll 1/\bar\rho$ and the massless Goldstone pions. Let us formulate the theory to which
QCD is reduced at low momenta  $k\le 1/\bar\rho$: it will involve only quarks with dynamical
mass, interacting with pions. 

Bosonizing the instanton-induced interaction \ur{Y2} and neglecting all heavy fields
one arrives to the effective theory to which QCD is reduced at low momenta. 
The theory is defined by the partition function \cite{DP2b,DP3} 

\[
{\cal Z}=\int\!D\pi^A\int\!D\psi^\dagger D\psi
\exp\int\!d^4x \left\{\psi_f^\dagger(x) i\Dd \psi^f(x)
+i\int\!\frac{d^4kd^4l}{(2\pi)^8}e^{i(k-l,x)}\sqrt{M(k)M(l)}
\right.
\]
\[
\left.
\cdot\left[\psi_{f\alpha}^\dagger(k)\left(U^f_g(x)
\frac{1+\gamma_5}{2}+U^{\dagger f}_g(x)\frac{1-\gamma_5}{2}
\right)\psi^{g\alpha}(l)\right]\right\},\]
\beq
U^f_g(x)=\left(\exp i\pi^A(x)\lambda^A)\right)^f_g.
\la{Znash}\eeq
\Eq{Znash} shows quarks interacting with chiral fields $U(x)$, with
formfactor functions equal to the square root of the dynamical quark
mass attributed to each vertex where $U(x)$ applies. The matrix
entering in the parentheses is actually a $N_f\times N_f$ matrix in
flavour and a $4\times 4$ matrix in Dirac indices. It can be
identically rewritten as

\beq
U(x)\frac{1+\gamma_5}{2}
+U^\dagger (x)\frac{1-\gamma_5}{2}
=\exp\left(i\pi^A(x)\lambda^A\gamma_5\right)\equiv U^{\gamma_5}(x),
\la{pasha}\eeq
the industrious final abbreviation being due to Pavel Pobylitsa.

The formfactor functions $\sqrt{M(k\bar\rho)}$ for each quark
line attached to the chiral vertex automatically cut off momenta
at $k\ge 1/\bar\rho$. In the range of quark momenta $k\ll 1/\bar\rho$
(which we shall be mostly interested in) one can neglect this
non-locality, and the partition function \ur{Znash} is simplified
to a local field theory:

\beq
{\cal Z}=\int\!D\pi^A\int\!D\psi^\dagger D\psi\;
\exp\int\!d^4x\: \psi^\dagger(x)\left[i\Dd+iMU^{\gamma_5}(x)\right]
\psi(x).
\la{Zna}\eeq
One should remember, however, to cut the quark loop integrals
at $k\approx 1/\bar\rho\approx 600\,{\rm MeV}$. Notice that there is
no kinetic energy term for pions: it appears only after one integrates
over the quark loop, see below. Summation over colour is assumed in the
exponent of \eq{Zna}. \Eq{Zna} defines a simple and elegant local field theory although it is
still a highly non-trivial one. This theory has been reviewed in some detail in ref. \cite{D3}. 

\vskip 1true cm 

\section{Chromomagnetic pomeron}
\setcounter{equation}{0}
\def\theequation{\arabic{section}.\arabic{equation}}  

The soft part of hadron-hadron collisions at high energies is still
full of riddles. Why is the additive quark model so successful
phenomenologically, stating that hadron-hadron collision can be
viewed, to a good approximation, as an incoherent sum of collisions
of their constituent quarks \cite{LFLSAS,FF}? 
Why is the total $pp$ cross section around $40\,{\rm mb}$? This number,
as any other in strong interactions, needs to be obtained from the
transmutation of dimensions and expressed through $\Lambda$ \ur{Lambda}. 
Why is the differential elastic cross section approximately proportional to the 
fourth power of the electromagnetic formfactor? Why are spin effects 
represented, e.g., by the analyzing power and the polarization of the  
inclusively produced hyperons, so strong and independent of energy  \cite{revpol}? 

The traditional view on a pomeron is that it is basically a
two-gluon exchange {\it \`a la} Low-Nussinov \cite{LN} or its more
sophisticated Balitsky-Fadin-Kuraev-Lipatov sibling \cite{BFKL},
see the recent book \cite{DDLN} for a review.
Both versions are infrared-unsafe, in the sense that the transverse
momenta carried by gluons are cut from below not ``by themselves'' but
rather by the inverse sizes of the colliding hadrons. [In the BFKL case
one eventualy runs into the infrared region even if the hadron sizes
are vanishing.] In the Nachtmann--Dosch model \cite{ND} where
non-perturbative scattering of strings between quarks inside hadrons is
considered, the characteristic transverse scale is also set by hadron
sizes; both models are opposite in spirit to the additive quark model,
so that e.g. the famous Levin--Frankfurt ratio \cite{LFLSAS}
$\sigma_{NN}:\sigma_{\pi N}:\sigma_{\pi \pi}=9:6:4$ appears to be
accidental. None of the pomeron models suggested so far are able to
explain coherently the riddles of large spin effects in high energy
production \cite{Soffer}.

In this section we would like to suggest another point of view on the soft
pomeron~\footnote{This section has been prepared in co-authorship 
with Maxim Polyakov. We thank V. Petrov, M. Ryskin and M. Strikman for 
discussing these matters.}. From the instantons' point of view, the leading
{\it soft} gluon exchange flips quark's helicity and not conserves it, as
in perturbation theory. An interference between the two processes may
lead to the interesting zoo observed in spin physics in high-energy
hadron reactions \cite{Soffer}.   

We know already from the previous sections that instantons induce very 
strong helicity-flip interactions of quarks, known as 't Hooft interactions. 
Being point-like (more precisely, of the range of $\bar\rho$), they are irrelevant
at high energies. However, since 't Hooft interactions are induced by
instantons that are lumps of gluon field, there exist more complicated helicity-flip 
processes, with the classical gluon field of instantons radiated from quark vertices, 
see Fig. 9. That kind of amplitudes have been studied in relation to baryon number 
violating processes in electroweak theory \cite{barviol,DPol}.

The simplest helicity-flip amplitude induced by instantons is
presented in Fig. 9a. Its first term in the expansion in the gluon
field is the quark {\em anomalous chromomagnetic moment} first introduced
by Kochelev \cite{Koch}:

\beq
\la{Lagr}
{\cal L}^{{\rm magn}}
=-i\,\frac{g\mu}{2M}\,(\bar\psi\,\sigma_{\mu\nu}\,t^a\,\psi)\,G^a_{\mu\nu},
\qquad \sigma_{\mu\nu}=\frac{1}{2}[\gamma_\mu\gamma_\nu],
\eeq
where $G^a_{\mu\nu}$ is the gluon field strength, $\psi$ is
the quark field, $M=M(0)$ is the dynamical quark mass at zero
virtuality and $\mu$ is the anomalous chromomagnetic moment induced by instantons,

%%%%%%%%%% % FIGURE 9 %%%%%%%%%%
\begin{figure}
%%\vspace*{}
\centerline{\epsfxsize8.0cm\epsffile{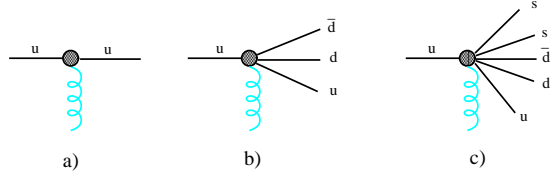}}
%%\vspace*{}
\caption[]{Helicity-flip quark-gluon vertices.}
\end{figure}

\beq
\la{anommom}
\mu=-\frac{1}{2}\,\frac{2\pi}{\alpha_s}\,(M\bar\rho)^2\,
\frac{N_c}{N_c^2-1}\approx -0.744.
\eeq

Although we agree qualitatively with ref. \cite{Koch}, our formula for
$\mu$ is different and the value is several times bigger because of 
a too crude estimate used there. In our estimates we use the
formalism and the instanton parameters of refs. \cite{Sh1,DP2b,DPW}:
$M\approx 345\,{\rm MeV}, \bar\rho^{-1}\approx 600\,{\rm MeV},
2\pi/\alpha_s\approx 12,\, n\approx (200\,{\rm MeV})^4$, where $n=N/V$ 
is the instanton density. We note that $\mu$ is stable at large number of
colours, which is also the property of the perturbation theory. The
fact that $\mu$ is large is not accidental but follows from the relation

\beq 
4\pi^2n\bar\rho^4\frac{2\pi}{\alpha_s}\frac{1}{N_c^2-1}\sim 1
\la{unity}\eeq
being the condition for the instanton ensemble to stabilize itself \cite{DPW}.

The chromomagnetic vertex \ur{Lagr} can be obtained from more general
instanton-induced vertices of Fig.~9b,c by integrating out quark
fields. The general vertex in case of $N_f$ flavours is
\bea
\n
{\cal L}^{{\rm magn}}\!
&\!=\!&\!-i\frac{g\mu}{2M}\!\left(\frac{2M}{n}\right)^{N_f-1}\!
\left(\!\bar\psi_f\sigma_{\mu\nu}\frac{1+\gamma_5}{2}t^a\psi_g\!\right)\!
J_+^{\prime\,fg}\\
\la{LagranyNf}
&+& (\gamma_5\to -\gamma_5)
\eea
where $J_\pm^{\prime\,fg}$ denotes the $(fg)$ minor of the flavour matrix
$J^{fg}_{\pm}=\bar\psi_f \frac{1\pm \gamma_5}{2}\psi_g$. The summation
over flavour indices $f,g$ and colour is understood.
After bosonization at low momenta \cite{DP3,D3} the general $2N_f$-fermion vertex
with an additional gluon \ur{LagranyNf} may be presented in a multi-Goldstone form
\cite{Balla}:

\beq
{\cal L}^{{\rm magn}}
=-i\,\frac{g\mu}{2M} \bar\psi_f\sigma_{\mu\nu}
\left[e^{i \pi^A \lambda^A \gamma_5 / F_\pi} \right]_{fg}t^a\psi_g
\,G^a_{\mu\nu}.
\la{LagranyNfpion}\eeq
In this form the vertex is explicitly chirally invariant.

The large anomalous chromomagnetic moment gives rise to a sizable
(possibly dominant) contribution to the hadron-hadron scattering at
high energies. We would like to stress that the quark-gluon vertex
\ur{Lagr} is gauge-invariant -- in sharp contrast to the usual
perturbative vertex $\bar\psi\gamma_\mu A_\mu\psi$. To get a gauge-invariant
result, that vertex has to be inserted in all possible
quark lines belonging to a `colourless' hadron. In other words, the
$\gamma_\mu$ vertex interacts with the quarks' colour charge, and one
must take into account that hadrons have zero colour charge. That is why the
Low--Nussinov--BFKL pomeron is sensitive to the size of hadrons. For the
chromomagnetic vertex,  the colour interference is not inevitable. Moreover,
it is suppressed  as $\sim 1/(k_\perp R)$ where $R$ is the hadron size
and $k_\perp$  is the typical momentum transfer through instantons,
which is about  $1\,{\rm GeV}$, as we shall see. Therefore, one can
neglect, to the zero approximation in this small parameter, the
interference of diagrams where chromomagnetic vertices are attached to
different quarks in the amplitude and in the conjugate amplitude. This
immediately leads to the conclusion  that hadron cross sections are
incoherent sums of the quark-quark cross  sections, and that is the
basis of the additive quark model.

%%%%%%%%%%
% FIGURE 10
%%%%%%%%%%
\begin{figure}
%%\vspace*{}
\centerline{\epsfxsize8.0cm\epsffile{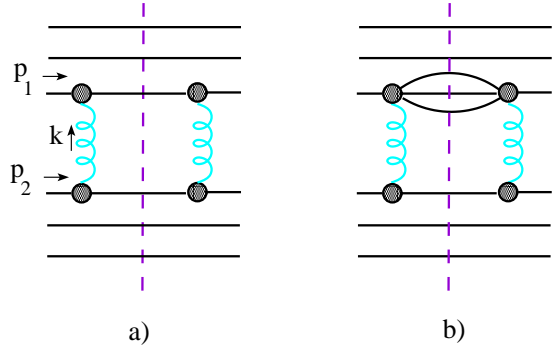}} %%\vspace*{}
\caption[]{Leading contributions to the inelastic cross section at intermediate energies.}
\end{figure}

Let us estimate the imaginary part of the forward scattering amplitude
corresponding to the simplest diagram of Fig. 10a. The gluon exchange
in the $t$-channel is in fact a complicated object as it corresponds to the
interaction of two coherent classical gluon fields of (anti)instantons in a
non-linear Yang--Mills theory. The stringy line in Fig.~10 is actually the
connected part of the correlation function
$<G^a_{\mu\nu}(x)G^b_{\rho\sigma}(y)>$ where both field strengths are
superpositions of (anti) instantons belonging to the two vertices.

Assuming the sum Ansatz for the two instanton fields the correlation function
can be directly calculated at all $k$ using formulas from ref. \cite{DP1}. 
A cleaner way to calculate it would be to use the unitarity and analiticity 
method of refs. \cite{DPol,DP4} but that has not been done. For an estimate, 
we use the simplest model, replacing each field strength just by the derivative of the
instanton potential. The model can be conveniently written in the form of the
anomalous chromomagnetic vertex supplemented by the instanton
formfactor:
\beq
{\rm
Magnetic\;vertex}=\frac{g\mu}{M}\,A(k\rho)\,\sigma_{\mu\nu}k_\nu\,t^a,
\la{mv}\eeq
\beq
A(z)=2\left[\frac{2}{z^2}-K_0(z)-\frac{2}{z}K_1(z)\right],
\la{ff}\eeq
where $K(z)$ are the modified Bessel functions, $A(0)=1$.
The fact that there must be some formfactor is model-independent as one
cannot transfer momenta larger than $1/\rho$ to a lump of coherent field of
size $\rho$ without destroying it. 

Using the optical theorem, $\sigma_{\rm tot}={\rm Im}\,A(s,t\!=\!0)/s$,
and neglecting the constituent quark masses at large energies $s$ we estimate
the total $qq$ cross section arising from Fig.~10a as \footnote{More precisely,
the cut in Fig.~10a corresponds to the inelastic cross section.}  

\beq
\sigma^{\rm qq}_{\rm tot}=\frac{1}{(2s)^2}\left(\frac{g\mu}{M}\right)^4
\int\!\frac{d^2k_\perp}{(2\pi)^2}
\,\left(\frac{A^2(k_\perp\rho)}{k^2_\perp}\right)^2\cdot 4s^2k^4_\perp,
\la{s1}\eeq
where the last factor comes from contracting the spin traces
over unpolarized quarks,

\beq
\half\Tr[\Dpp_1\sigma_{\mu\nu}k_\nu(\Dpp_1+\Dk)
\sigma_{\alpha\beta}k_\beta]
\cdot
\half\Tr[\Dpp_2\sigma_{\mu\kappa}k_\kappa(\Dpp_2-\Dk)
\sigma_{\alpha\lambda}k_\lambda],
\la{spur}\eeq
and using $s\gg M^2, k^2$ and $k^2\simeq -k^2_\perp$. We see that
the cross section is constant in energy which is of course due to the spin 1
exchange. Despite the additional factors of $k_\alpha$ in magnetic
vertices as compared to the $\gamma_\mu$ ones, integration over
$k_\perp$ converges at $k_\perp\sim 1/\rho$ owing to the instanton formfactor
which we have modelled by the instanton field itself.

\Eq{s1} should be multiplied by the colour factor $\frac{N_c^2}{4(N_c^2-1)}$
arising from contracting twice the colour matrices $t^a$ in a colourless meson
or baryon. It will then become the `true' quark-quark cross section which one
can multiply by 4, $2N_c$ or $N_c^2$ to get the meson-meson,
meson-baryon or baryon-baryon cross sections, respectively.

Integrating over $k_\perp$ we obtain

\beq
\sigma^{\rm qq}_{\rm tot}=\frac{1}{2\pi\rho^2}\left(\frac{g\mu}{M}\right)^4
\frac{N_c^2}{4(N_c^2-1)}\cdot 0.572
\approx 2.73\,{\rm mb}.
\la{s2}\eeq
The last (numerical) factor is the dimensionless integral over the specific
instanton formfactor \ur{ff}; it is of the order of unity but its concrete
value may change in a more precise theory. Also, one has to add the
multi-quark processes of Fig.~9b,c leading to Fig.~10b and the kind.
Although the cross section of Fig.~10b is $N_c=3$ times suppressed as
compared to that of Fig.~10a, a 30\%-correction squared moves the
estimate \ur{s2} close to the $qq$ cross section of $4\,{\rm mb}$
needed to explain the $36\,{\rm mb}$ of the inelastic $pp$ and $\bar pp$
cross sections, before they start to rise. 

In a wide range of $\sqrt{s}\sim 5-50\,{\rm GeV}$ where the total
cross sections are approximately constant and the rapidity plateau is
not fully developed, the inelastic collisions are presumably dominated
by the decays of hadrons whose constituents experience the $qq$
scattering shown in Fig.~10. At higher energies, the rapidity plateau
develops, possibly due to the instanton-ladder mechanism of ref.
\cite{instpom}, and that is accompanied by a weak (logarithmic or
small-power) increase of the cross section. As shown in ref.
\cite{DP4}, the point-like instanton-induced cross section for
multi-gluon production rises at low and decreases at high invariant masses;
the interpolation of the two curves inside the intermediate mass range
gives the position and the height of the maximum suspiciously close to 
what corresponds to the  production of the spherically  symmetric `sphaleron'
configuration of gluons with the invariant mass $m_{\rm sph}$  such that~\cite{DP4}

\beq
\frac{\alpha_s}{2\pi}m_{\rm sph}\,\rho=\frac{3}{8},
\la{sphal}\eeq
from where one can estimate $m_{\rm sph}\approx 2.7\,{\rm GeV}$
and predict its decay spectrum. A possible relevance of sphaleron production
in high-energy heavy ion collisions has been stressed by Shuryak \cite{ShHI}.
However, it is not evident that the same sphaleron production survives in the ladder
kinematics in the $pp$ collisions as assumed in ref. \cite{instpom}, because of the possible
gluon interference which has not been fully investigated. Whatever the mechanism of the
plateau formation at very high energies, it gives a relatively weak rise of the total cross
section -- on top of what? This section attempts to answer this basic question.

To get the elastic hadron cross section or, more generally, the
diffractive dissociation, one needs to square the diagrams of Fig.~10
multiplying it by the probability that one or both colliding hadrons
collect back the constituent quarks that have been struck. Since the
size of hadrons is much larger than the range $\bar\rho$ of the
constituent quarks' interaction, one can roughly consider it as
point-like.  Therefore, the probability that a hadron remains one
hadron after a $qq$ collision is basically given by the
same formfactor as in a true point-like electromagnetic  collision. We
get, thus, a natural explanation why the differential elastic cross
section is proportional to the fourth power of the electromagnetic
formfactor \cite{FF}. Deviation from this remarkable rule is
expected at momentum transfers greater than $\sim 0.6\,{\rm GeV}$, and
indeed the elastic cross section is decreasing faster than $F^4(t)$ at
large $t$. \\

We now briefly discuss the ensueing spin effects. Despite the unusual
spin structure of the gluon exchange the dominant structure of the
pomeron as a whole has the usual $\Dpp_2\otimes \Dpp_1$ form.
However, there is always an interference between helicity-flip and
non-flip amplitudes. A large energy-independent non-flip amplitude is,
again, induced by instantons, this time by the part of the fermion
propagator in the instanton background that does not contain chiral
zero modes \cite{BCCL}. Estimating the corresponding non-flip vertex on the
quark `mass shell', $p^2\simeq M^2$, we get $\sim g(\gamma_\mu t^a)\cdot
4\pi^2n\bar\rho^4(2\pi/\alpha_s)/N_c^2$ which is of the same order as
the perturbative vertex owing to the estimate \ur{unity} but
effectively $(M\bar\rho)^2\approx 1/3$ of the magnetic vertex
\ur{Lagr}. It is also accompanied by some formfactor $B(k\rho)$.

The enhancement of spin-flip amplitudes as compared to the
non-flip ones is of a general nature: The former are related to the
{\em spontaneous} symmetry breaking and are therefore {\em nonanalytic} 
in the (small) instanton density, namely they are proportional to the
{\em square root} of the density $N/V$ \cite{DP2b}, whereas the latter,
being `normal', are naturally linear in the density. It suggests that at small
momentum transfer the spin-flip chromomagnetic exchange may be quite large, 
if not dominant. The above estimate of the total cross section demonstrates
that it can well explain the bulk of the total cross section. 

At the same time, the non-flip is not negligible, and
the interference leads to sizable and energy-independent polarization
effects. Speaking generally, we see that as much as a
$(M\bar\rho)^2\approx 30\%$ polarization of quarks can be expected in
experiments with unpolarized hadrons.

As correctly stressed by Soffer \cite{Soffer}, the polarization is a
purely quantum-mechanical effect: it is nonzero only if the relative
phase of the spin-flip and non-flip amplitudes is nonzero. Therefore,
one has to go beyond the Born approximation of Fig.~10. The first
diagram leading to quark polarization is with {\em three} gluon exchanges. 
If one takes one of the three vertices to be `electric' (i.e. spin non-flip)
and two `magnetic' (i.e. spin-flip) such a diagram will represent an
interference between a real (Born) spin-flip amplitude and almost purely imaginary 
(box) non-flip amplitude, or {\it vice versa}, so that the relative phase is maximal
possible.  

We estimate the quark polarization arising from three-gluon exchange diagrams to be

\beq
P=\pi\,|p_\perp|\bar\rho\, C(p_\perp\bar\rho),
\la{pol}\eeq
where $p_\perp$ is the transverse momentum of the quark whose
polarization is studied and $C$ is a combination of instanton formfactors
$A,B$ and their integrals. At present we have only a model-dependent
$C(p_\perp\bar\rho)$ which is not very useful. However, the message is clear:
The polarization rises  linearly in $p_\perp$, is independent of energy and at
$p_\perp\sim  0.6\,{\rm GeV}$ reaches its maximal value of more than 10\%. 
Qualitatively, it seems  to describe correctly the main striking features of the 
polarization experiments \cite{revpol}.

If a quark of new flavour (not present in the incoming hadron) is
produced, like in the $pp\to\Lambda$ polarization experiment, it goes
via the many-flavour version of the chromomagnetic coupling shown in
Fig.~9b,c. In the bosonized language of \eq{LagranyNfpion}, it means
that one must insert twice the $\gamma_5$ matrix in the quark line,
which inverses the sign of the polarization with respect to the case
when flavour is not changed! This observation may explain why $\Lambda$
polarizations in $K^-p$ and $pp$ production have opposite sign, and
possibly other mysteries of the polarization / analysing power zoo
\cite{revpol,Soffer}. \\

Finally, it is natural to assume that the same process that dominates
soft collisions is responsible for the gluon distribution in hadrons,
at least at low virtuality $Q\approx 0.6\,{\rm GeV}$ where the hard
gluon bremsstrahlung has not yet set in.  Substituting the standard DGLAP
calculation by that with the anomalous magnetic vertex \ur{mv} we
obtain the gluon distribution  in the constituent quark as function of
Bjorken's $x$:
\bea
\n
g(x)&=&
\frac{1}{8\pi^2}\frac{N_c^2-1}{2N_c}\left(\frac{g\mu}{M}\right)^2
\frac{1}{x}\int\!dk_\perp\,k_\perp\,A^2\left(\frac{k_\perp\rho}{\sqrt{1-x}}
\right)\\
\la{gx}
&=&\frac{1-x}{x}\,\mu\cdot 2
\approx 1.5 \frac{1-x}{x},
\eea
the polarized gluon distribution $\Delta g(x)$ being nonsingular at
small $x$ and additionally suppressed as $(M\bar\rho)^2$. Taken
literally, \eq{gx} says that 75\% of the nucleon momentum is carried by
glue, which is a factor of 2 too much, but that factor arises from the
model instanton formfactor used here and may be an artifact. The $x$-dependence
of the gluon distrubution is more interesting. We get the $1/x$ behaviour at small $x$,
which is consistent with phenomenological fits at low virtuality. \\

To conclude this section: Knowing the microscopic mechanism of the SCSB 
one can reveal the non-perturbative glue inside the constituent quark. 
The largest coupling of the `soft' glue to the massive constituent quark 
appears to be via its anomalous chromomagnetic moment. Therefore, the 
magnetic gluon exchange is expected to be large if not dominant in high energy quark
scattering. It seems that this picture is supported by many facts we know about high energy
collisions. \\

I would like to remark that very recently an interesting suggestion on the role of
instantons in the transition from hard to soft physics at small x has been made in 
refs. \cite{S,SU}.  Instantons can induce specific events in deep inelastic scattering
\cite{DIS} which have been recently studied experimentally \cite{HERA}.

%%%%%%%%%%%%%%
%% FIGURE 11 %%
%%%%%%%%%%%%%%
\begin{figure}[t]
\hskip 4true cm 
\epsfig{file=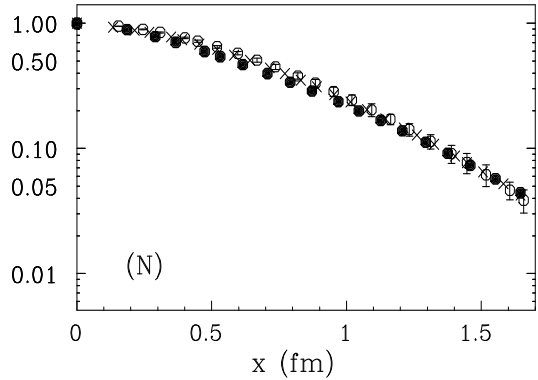,height=5cm,width=7cm}
% \vspace{-.25cm}
\caption{Density-density correlation function
in the nucleon \cite{CGHN}. Filled circles are measurements in the
full gluon vacuum (corresponding to Fig. 3a,b) while open circles
are measured in the vacuum with instantons only (Fig. 3c,d).
Despite that linear confining potential is absent in the instanton
vacuum the nucleon structure seems to be very well reproduced.
\label{Ncorr}}
\end{figure}

%\vskip 1.5true cm 

\section{Baryons}
\setcounter{equation}{0}
\def\theequation{\arabic{section}.\arabic{equation}}  

There is a remarkable evidence of the importance of instantons for the baryon structure. In
ref. \cite{CGHN} the so-called density-density correlation function inside the nucleon has
been measured on the lattice both in the full vacuum and in the instanton vacuum resulting
from the full one by means of cooling. The correlation in question is between the
densities of $u$ and $d$ quarks separated by a distance $x$ inside the
nucleon which is created at some time and annihilated at a later time.
The two correlators (`full' and `instanton') are depicted in Fig. 11: one
observes a remarkable agreement between the two, up to $x=1.7\,{\rm fm}$.

It must be stressed that neither the one-gluon exchange nor the
linear confining potential present in the full gluon vacuum
survive the smoothing of the gluon field shown in the lower part of Fig. 3.
Nevertheless, quark correlations in the nucleon remain basically
unaltered! It means that neither the one-gluon exchange nor the
linear confining potential are important for the quark binding
inside the nucleon. As a matter of fact, the same remark can be
addressed to the lightest mesons $\pi$ and $\rho$ since the
density-density correlators for these hadrons also remain basically
unchanged as one goes from the full glue to the reduced instanton
vacuum \cite{CGHN}. Therefore, one must be able to explain at least
the lightest $\pi,\rho,N$ on the basis of instantons only.

The dynamics remaining in the instanton vacuum is the SCSB, the
appearance of the dynamical quark mass $M(p)$, and quark interactions
induced by the possibility that they scatter off the same instanton.
Actualy these interactions named after 't Hooft, are quite strong,
see section 5. They are in fact so strong that for quark and antiquark
in the pion channel they eat up the $700\,{\rm MeV}$ of twice the
constituent quark mass to nil, as required by the Goldstone theorem.
In the vector meson channel 't~Hooft interactions are suppressed,
and that is why the $\rho$ mass is roughly twice the constituent quark
mass. In the nucleon they are fully at work but in a rather peculiar way:
instanton-induced interactions can be iterated as many times as one
wishes in the exchanges between quarks, see Fig. 12, left. It can be easily
verified that the diagram in Fig. 12, left, can be drawn as three continuous
quark lines going from the l.h.s of the diagram to its r.h.s., without
adding closed loops. Therefore, that kind of interaction arises
already in the so-called quenched approximation. At the same time, it
yields plenty of Z-graphs absent in ``valence QCD'' but which are
necessary to reproduce hadron properties \cite{Liu}.

%%%%%%%%%%%%%%%%%%%%%%%%%%
% FIGURE 12               %
%%%%%%%%%%%%%%%%%%%%%%%%%%

\begin{figure}[t]
%\hspace{2cm}
\epsfig{file=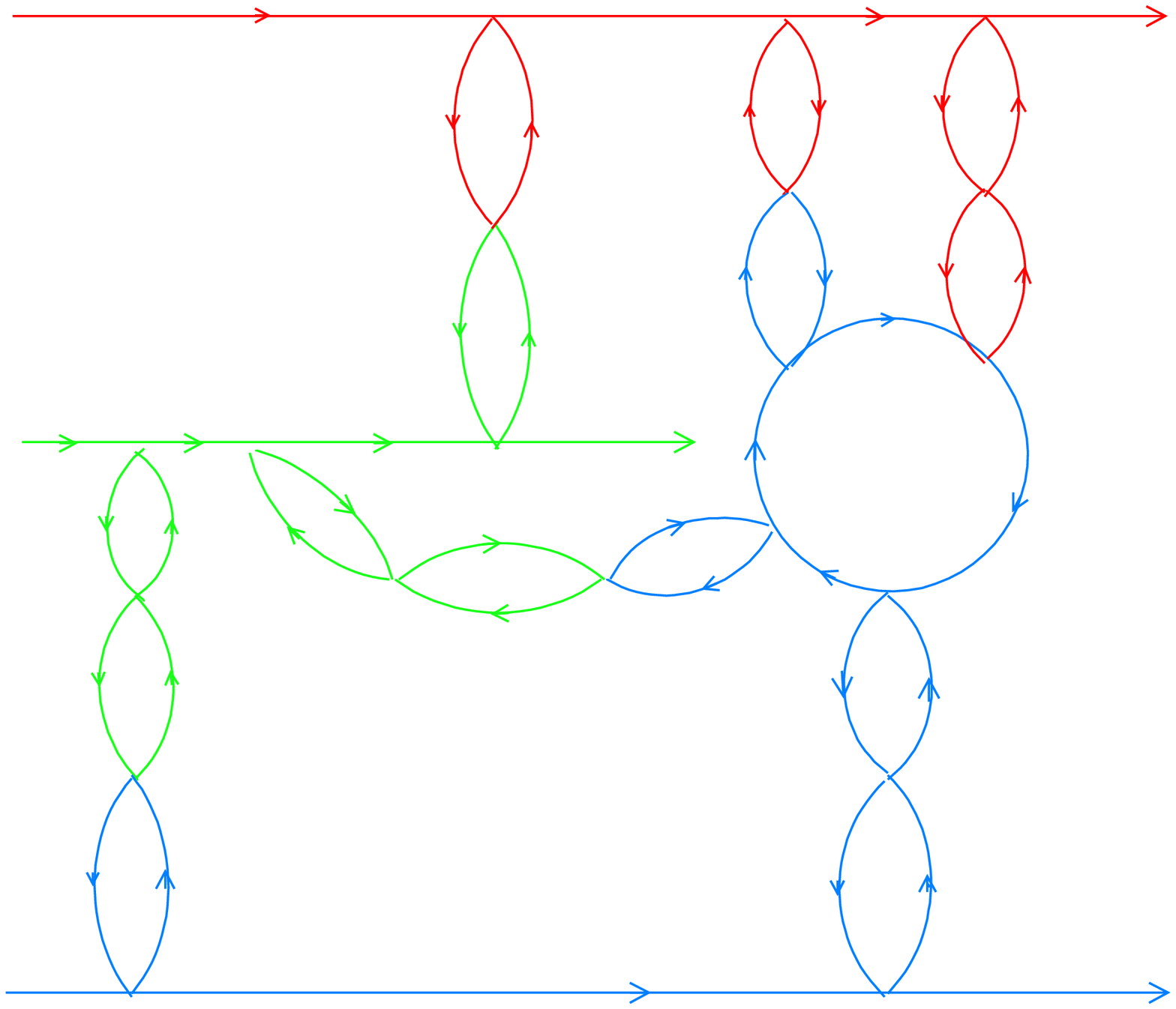,height=6cm,width=7cm}
% %\vskip -2.2true cm
\hspace{.3cm}
\epsfig{file=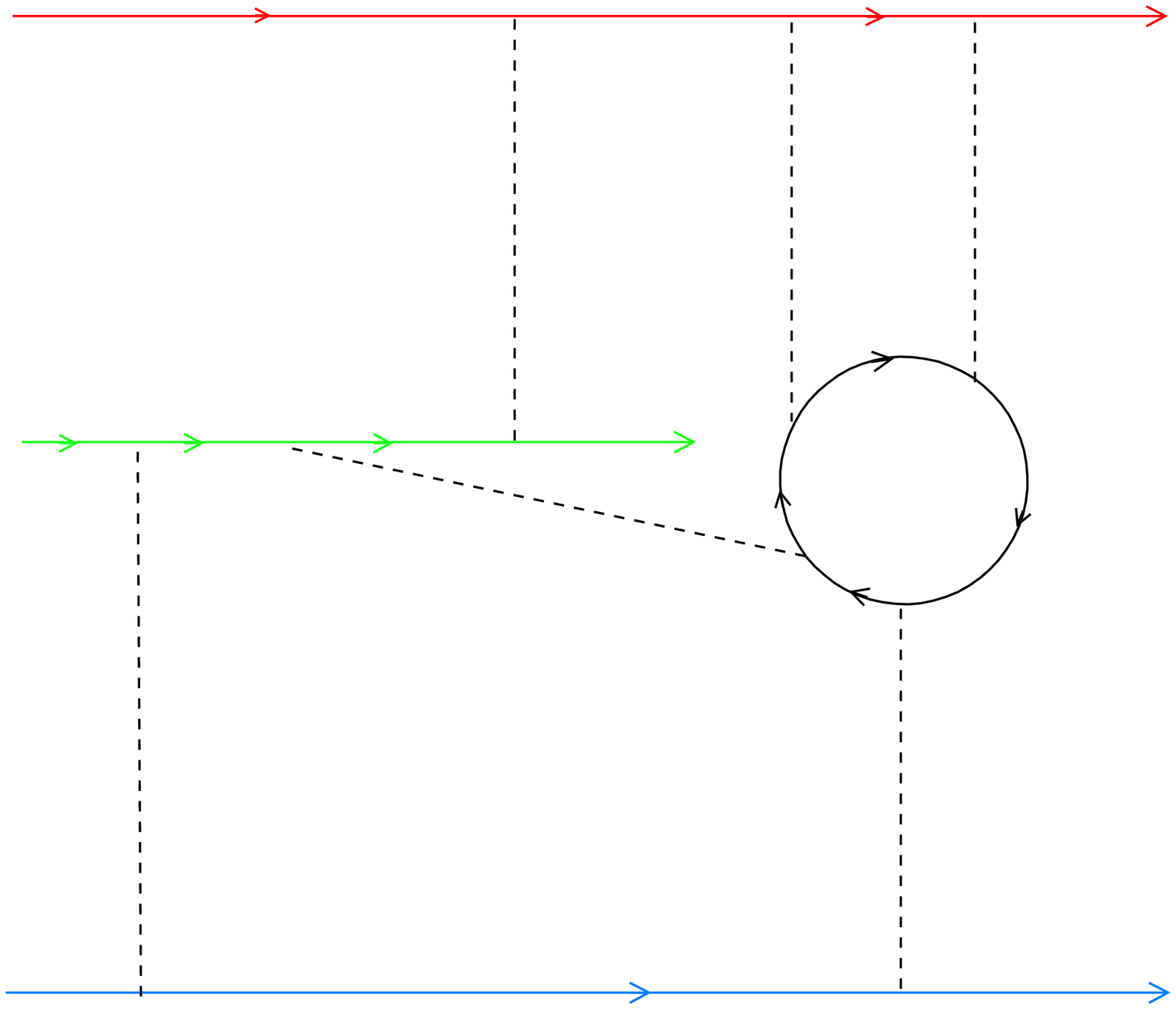,height=6cm,width=7cm}
% \vspace{-.88cm}
\caption{'t Hooft interactions in the nucleon (left)
essentially come to quarks interacting via pion fields (right).
\label{nucl2}}
\end{figure}

Summing up all interactions of the kind shown in Fig. 12, left, seems
to be a hopeless task. Nevertheless, the nucleon binding problem can be
solved {\em exactly} when two simplifications are used. The first exploits
the fact that in the instanton vacuum there are two lightest degrees
of freedom: pions (since they are the Goldstone bosons) and quarks
with the dynamical mass $M$. All the rest collective excitations of
the instanton vacuum are much heavier, and one may wish to neglect them.
Pions arise from summing up the $q\bar q$ bubbles schematically shown
in Fig. 12, left. The resulting effective low-energy theory takes the
form of the non-linear $\sigma$-model introduced in section 6:

\beq
{\cal L}_{\rm eff}=\bar q\,
\left[i\Dd-M\exp(i\gamma_5\pi^A\tau^A/F_\pi)\right]\,q.
\la{d2}\eeq
The absence of the explicit kinetic energy term for pions (which
would lead to the double counting) distinguishes it from the
Manohar--Georgi model \cite{MG}. Expanding the exponent to the
first power in $\pi^A$ we find that the dimensionless pion--constituent
quark coupling,

\beq
g_{\pi qq}=\frac{M}{F_\pi}\approx 4,
\la{coupl}\eeq
is quite strong. The domain of applicability of the low-energy
effective theory \ur{d2} is restricted by momenta
$k<1/\bar\rho=600\,{\rm MeV}$, which is the inverse size of constituent
quarks. At higher momenta one starts to feel the internal structure of
constituent quarks, and the two lightest degrees of freedom of \eq{d2}
become insufficient. However, the expected typical momenta of quarks in
the nucleon are of the order of $M\approx 345\;{\rm MeV}$, which is inside
the domain of applicability of the low-momentum effective theory.

The chiral interactions of constituent quarks in the nucleon,
following from the effective theory \ur{d2}, are schematically shown
in Fig. 12, right, where quarks are denoted by lines with arrows. Notice that,
since there is no explicit kinetic energy for pions in \eq{d2}, the pion
propagates only through quark loops. Quark loops induce also many-quark
interactions indicated in Fig. 12 as well. We see that the emerging picture is
rather far from a simple one-pion exchange between the constituent quarks: the
non-linear effects in the pion field are essential.

The second simplification is achieved in the limit
of large $N_c$. For $N_c$ colours the number of constituent quarks in a
baryon is $N_c$ and all quark loop contributions are also proportional
to $N_c$. Therefore, at large $N_c$ one can speak about a
{\em classical self-consistent pion field} inside the nucleon:
quantum fluctuations about the classical field are suppressed as $1/N_c$.
The problem of summing up all diagrams of the type shown in Fig. 12
is thus reduced to finding a classical pion field pulling $N_c$ massive
quarks together to form a bound state.

%\vskip 1.5true cm 

\section{Chiral Quark--Soliton Model}
\setcounter{equation}{0}
\def\theequation{\arabic{section}.\arabic{equation}}  

Let us imagine  a classical time-independent pion
field which is strong and spatially wide enough to form a bound-state
level in the Dirac equation following from \eq{d2}. The background
chiral field is colour-neutral, so one can put $N_c$ quarks
on the same level in an antisymmetric state in colour, i.e.
in a colour-singlet state. Thus we obtain a baryon state, as compared
to the vacuum.

One has to pay for the creation of this trial pion field, however.
Since there are no terms depending directly on the pion field in the
low-momentum theory \ur{d2} the energy of the pion field is actually
encoded in the shift of the lower negative-energy Dirac sea of quarks,
as compared to the free case with zero pion field. The baryon mass
is the sum of the bound-state energy and of the aggregate energy
of the lower Dirac sea, see Fig. 13. It is a functional of the trial pion field;
one has to minimize it with respect to that field to find the
self-consistent pion field that binds quarks inside a baryon. It is
a clean-cut problem, and can be solved numerically or, approximately,
analytically.  The description of baryons based on this construction
has been named the Chiral Quark--Soliton Model (CQSM) \cite{KRS,BB,DP5}.

%%%%%%%%%%%%%%%%%%%%%%%%%%
% FIGURE 13               %
%%%%%%%%%%%%%%%%%%%%%%%%%%

\begin{figure}[t]
\hspace{4cm}\epsfig{file=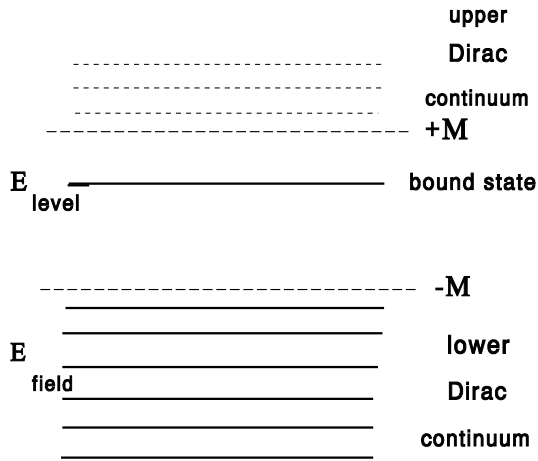,height=6cm,width=7cm}
% %\vskip -2.2true cm
\caption{The nucleon mass is a sum of the energy of three `valence' quarks 
$E_{\rm level}$ and of the infinite number of `sea' quarks whose energy is the aggregate
energy of the lower Dirac continuum, $E_{\rm field}$. Solid lines are occupied states in the
pion background field, each $N_c$ times degenerate.}
\label{nucfig3}
\end{figure}

The model reminds the large-$Z$ Thomas--Fermi atom where
$N_c$ plays the role of $Z$. Fortunately, corrections to the model go as
$1/N_c$ or even as $1/N_c^2$ and have been computed for many observables.
In the Thomas--Fermi model of atoms corrections to the self-consistent
(electric) field are of the order of $1/\sqrt{Z}$ and for that reason
are large unless atoms are very heavy.

In the end of the 80's and the beginning of the 90's dozens of baryon
characteristics have been computed in the CQSM, including masses,
magnetic moments, axial constants, formfactors, splittings inside the
mutliplets and between multiplets, polarizability, fraction of nucleon
spin carried by quarks, etc. --
see \cite{Review,D3,Rip,DP6} for reviews and references therein. 

Starting from 1996 a new class  of problems have been addressed, namely 
parton distributions in the
nucleon at low virtuality \cite{SF}. Parton distributions are a snapshot of
the nucleon in the infinite momentum frame. One needs an inherently
relativistic model in order to describe them consistently. For
example, a bag model or any other nonrelativistic model with three quarks in a
bound state, being naively boosted to the infinite-momentum frame gives a
{\it negative} distribution of antiquarks, which is nonsense. On the contrary,
being a relativistic field-theoretic model CQSM predicts parton distributions
that satisfy all general requirements known in full QCD, like positivity and
sum rules constraints.

Numerous parton distributions have been computed in the CQSM, mainly
by the Bochum group \cite{SF,PPGWW,DGPW}, see also 
\cite{WGRS,RA,Wak} and references therein. There have been a number of
mysteries from naive quark models' point of view: the large number of
antiquarks already at a low virtuality, the `spin crisis' \cite{WY}, the large flavour
asymmetry of {\it anti}quarks, etc. The CQSM explains all those `mysteries'
in a natural way as it incorporates, together with valence quarks bound by the
isospin-1 pion field, the negative-energy Dirac sea. Furthermore, the CQSM
predicts nontrivial phenomena that have not been observed so far: large
flavour asymmetry of the {\em polarized} antiquarks \cite{DGPW}, transversity
distributions \cite{PP},  peculiar shapes of the so-called skewed parton
distributions \cite{PPPBGW} and other phenomena in hard exclusive
reactions~\cite{GPV}. Baryon dynamics is rich and far from naive
``three quarks'' expectations. \\

Finally, I would like to stress that physical quantities discussed in this
review (and many more remaining beyond) are all expressed, via the
`transmutation of dimensions', through the only dimensionful quantity
there is in QCD, that is $\Lambda$, as it should be in the ultimate theory. 
Instantons serve as a bridge from the first principles of QCD -- via the microscopic
mechanism of spontaneous chiral symmetry breaking -- to the observables. Once
the basic properties of the instanton vacuum are established one does
not need to play with parameter fitting, going from one observable to another.
Instantons provide a coherent picture of very different phenomena in strong interactions, 
and I find it quite remarkable. \\

I am most grateful to Victor Petrov, Pavel Pobylitsa and Maxim Polyakov for a long-time
collaboration on topics presented here.  
 
\vskip 1true cm

\section{Summary}

\hskip .7true cm
1. The would-be linear confining potential of the pure glue
world is necessarily screened by pion production at very moderate separations
between quarks. Therefore, light hadrons need not be sensitive to
confinement forces but rather to the dynamics of the spontaneous chiral
symmetry breaking (SCSB). \\

2. Very likely, the SCSB is driven by instantons -- large non-perturbative
fluctuations of the gluon field having the meaning of tunneling. The SCSB is
due to `hopping' of quarks from one randomly situated instanton to another,
each time flipping the helicity. The instanton theory of the SCSB is in
agreement with the low-energy phenomenology ({\it cf.} the chiral condensate
$<\bar\psi\psi>$, the dynamical quark mass $M(p)$, $F_\pi$, $m_{\eta^\prime}$...)
and seems to be confirmed by direct lattice methods.
Furthermore, lattice simulations indicate that instantons alone are
responsible for the properties of lightest hadrons $\pi,\rho,N,...$. \\

3. Instantons induce not only very strong non-perturbative quark interactions
but also new and interesting vertices with an additional gluon emission. In
particular, they induce a large anomalous chromomagnetic moment which
can play an important role in soft high-energy hadron scattering, {\it e.g.}
in spin phenomena. \\

4. Summing up instanton-induced quark interactions in baryons
leads to the Chiral Quark--Soliton Model where baryons appear to
be bound states of constituent quarks pulled together by the
chiral field. The model enables one to compute numerous parton distributions,
as well as `static' characteristics of baryons -- with no fitting parameters
whatsoever. \\

5. For highly excited baryons ($m\!=\!1.5\!-\!3\,{\rm GeV}$) the relative
importance of confining forces {\it vs.} those of the SCSB may be reversed.
One can view a large-spin $J$ resonance as due to a short-time stretch of an
unstable string or, alternatively, as a rotating elongated pion cloud
\cite{DPRegge}. What picture is more adequate is a question to experiment. In the
first case the dominant decay is on the average of the type ${\rm Bar}_{J}\to
{\rm Bar}_{\sim J/2} +{\rm Mes}_{\sim J/2}$; in the second case it is
mainly a cascade ${\rm Bar}_{J}\to {\rm Bar}_{J-1}+\pi
\to {\rm Bar}_{J-2}+\pi\pi\to ...$ Studying resonances can
elucidate the relation between chiral and confining forces.

\vskip1.5true cm


\newpage

\begin{thebibliography}{99}
\itemsep -2pt 

\bibitem{BPST}  %1
A. Belavin, A. Polyakov, A. Schwartz and Yu. Tyupkin, {\it Phys. Lett.} {\bf 59} (1975) 85

\bibitem{Pol} %2
A. Polyakov, {\it Nucl. Phys.}  B 120 (1977) 429

\bibitem{tH} %3
G. 't Hooft, {\it Phys. Rev. Lett.}  37 (1976) 8; \\ 
{\it Phys. Rev.}  D 14 (1976) 3432; Erratum: {\it ibid.}  D 18 (1978) 2199 

\bibitem{CDG} %4
C. Callan, R. Dashen and D. Gross, {\it Phys. Rev.}  D 17 (1978) 2717

\bibitem{Sh1} %5
E. Shuryak, {\it Nucl. Phys.} B 203 (1982) 93, 116, 140  

\bibitem{IMP} %6
E.M. Ilgenfritz and  M. M\"uller-Preussker, {\it Nucl. Phys.} B184 (1981) 443

\bibitem{DP1} %7
D. Diakonov and V. Petrov, {\it Nucl. Phys.} B 245 (1984) 259 

\bibitem{SVZ} %8
M. Shifman, A. Vainshtein and V. Zakharov, {\it Nucl. Phys.} B 147 (1979) 385

\bibitem{Akas} %9
D. Diakonov, {\it The U(1) problem and instantons}, in: Gauge Theories of the Eighties,
Lecture Notes in Physics, Springer-Verlag (1983) p. 207

\bibitem{DP2a} %10
D. Diakonov and V. Petrov, {\it Phys. Lett.}  B 147 (1984) 351

\bibitem{DP2b} %11
D. Diakonov and V. Petrov,  {\it Sov. Phys. JETP} 62 (1985) 204;  
{\it Sov. Phys. JETP} 62 (1985) 431; {\it Nucl. Phys.}  B 272 (1986) 457

\bibitem{DP3} %12
D. Diakonov and V. Petrov, {\it Spontaneous breaking of chiral symmetry in the instanton
vacuum}, preprint LNPI-1153 (1986), in: Hadron matter under extreme
conditions, Kiew (1986) p. 192; \\ 
D. Diakonov and V. Petrov,  in: Quark Cluster Dynamics, Lecture Notes in Physics,
Springer-Verlag (1992) p. 288

\bibitem{Pob} %13
P. Pobylitsa, {\it Phys. Lett.} B 226 (1989) 387  

\bibitem{DPP1} %14
D. Diakonov, V. Petrov and P. Pobylitsa, {\it Phys. Lett.} B 226 (1989) 372

\bibitem{D1} %15
D. Diakonov, in: Skyrmions and Anomalies, World Scientific (1987) p. 27  

\bibitem{BSW} %16
G. Bali, K. Schilling and A. Wachter, in:  Confinement 95,
eds. H. Toki {\it et al.}, World Scientific (1995) p. 82,
{\tt hep-lat/9506017}

\bibitem{KLP} %17
F. Karsch, E. Laermann and A. Peikert, {\it Nucl. Phys.}
{\bf B605}, 579 (2001), {\tt hep-lat/0012023}

\bibitem{DPRegge} %18
D. Diakonov and V. Petrov, {\it Rotating chiral solitons lie on linear
Regge trajectories}, preprint LNPI-1394 (1988); see also D. Diakonov,
{\it Acta Phys. Polon.} B 25 (1994) 17. 

\bibitem{Sh2} %19
E. Shuryak and J. Verbaarschot, {\it Nucl. Phys.}  B 410 (1993) 55;\\ 
T. Sch\"{a}fer, E. Shuryak and J. Verbaarschot, {\it Nucl. Phys.}  B 412 (1994) 143;\\ 
T. Sch\"{a}fer and E. Shuryak, {\it Phys. Rev.}  D50 (1994) 478

\bibitem{Sh3} %20
E. Shuryak, {\it Rev. Mod. Phys.} 65 (1993) 1 

\bibitem{CGHN} %21
M.-C. Chu, J. Grandy, S. Huang and J. Negele, {\it Phys. Rev. Lett.}  70 
(1993) 225; {\it Phys. Rev.}  D49 (1994) 6039;\\
J. Negele, {\it Nucl. Phys. Proc. Suppl.}  73 (1999) 92, {\tt hep-lat/9810053} 

\bibitem{Sh4} %22
T. Sch\"{a}fer and E. Shuryak, {\it Rev. Mod. Phys.} 70 (1998) 323, {\tt hep-ph/9610451} 

\bibitem{Fad} %23
L.D. Faddeev, {\it Looking for multi-dimensional solitons} in: Non-local Field Theories, Dubna
(1976)  

\bibitem{JR} %24
R. Jackiw and C. Rebbi, {\it Phys. Rev. Lett.} 37 (1976) 172  

\bibitem{JO} %25
J.D. Jackson and L.B. Okun, {\it Rev. Mod. Phys.} 73 (2001) 663, {\tt hep-ph/0012061}

\bibitem{NSVZ}  %26
V. Novikov, M. Shifman, A. Vainshtein and V. Zakharov, {\it Nucl. Phys.} B 191 (1981) 301

\bibitem{DPW} %26
D. Diakonov, M. Polyakov and C. Weiss, {\it Nucl. Phys.}  B 461 (1996) 539, 
{\tt hep-ph/9510232}

\bibitem{B} %27
C. Bernard, {\it Phys. Rev.}  D 19 (1979) 3013

\bibitem{HH}
A. Hasenfratz and P. Hasenfratz, {\it Nucl. Phys.} B 193 (1981) 210

\bibitem{VZNS} %28
A. Vainshtein, V. Zakharov, V. Novikov and M. Shifman, {\it Sov. Phys. Uspekhi} 
136 (1982) 553  

\bibitem{DPol} %29
D. Diakonov and M. Polyakov, {\it Nucl. Phys.} B 389 (1993) 109

\bibitem{DP4} %30 
D. Diakonov and V. Petrov, {\it Phys. Rev.} D 50 (1994) 266 
  
\bibitem{FFS} %31
V. Fateev, I. Frolov and A. Schwartz, {\it Nucl. Phys.} B 154 (1979) 1; \\
V. Fateev, I. Frolov and A. Schwartz, {\it Sov. J. Nucl. Phys.} 30 (4) (1979) 590

\bibitem{BL} %32
B. Berg and M. L\"uscher, {\it Comm. Math. Phys.} 69 (1979) 57  

\bibitem{DM} %33
D. Diakonov and M. Maul, {\it Nucl. Phys.} B 571 (2000) 91, {\tt hep-th/9909078}

\bibitem{deF} %34
P. de Forcrand, M. G. Perez,  I.-O. Stamatescu, {\it Nucl. Phys.} B 499 (1997) 409,
{\tt hep-lat/9701012}  

\bibitem{DGHK} %35
T. DeGrand, A. Hasenfratz and T. Kovacs, {\it Nucl. Phys.} B 505 (1997) 417,
{\tt hep-lat/9705009} 

\bibitem{DPpot} %36
D. Diakonov and V. Petrov, in: Non-perturbative approaches to Quantum
Chromodynamics, Proc. int. workshop at ECT*, Trento, 1995, ed. D. Diakonov,
Gatchina (1995) p. 239;\\
D. Diakonov and V. Petrov, {\it Phys. Scripta} 61 (2000) 536, {\tt hep-lat/9810037}

\bibitem{AFF} %37
V. de Alfaro, S. Fubini and G. Furlan, {\it Phys. Lett.} B 65 (1976) 163

\bibitem{CDG2} %38
C. Callan, R. Dashen and D. Gross, {\it Phys. Rev.}  D 19 (1979) 1826

\bibitem{KvB} %39
T.C.~Kraan and P.~van Baal, {\it Phys. Lett}  B 428 (1998) 268,
{\tt hep-th/9802049}; {\it Nucl. Phys.} B 533 (1998) 627, {\tt hep-th/9805168}

\bibitem{LL} %40
K.~Lee and C.~Lu, {\it Phys. Rev.}  D 58 (1998) 025011, {\tt hep-th/9802108}

\bibitem{Bog} %41
E.B. Bogomolnyi, {\it  Yad. Fiz.} 24 (1976) 861 [{\it  Sov. J. Nucl. Phys} 24 (1976) 449]

\bibitem{PS} %42
M.K. Prasad and C.M. Sommerfeld, {\it Phys. Rev. Lett.} 35 (1975) 760

\bibitem{tHvor} %43
G. \t Hooft, {\it Nucl. Phys.}  B 138 (1978) 1

\bibitem{Ma} %44
G. Mack, Carg{\`e}se lectures (1979)

\bibitem{MN} %45
A. Montero, J. Negele, {\it Phys. Lett.} B 533 (2002) 322, {\tt hep-lat/0202023}

\bibitem{LY} %46
K. Lee and P. Yi, {\it Phys. Rev} D 56 (1997), {\tt hep-th/9702107}

\bibitem{DPSUSY} %47
D. Diakonov and V. Petrov, {\tt hep-th/0212018}

\bibitem{DHK} %48
N.M. Davies, T.J. Hollowood and V.V. Khoze, {\tt hep-th/0006011}

\bibitem{Zar} %49
K. Zarembo, {\it Nucl. Phys.} B 463 (1996) 73, {\tt hep-th/9510031}

\bibitem{GPY} %50
D.J. Gross, R.D. Pisarski and L.G. Yaffe, {\it Rev. Mod. Phys.} 53 (1981) 43

\bibitem{NW} %51
N. Weiss, {\it Phys. Rev.} D 24 (1981) 475;  {\it Phys. Rev.} D 25 (1982) 2667

\bibitem{HS} %52
B.J. Harrington and H.K. Shepard, {\it Phys. Rev.} D 17 (1978) 2122; 
{\it Phys. Rev.} D 18 (1978) 2990

\bibitem{DMir} %53
D. Diakonov and A. Mirlin, {\it Phys. Lett.} B 203 (1988) 299

\bibitem{DHKM} %54
N.M. Davies, T.J. Hollowood, V.V. Khoze and M.P. Mattis,
{\it Nucl. Phys.} B 559 (1999) 123, {\tt hep-th/9905015}

\bibitem{NSVZsusy} %55
V.A. Novikov, M.A. Shifman, A.I. Vainshtein and V.I.~Zakharov,
{\it Nucl. Phys.} B 229 (1983), 394, 407

\bibitem{IMMPSV} %56
E.M. Ilgenfritz, B.V. Martemyanov, M. Muller-Preussker, S.~Shcheredin and A.I.~Veselov,  
{\it Phys. Rev.} D 66 (2002) 074503, {\tt hep-lat/0206004} 

\bibitem{Gatt} %57
C. Gattringer, {\tt hep-lat/0210001}; C. Gattringer and S. Schaefer, {\tt hep-lat/0212029} 

\bibitem{GA} %58
A. Gonzalez-Arroyo, in: Advanced school on non-perturbative
quantum field physics, M. Asorey and A. Dobado, eds., World Scientific (1998) p. 57,
{\tt hep-th/9807108};\\
A. Gonzalez-Arroyo, A. Montero, {\it Phys.Lett.} B 442 (1998) 273, {\tt hep-th/9809037};\\
M. Garcia Perez, A. Gonzalez-Arroyo, A. Montero, C. Pena and P. van Baal,
{\it  Nucl. Phys. Proc. Suppl.} (2000) 83. {\tt hep-lat/9909112} 

\bibitem{Green} %59
J. Greensite, submitted to {\it Prog. Part. Nucl. Phys.}, {\tt hep-lat/0301023} 
 
\bibitem{Dvort} %60
D. Diakonov, {\it Mod. Phys. Lett.} A 14 (1999) 1909, {\tt hep-th/9908069}

\bibitem{DMvort} %61
D. Diakonov and M. Maul, {\it Phys. Rev.} D 66 (2002) 096004, {\tt hep-lat/0204012}

\bibitem{Bord} %62
M. Bordag, {\tt hep-th/0211080}

\bibitem{LT} %63
B. Lucini and M. Teper, {\it JHEP} 06 (2001) 050, {\tt hep-lat/0103027};
{\it Phys. Rev.} D 64 (2001) 105019, {\tt hep-lat/0107007} 

\bibitem{DDPRV} %64
L. Del Debbio, H. Panagopoulos, P. Rossi and E. Vicari, {\it JHEP} 01 (2002) 009,
{\tt hep-th/0111090}

\bibitem{DDD} %65
L. Del Debbio and D. Diakonov, {\it Phys. Lett.} B 544 (2002) 202, {\tt hep-lat/0205015}

\bibitem{SNc} %66
T. Schafer, {\it Phys. Rev.} D 66 (2002) 076009, {\tt hep-ph/0206062},
{\tt hep-ph/0204026} 

\bibitem{Ddual} %67
D. Diakonov, invited talk at the 5th International Conference on Quark Confinement and the
Hadron Spectrum, Gargnano, Italy, Sep 2002, {\tt hep-th/0212187} 

\bibitem{BC} %68
T. Banks and A. Casher, {\it Nucl. Phys.} B 169 (1980) 103  
 
\bibitem{D2} %69
D. Diakonov, Habilitation thesis (LNPI, 1986) (unpublished); in: 
Selected Topics in Nonperturbative QCD, A. Di Giacomo and D. Diakonov, eds.,
IOS Press (1996) p.397, {\tt hep-ph/9602375}   

\bibitem{JNPZ} %70
R.A. Janik, M.A. Nowak, G. Papp, and I. Zahed, {\it Phys. Rev. Lett.} 81 (1998) 264 

\bibitem{OTV} %71
J.C. Osborn, D. Toublan and J.J. Verbaarschot, {\it Nucl. Phys.} B 540 (1999) 317

\bibitem{Now} %40
M. Nowak, {\tt hep-ph/0112296}

\bibitem{N2}  %41
J. Negele, {\it Nucl. Phys. Proc. Suppl.} 73 (1999) 92, {\tt hep-lat/9810053}

\bibitem{Da} %42
T.~DeGrand and A.~Hasenfratz,
{\it Phys. Rev.}  D 64 (2001) 034512, {\tt hep-lat/0012021}; \\
M.~G\"ockeler, P.~Rakow, S.~Schaefer, A.~Sch\"afer,
{\it Nucl.~Phys.} B 617 (2001) 101, {\tt hep-lat/0107016}; \\
T.~Blum {\it et al.}, {\it Phys.~Rev.} D 65 (2002) 014504, {\tt hep-lat/0105006}; \\
R.~G.~Edwards and U.~M.~Heller, {\it Phys.~Rev.} D 65 (2002) 014505, 
{\tt hep-lat/0105004}; \\
T.~Lippert, H.~Neff, K.~Schilling and W.~Schroers, {\it Phys.~Rev.} D 65 (2002) 014506,  
{\tt hep-lat/0105001}; \\
C.~Gattringer, {\it Phys. Rev. Lett.} 88 (2002) 221601, {\tt hep-lat/0202002}

\bibitem{Net} %43
I.~Horv\'ath {\it et al.}, {\it Phys. Rev.} D 66 (2002) 034501, {\tt hep-lat/0201008}

\bibitem{Mlat} %44
P. Bowman, U. Heller, D. Leinweber and A. Williams, submitted to 20th International
Symposium on Lattice Field Theory (LATTICE 2002), {\tt hep-lat/0209129}

\bibitem{D3}  %45
D. Diakonov, in: Advanced School on Non-perturbative Quantum Field Theory,
M.~Asorey and A.~Dobado, eds., World Scientific (1998) p. 1, {\tt hep-th/9802298}

\bibitem{W} %46
E. Witten, {\it Nucl. Phys.} B 156 (1979) 269

\bibitem{V} %47
G. Veneziano, {\it Nucl. Phys.} B 159 (1979) 231

\bibitem{NVZ} %48
M. Nowak, J. Verbaarschot and I. Zahed, {\it Nucl. Phys.} B 324 (1989) 1

\bibitem{LS}
H. Leutwyler and A. Smilga, {\it Phys. Rev.} {\bf D 46} (1992) 5607

\bibitem{DE}  %49
D. Diakonov and M. Eides, {\it Zh. Eksp. Teor. Fiz.} {\bf 81} (1981) 434 
[{\it Sov. Phys. JETP} {\bf 54} (1981) 232 

\bibitem{NJL} %50
Y. Nambu and G. Jona-Lasinio, {\it Phys. Rev. } 122 (1961) 345; 124 (1961) 246

\bibitem{Mus} %51
M. Musakhanov,  {\it Nucl. Phys.} A 699 (2002) 340, {\tt hep-ph/0206233}  

\bibitem{BLa} %52
R.G. Betman and L.V. Laperashvili, {\it Sov. J. Nucl. Phys.} 41 (1985) 295 

\bibitem{DFL} %53
D. Diakonov, H. Forkel and M. Lutz, {\it Phys. Lett.} B 373 (1996) 147,
{\tt hep-ph/9512385}

\bibitem{RSSV} %49
R. Rapp, T. Sch\"afer, E. Shuryak and M. Velkovsky, {\it Phys. Rev. Lett.} 81 (1998) 53,
{\tt hep-ph/9711396}

\bibitem{ARW} %50
M. Alford, K. Rajagopal and F. Wilczek, {\it Phys. Lett.} B 422 (1998) 247, {\tt
hep-ph/9711395}

\bibitem{CD1} %51
G. Carter and D. Diakonov, {\it Nucl. Phys.} A 642 (1998) 78; 
{\it Phys. Rev.} D 60 (1999) 016004, {\tt hep-ph/9812445}

\bibitem{CD2} %52
G. Carter and D. Diakonov, {\it Nucl. Phys.} B 582 (2000) 571, {\tt hep-ph/0001318}

\bibitem{LFLSAS} %53
E.M. Levin and L.L. Frankfurt, {\it JETP Lett.}  2 (1965) 65; \\
H.J. Lipkin and F. Scheck, {\it Phys. Rev. Lett.} 16 (1966) 71; \\
V.V. Anisovich and V.M. Shekhter, {\it Nucl. Phys.} B 55 (1973) 455

\bibitem{FF} %54
See e.g. V.V. Anisovich, M.N.Kobrinsky, J. Nyiri and Yu.M. Shabelski,
{\it Quark Model and High Energy Collisions}, World Scientific (1985), 
and references therein  

\bibitem{revpol} %55
For a review see, L.G. Pondrom, {\it Phys. Rept.}
122 (1985) 57; \\
J.~Lach, {\it Hyperon polarization: an experimental
overview}, FERMILAB-CONF-92-378

\bibitem{LN} %56
F.E. Low, {\it Phys. Rev.} D 12 (1975) 163; \\
S. Nussinov, {\it Phys. Rev. Lett.} 37 (1975) 1286

\bibitem{BFKL} %57
E.A. Kuraev, L.N. Lipatov and V.S. Fadin, {\it Sov. Phys. JETP}  45 (1978)
199; \\
Ya.Ya.~Balitsky and L.N.~Lipatov, {\it Sov. J. Nucl. Phys.} 28 (1978) 22

\bibitem{DDLN}
S. Donnachie, G. Dosch, O. Nachtmann and P. Landshoff,
{\it Pomeron physics and QCD},  Cambridge Univ. Pr. (2002) 347 p.

\bibitem{ND}  %58
O. Nachtmann, {\it Ann. Phys. (NY)} 209 (1991) 436; \\
H.~-G.~Dosch, E.~Ferreira and  A.~Kramer, {\it Phys. Lett.} B 289 (1992) 153.

\bibitem{Soffer} %59
J. Soffer, in: Batavia 1999,
{\it Hyperon Physics}, p.121-126,  {\tt hep-ph/9911373}

\bibitem{barviol} %60
A. Ringwald, {\it Nucl. Phys.} B 330 (1990) 1;\\
O. Espinosa, {\it Nucl. Phys.} B 343 (1990) 310;\\
L. McLerran, A. Vainshtein and M. Voloshin, {\it Phys. Rev.} D 42 (1990) 171,180; \\
V. Zakharov, {\it Nucl. Phys.} B 353 (1991) 683; \\
A. Mueller, {\it Nucl. Phys.} B 381 (1992) 597

\bibitem{Koch} %61
N.I. Kochelev, {\it Phys. Lett.}  B 426 (1998) 149; see also {\tt hep-ph/9707418} 

\bibitem{Balla} %62
J. Balla, M. Polyakov and C. Weiss, {\it Nucl. Phys.} B 510 (1998) 327

\bibitem{instpom} %63
E. Shuryak, {\it Phys. Lett.} B 486 (2000) 378; \\
D. Kharzeev, Yu. Kovchegov and E. Levin, {\it Nucl. Phys.} A 690 (2001) 621;\\
M. Nowak, E. Shuryak and I. Zahed, {\it Phys. Rev.}  D 64 (2001) 034008

\bibitem{ShHI}
E. Shuryak, {\tt hep-ph/0205031}

\bibitem{BCCL} %64
L.S. Brown, R.D. Carlitz, D.B. Creamer and C. Lee,
{\it Phys. Rev.} D 17 (1978) 1583

\bibitem{S} %65
F. Schrempp, {\it J. Phys.} G 28 (2002) 915, {\tt hep-ph/0109032}

\bibitem{SU} %66
F. Schrempp and A. Utermann, {\it Phys. Lett.} B 543 (2002) 197, {\tt hep-ph/0207300} 

\bibitem{DIS} %67
I. Balitsky and V. Braun, {\it Phys. Lett.} B 314 (1993) 237; {\it Phys. Rev.} D 47 (1993)
1879;\\
A. Ringwald and F. Schrempp,  in: Quarks '94, Proc. 8th Int. Seminar, Vladimir,
    Russia, 1994, ed. D. Grigoriev et al., World Scientific (1995) p. 170; \\ 
S. Moch, A. Ringwald and F. Schrempp,  {\it Nucl. Phys.}  B 507  (1997) 134,     
{\tt hep-ph/9609445}; \\ 
A. Ringwald and F. Schrempp, {\it Phys. Lett.} B 438 (1998) 217, {\tt hep-ph/9806528}; \\ 
A. Ringwald and F. Schrempp, {\it Phys. Lett.} B 503 (2001) 331, {\tt  hep-ph/0012241} 

\bibitem{HERA} %68
C. Adloff et al., {\it Eur. Phys. J.} C 25 (2002) 495,  {\tt hep-ex 0205078} 

\bibitem{Liu} %69
K.F. Liu  et al., {\it Phys. Rev.}  D 59 (1999) 112001, {\tt hep-ph/9806491}

\bibitem{MG} %70
A. Manohar and H. Georgi, {\it Nucl. Phys.}  B 234 (1984) 189

\bibitem{KRS} %71
S. Kahana, G. Ripka and V. Soni, {\it Nucl. Phys.} A 415 (1984) 351;\\
S. Kahana and G. Ripka, {\it Nucl. Phys.} A 429 (1984) 462

\bibitem{BB} %72
M.S. Birse and M.K. Banerjee, {\it Phys. Lett.} B 136 (1984) 284

\bibitem{DP5} %73
D. Diakonov and V. Petrov, {\it Sov. Phys. JETP Lett.}  43 (1986) 57; \\
D.~Diakonov, V.~Petrov and P.~Pobylitsa, in: Proc. 21st PNPI Winter
School, Leningrad (1986) p. 158; {\it Nucl. Phys.} B 306 (1988) 809

\bibitem{Review} %74
C.~Christov~ et~al.,
%, A. Blotz, H.-C. Kim, P. Pobylitsa, T. Watabe, Th. Meissner,
%E.~Ruiz~Arriola and K.~Goeke,
{\it Prog. Part. Nucl. Phys.}  37 (1996) 91, {\tt hep-ph/9604441}

\bibitem{Rip}
G. Ripka, {\it Quarks bound by chiral fields}, Clarendon Press, Oxford (1997), 222 p.

\bibitem{DP6} %75
D. Diakonov and V. Petrov, in: At the Frontiers of Particle Physics (Handbook of QCD),
ed. M. Shifman, World Scientific (2001) vol.1 p. 359, {\tt hep-ph/0009006}

\bibitem{SF} %76
D. Diakonov,  V. Petrov, P. Pobylitsa, M. Polyakov and C.~Weiss,
{\it Nucl. Phys.} B 480 (1996) 341, {\tt hep-ph/9606314};
{\it Phys. Rev.} D 56 (1997) 4069, {\tt hep-ph/9703420}

\bibitem{PPGWW} %77
P. Pobylitsa, M. Polyakov, K. Goeke, T. Watabe and C.~Weiss,
% ISOVECTOR UNPOLARIZED QUARK DISTRIBUTION IN THE
% NUCLEON IN THE LARGE N(C) LIMIT.
{\it Phys. Rev.}  D 59 (1999) 034024, {\tt hep-ph/9804436}

\bibitem{DGPW} %78
B. Dressler, K. Goeke, M. Polyakov and C.~Weiss,
% FLAVOR ASYMMETRY OF POLARIZED ANTI-QUARK DISTRIBUTIONS AND
% SEMIINCLUSIVE DIS.
{\it Eur. Phys. J.} C 14 (2000) 147, {\tt hep-ph/9909541}

\bibitem{WGRS}
H. Weigel, L. Gamberg, H. Reinhardt and  O. Schroeder,
{\it Nucl. Phys. Proc. Suppl.} (1999) 74, {\tt hep-ph/9807506}

\bibitem{RA}
E. Ruiz Arriola, in: Hadrons as solitons, Bled (1999) p. 5, {\tt hep-ph/9910382}

\bibitem{Wak}
M. Wakamatsu, {\tt hep-ph/0209011, hep-ph/0212356} 

\bibitem{WY}
M. Wakamatsu and H. Yoshiki, {\it Nucl. Phys.} A 524 (1991) 561 

\bibitem{PP} %79
P. Pobylitsa and M. Polyakov, {\it Phys. Lett.} B 389 (1996) 350, {\tt hep-ph/9608434}

\bibitem{PPPBGW} %80
V. Petrov, P. Pobylitsa, M. Polyakov, I. B\"ornig, K.~Goeke and C.~Weiss,
{\it Phys. Rev.} D 57 (1998) 4325, {\tt hep-ph/9710270}
% OFF - FORWARD QUARK DISTRIBUTIONS
% OF THE NUCLEON IN THE LARGE N(C) LIMIT.

\bibitem{GPV} %81
K. Goeke, M. Polyakov and  M. Vanderhaeghen,
% HARD EXCLUSIVE REACTIONS AND THE STRUCTURE OF HADRONS.
{\it Prog. Part. Nucl. Phys.}  47 (2001) 401, {\tt hep-ph/0106012}

\end{thebibliography}
\end{document}